\documentclass[usenatbib,usegraphicx, iop]{emulateapj}
\usepackage{amsmath}
\usepackage[]{hyperref}
\usepackage{graphicx}
\usepackage{color}
\usepackage{units}
\usepackage{ulem}
\usepackage{fixltx2e}
\DeclareGraphicsExtensions{.eps,.pdf,.png,.jpg}

\hypersetup{
    colorlinks=true,
    linkcolor=black,
    citecolor=black,
    filecolor=black,
    urlcolor=black,
}

\shorttitle{Overmassive Brown Dwarfs}
\shortauthors{Forbes and Loeb}

\begin{document}

\title{{\bf On the Existence of Brown Dwarfs More Massive than the Hydrogen Burning Limit}}

\author{John C. Forbes and Abraham Loeb}
\affil{Astronomy Department, Harvard University, 60 Garden St., Cambridge MA 02138, USA; \\ john.forbes@cfa.harvard.edu, aloeb@cfa.harvard.edu }

\begin{abstract}
Almost by definition brown dwarfs are objects with masses below the hydrogen burning limit, around $0.07 M_\odot$. Below this mass, objects never reach a steady state where they can fuse hydrogen. Here we demonstrate, in contrast to this traditional view, that brown dwarfs with masses greater than the hydrogen burning limit may in principle exist in the universe. These objects, which we term ``overmassive brown dwarfs'' form a continuous sequence with traditional brown dwarfs in any property (mass, effective temperature, radius, luminosity). To form an overmassive brown dwarf, mass must be added sufficiently slowly to a sufficiently old traditional brown dwarf below the hydrogen burning limit. We identify mass transfer in binary brown dwarf systems via Roche lobe overflow driven by gravitational waves to be the most plausible mechanism to produce the bulk of the putative overmassive brown dwarf population.
\end{abstract}

\keywords{binaries: general, brown dwarfs, stars: evolution, stars: low-mass }

\section{Introduction}

The process which forms stars out of molecular clouds produces objects over a wide range of masses, with an abundance peak around 0.2 $M_\odot$ \citep[e.g.][]{chabrier_galactic_2003}. In the course of their formation, stars contract gravitationally, heating up their interiors until the conditions are right for the fusion of hydrogen. At this point (modulo interim phases involving deuterium fusion), the contraction is halted and the star settles into a steady state structure, which changes very gradually as the star converts hydrogen to heavier elements. Below a critical mass, which we shall denote $M_\mathrm{HBL}$ for the ``hydrogen burning limit,'' this process of collapse never yields conditions for which the power generated by nuclear fusion can supply the star's surface luminosity. The core of the star becomes supported by electron degeneracy pressure, and the sub-stellar object (a brown dwarf) cools and contracts on a thermal timescale \citep{kumar_structure_1963, hayashi_evolution_1963}.

Since the first studies in the 1960s, research on brown dwarfs was re-energized in the 1990s by their observational discovery \citep{rebolo_discovery_1995}. Continuous comparison with data has led to improving evolutionary models \citep{laughlin_luminosity_1993, burrows_expanded_1993, saumon_cool_1994, baraffe_new_1995, tsuji_evolution_1996, burrows_nongray_1997, allard_model_1997, chabrier_structure_1997, chabrier_evolutionary_2000, burrows_theory_2001, baraffe_evolutionary_2003, saumon_evolution_2008, macdonald_structural_2009, baraffe_new_2015} with much of the modeling uncertainty associated with the atmospheres and the influence of clouds \citep[e.g.][]{ackerman_precipitating_2001, cushing_atmospheric_2008,  morley_neglected_2012, marley_cool_2015}.

The distinction between brown dwarfs and low-mass stars is often taken to be set by mass. There are a number of difficulties with this definition, however. First, masses are difficult to derive observationally. Second, the hydrogen burning limit has a well-known dependency on metallicity \citep[e.g.][]{chabrier_structure_1997}, with higher metallicity corresponding to a lower $M_\mathrm{HBL}$. Third, as we will show in this work, objects that behave like brown dwarfs may in principle exist with masses exceeding $M_\mathrm{HBL}$. We therefore recommend that brown dwarfs be defined more by their composition: hydrogen-rich substellar objects that do not fuse hydrogen.

Brown dwarfs with masses exceeding $M_\mathrm{HBL}$ were suggested several decades ago (variously called pristine white dwarfs or beige dwarfs) as a possible substantial contributor to dark matter \citep{salpeter_minimum_1992, hansen_origin_1999, lynden-bell_russell_2001} falling under the umbrella of Massive Compact Halo Objects (MACHOs). This category of dark matter candidates has fallen out of favor since then, with lensing surveys \citep{alcock_macho_2000, tisserand_limits_2007} placing stringent upper bounds on the fraction of dark matter that could reside in MACHOs, and the formation of structure at high redshift favoring a type of dark matter that could form early in the history of the universe (i.e. likely not MACHOs, and certainly not brown dwarfs). In this work we aim to explore the possibility that brown dwarfs exceeding the hydrogen burning limit exist even if they cannot account for the dark matter. We argue that the most likely scenario to produce these objects is a brown dwarf in a binary accreting from a companion brown dwarf via Roche lobe overflow. 

In this work we will proceed as follows. First, we present a qualitative and quantitative description of main sequence stars just above $M_\mathrm{HBL}$ in section \ref{sec:HBL}. Then in section \ref{sec:SMBD} we show that this picture, based on a simple analytic model and confirmed with numerical simulations has an interesting consequence, namely that brown dwarfs may exist with masses exceeding $M_\mathrm{HBL}$. In this section we make predictions for the properties of these objects. We discuss the implications and prospects for observational tests in section \ref{sec:discussion} then summarize our conclusions in section \ref{sec:summary}.

\section{The Hydrogen Burning Limit}
\label{sec:HBL}

The study of low-mass stars and brown dwarfs relies heavily on numerical simulations of varying sophistication, dating all the way back to the papers that initially pointed out that there was a particular mass where these objects transitioned from long-lived hydrogen burning stars to substellar degenerate objects that burned very little hydrogen. Indeed, detailed stellar models form the bedrock of much of modern astrophysics, allowing precision measurement of individual stars and whole stellar populations, enabling the study of exoplanets, stellar clusters, and galaxies. While there is obvious import and power in numerical modeling, it is also worth considering simplified models of low-mass stars, not for their quantitative predictions, but for their guide to understanding how properties of the stars scale, the hydrogen burning limit, and how brown dwarfs may exceed the hydrogen burning limit, as we will show.

Modeling low-mass stars is made easier by the fact that energy in objects within the relevant mass range is nearly entirely transported by convection as opposed to radiation (or conduction). This means that the star can be modeled as a constant-entropy polytrope with index $n=1.5$, plus some additional accounting for the change in entropy resulting from the transition between the partially ionized core and the molecular photosphere. Explicit formulae for a star's radius, effective surface temperature, core temperature, core density, and hence the surface and nuclear luminosities ($L_S$ and $L_N$ respectively) may be derived under these assumptions for fixed values of the star's composition, mass, and $\psi = k_B T/\mu_F$, the ratio of the thermal energy to the Fermi energy at the center of the star.

Recently \citet{auddy_analytic_2016} updated the classical analytic model for the hydrogen burning limit from \citet{burrows_science_1993}. These models are constructed with the goal of explicitly writing down a formula for the hydrogen burning limit. By applying the formulae mentioned in the previous paragraph, for a fixed composition one can identify a locus of points in $M-\psi$ space where the surface and nuclear luminosities are equal. This $M(\psi)$ curve has a minimum, which can be identified with the hydrogen burning limit. For concreteness, we explicitly write the formulae used in deriving the hydrogen burning limit in this way in the appendix. We follow \citet{auddy_analytic_2016} with some minor updates, acknowledging that particular details and assumptions in this model are at best rough approximations. Nonetheless, the shape of the $M(\psi)$ curve and the qualitative behavior we will discuss in the following sections are insensitive to the details of these approximations.

\subsection{The Main Sequence}

\begin{figure*}
\centering
\includegraphics[width=5.5in]{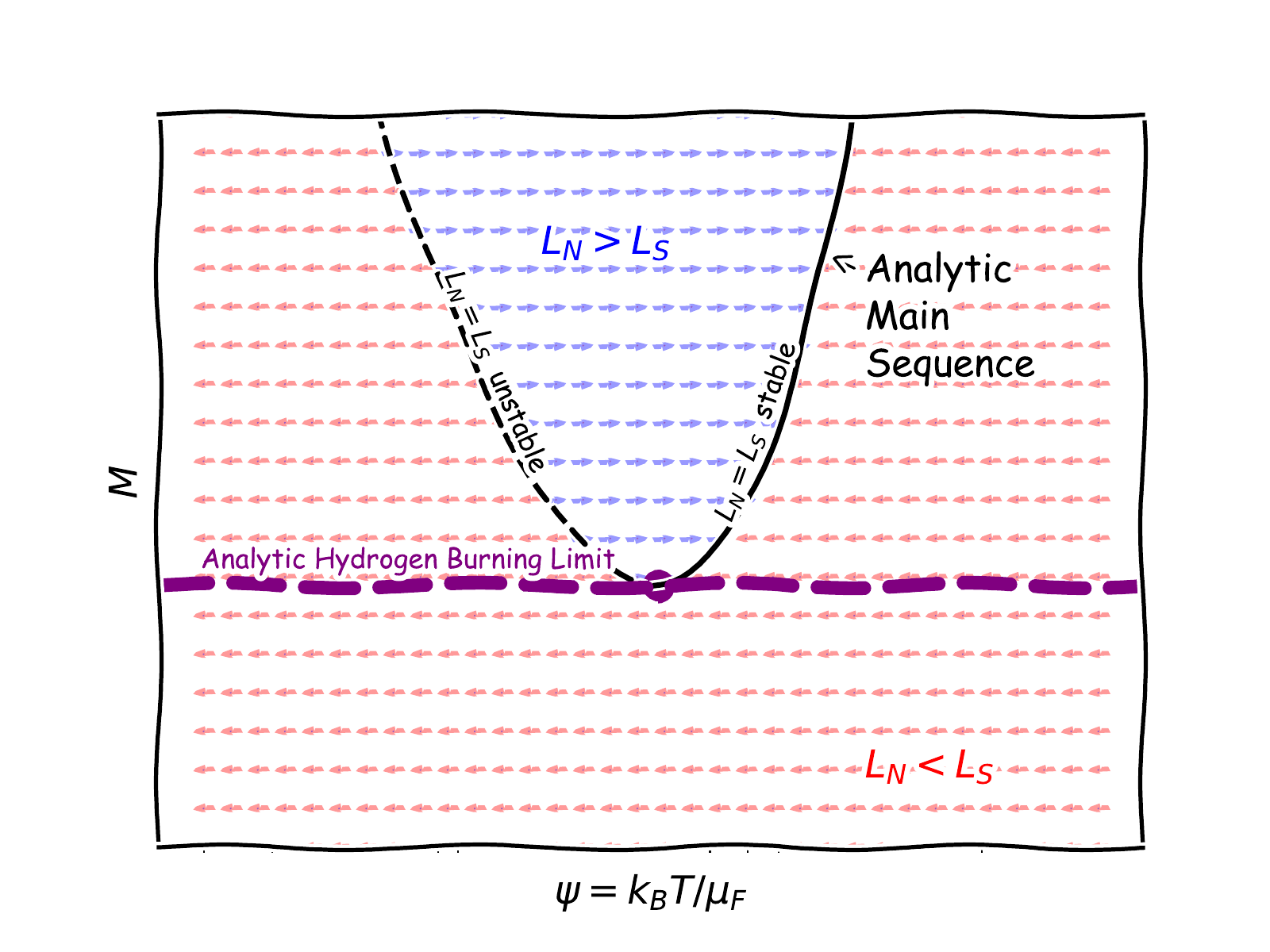}
\caption{A schematic diagram. The evolutionary arrows represent the evolution of stars or sub-stellar objects that happen to find themselves at a given point in this space, depending on the relative values of the nuclear luminosity $L_N$ and the surface luminosity $L_S$. The arrows are horizontal, since typically these objects should remain approximately the same mass throughout their lives. The curve where $L_N=L_S$ is shown in black. The minimum value of $M$ for this curve is the hydrogen burning limit, $M_\mathrm{HBL}$ according to the analytic model used to derive this curve. For $M>M_\mathrm{HBL}$, objects start from the right-hand side of the diagram and settle along the solid part of the $L_N=L_S$ curve, which is stable because perturbations off of the curve result in the object's return to the curve. The left-hand branch of the $L_N=L_S$ curve is unstable, but its existence implies that any object that finds itself to the left of that branch will cool, contract, and never reach the main sequence. It follows that objects that behave like brown dwarfs may in principle exist with masses exceeding $M_\mathrm{HBL}$ }
\label{fig:schematic}
\end{figure*}


Figure \ref{fig:schematic} illustrates the generic behavior of the analytic model. Above the $L_N=L_S$ line, given the very high power with which $L_N$ depends on $M$, the nuclear luminosity exceeds the surface luminosity. Below the $L_N=L_S$ line at any particular $\psi$, $L_N<L_S$. The mass at which the $L_N=L_S$ curve has a minimum is the canonical dividing line between brown dwarfs and stars. Below this line, objects may burn deuterium and even hydrogen briefly, but they never reach a steady state in which the nuclear luminosity is enough to power the energy being lost at the surface. Above the hydrogen burning limit, stars reach this steady state and may burn hydrogen in their cores for more than $10^{12}$ years.

In the absence of substantial accretion or winds, stars are restricted to horizontal trajectories in this diagram. Objects generally begin on the right-hand side of this diagram, i.e. $\psi \ga 1$. Before collapsing to form a stellar object, the gas pressure is thermal, and only once the collapse of the object is halted (or at least slowed to a thermal timescale) does the gas begin to be supported by degeneracy pressure. As the object cools, it moves leftwards in the diagram as indicated by the red arrows in Figure \ref{fig:schematic}. Objects above the hydrogen burning limit eventually reach the $L_N=L_S$ line. This is the main sequence of very low mass stars. Below the minimum of the $L_N=L_S$ curve, the objects continue to cool, moving leftward in the diagram.

The point at which stars reach the $L_N=L_S$ line represents an equilibrium point, but we should consider its stability. When $L_N>L_S$ the center of the star is unable to remove the excess of energy generated by the unbalanced $L_N$, so the center of the star must heat up, increasing $\psi$ and pushing the star rightward as indicated by the blue arrows in Figure \ref{fig:schematic}. Similarly when $L_N<L_S$, the star radiates energy more quickly than it can be generated, implying that the star should cool and move leftwards, as indicated by the red arrows. Thus at fixed mass the high-$\psi$ branch of the $L_N=L_S$ curve represents a stable equilibrium, whereas putative objects perturbed from the low-$\psi$ branch would tend to evolve away from the stationary point, making this branch unstable.

\begin{figure*}
\centering
\includegraphics[width=5.5in]{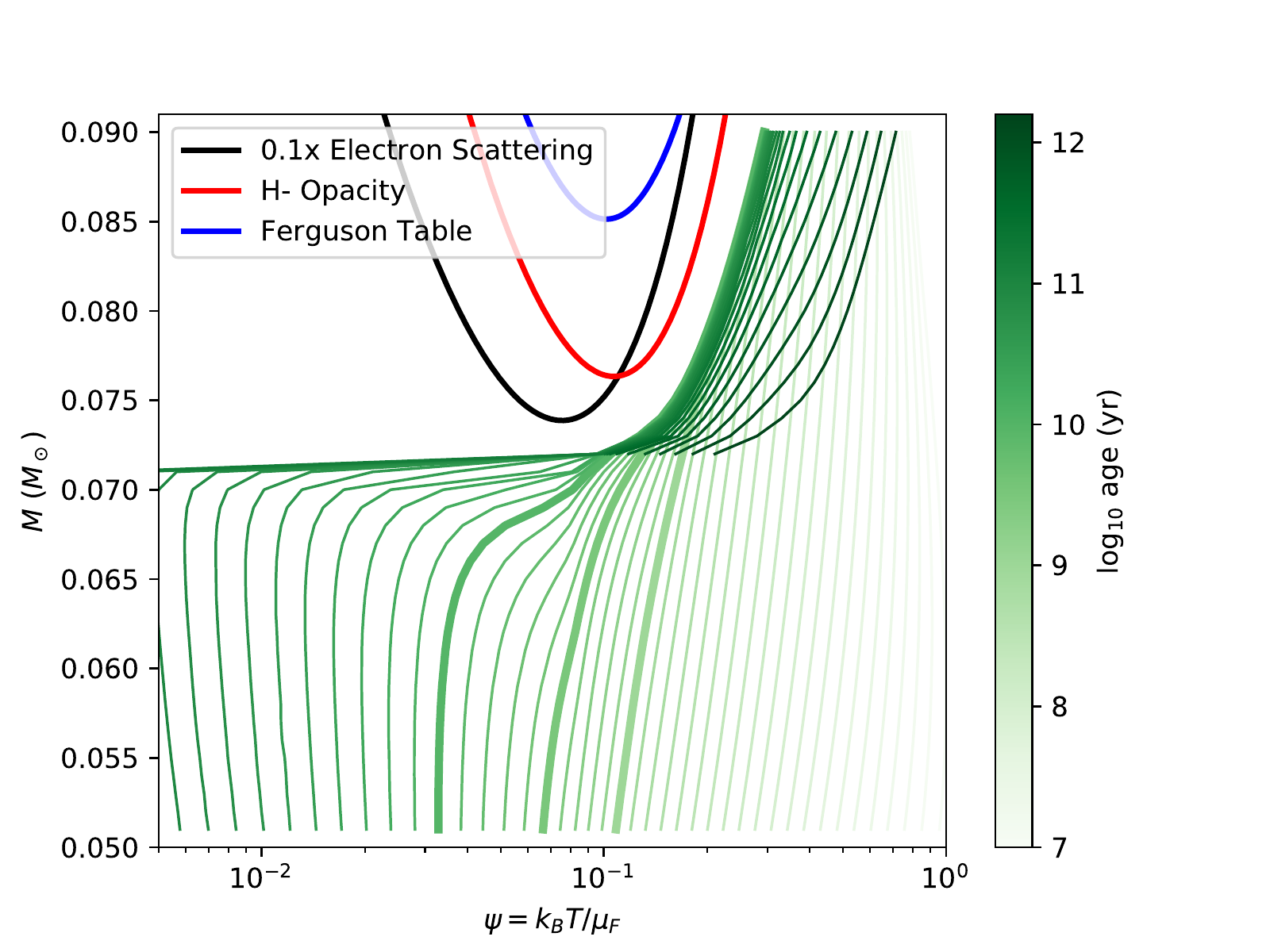}
\caption{MESA isochrones. Three different versions of the analytic model are shown in black, red, and blue -- everything is the same between them except the opacity law, as indicated. Plotted for comparison are isochrones spaced in 0.1 dex intervals of age extracted from a library of MESA models, with darker greens corresponding to older ages. Isochrones at 1, 3.1, and 10 Gyr are highlighted. The hydrogen burning limit according to the MESA models appears in this diagram as the mass dividing objects that continuously cool from those that settle on a curve resembling a half-parabola, just as expected from the schematic diagram of Figure \ref{fig:schematic}. While the locations of the main sequence and $M_\mathrm{HBL}$ do not agree perfectly with the analytic model, there is a remarkable qualitative agreement. }
\label{fig:isochrone}
\end{figure*}

\subsection{Simulations}
\label{sec:massLibrary}

To verify the analytic picture presented in the previous section, we have employed  MESA, the Modules for Experiments in Stellar Astrophysics \citep{paxton_modules_2011, paxton_modules_2013, paxton_modules_2015}, a modern, flexible, open, and modular stellar evolution code. In this work we use MESA essentially as-is\footnote{Release 10000}, i.e. we make no changes to the code itself. We also use the default values of all parameters with the following exceptions. 

The initial composition of the star is set via {\verb set_uniform_initial_composition = .true. }, with prescribed values of hydrogen, deuterium, and $^3$He and $^4$He. Metallicities and $^4$He abundances are related as discussed in the appendix. Throughout this work we assume solar metallicity. The $^2$H abundance is set to half the primordial value \citep{cooke_primordial_2016}, and the $^3$He abundance is set to a fixed fraction, $2.8 \times 10^{-5}$, of the $^4$He abundance. The parameters {\verb initial_mass }, {\verb initial_y }, and {\verb initial_z } are set to consistent values with the more detailed abundances. For running low-mass stars, we use {\verb which_atm_option = 'tau\_100\_tables' } as recommended in \citet{choi_mesa_2016}, and we use a nuclear reaction net that includes Deuterium burning via {\verb default_net_name = 'pp\_extras.net' }. Finally, the {\verb max_age } and {\verb max_model_number } are set to large enough values that they never cause the runs to terminate.

Essentially, we allow the model to evolve longer than the default since we are interested in the lifetimes of extremely long-lived stars; we use a nuclear reaction network that includes deuterium burning, since this is important at least in the early lives of these stars; and we use an outer boundary condition that avoids some of the difficulty in realistically modeling the atmospheres of very low-mass stars.

We begin by running a set of models closely-spaced in initial stellar mass until each model terminates, generally because it becomes too degenerate for MESA's equation of state. We run models for masses ranging from $0.05 M_\odot$ to $0.09 M_\odot$, encompassing the range of predicted values for the mass of the hydrogen burning limit. For comparison to the schematic diagram shown in Figure \ref{fig:schematic}, we extract isochrones in $\psi-M$ space by interpolating output values of $\psi$ and $M$ as a function of age (each model provides data at different age intervals depending on adaptive timestepping). 

Figure \ref{fig:isochrone} shows the evolution of these isochrones in intervals equally spaced in $\log(\mathrm{age})$. Models are not included in a given isochrone if they have already terminated at that age. As expected, models begin with large values of $\psi$. Brown dwarfs, below about $0.071 M_\odot$, cool steadily with a very regular leftward march in this diagram. Stars above this mass cool until they reach something resembling the analytic curve, where they pile up and slowly burn hydrogen. Eventually, as these stars deplete the hydrogen in their core (and elsewhere, since they are fully convective), they move back to the right as a result of the changing chemical composition. In particular $\psi \propto \mu_e^{2/3}$ and $\mu_e$ increases over time as hydrogen is consumed. For clarity, we only show isochrones up to $10^{12.2}$ years, but the models continue evolving through their post-main-sequence up to a maximum age of $7 \times 10^{12}$ years just above the hydrogen burning limit. This represents the ``end of the main sequence'' \citep{laughlin_end_1997}, i.e. the maximum age of stars, and the maximum age of these models agrees well with those calculations.

In addition to the isochrones shown in green, we include versions of the analytic model with several different opacity laws which are applied near the surface of the star. The black line assumes an opacity independent of density and temperature, namely $0.02(1+X)$ cm$^2$ g$^{-1}$, one tenth of the opacity appropriate for free electrons. This is a reasonable order of magnitude for the opacity, but the lack of dependence on temperature and density is not realistic. The blue line uses the same opacity law as MESA in this region of $T$-$\rho$ space, namely the tabulated results of \citet{ferguson_low-temperature_2005}. The red line, which comes closest to agreeing with the results of the MESA models, is an opacity similar to but not quite that of H-, which does indeed dominate the opacity at a slightly deeper layer in the star than the photosphere according to the simulations. In particular for this opacity we set $\kappa = 10^{-14} T^4$ cm$^2$ g$^{-1}$. 

These different versions of the analytic model do not quite agree with the MESA simulations in terms of the exact location of the hydrogen burning limit in $M$-$\psi$ space. Most surprisingly, the opacity table used by MESA itself, when applied in the analytic model, gives the worst agreement of the three. This is the result of other differences between the analytic model and the MESA simulations, likely the assumptions surrounding the phase transition between the ionized interior and molecular atmosphere of the low-mass stars (in the MESA simulations the change in entropy between the two regions tends to be small). There are also key differences in the treatment of the nuclear reactions that could account for this difference (in the analytic model, in order to integrate over the volume of the star's core, powerlaw approximations to the reaction rate are used). Despite these differences, the main sequence of the MESA models and the analytic model have the same smooth parabolic shape in $M-\psi$ space, with a well-defined boundary in mass between stars and brown dwarfs, strongly suggesting that the qualitative picture presented in Figure \ref{fig:schematic} is reasonable.

\section{Overmassive brown dwarfs}
\label{sec:SMBD}

Thus far we have reviewed the analytic derivation of the location of the hydrogen burning limit, the dividing line between brown dwarfs and stars. For both brown dwarfs and M dwarfs, the existence of the unstable low-$\psi$ branch of the $L_N=L_S$ curve shown in Figure \ref{fig:schematic} has been completely irrelevant. Brown dwarfs simply cool steadily, and M dwarfs reside on the high-$\psi$ branch before exhausting their supply of hydrogen and moving off to higher $\psi$. In other words, the typical evolutionary paths of low-mass stars and brown dwarfs never pass near the low-$\psi$ branch.

Despite what happens to the typical star, it is clear from Figure \ref{fig:schematic} that the analytic model makes the following prediction: {\it objects with some mass $M$ above the hydrogen burning limit behave like brown dwarfs if their value of $\psi$ is less than the lowest value of $\psi$ for which $L_N=L_S$ at mass $M$}. In other words, it is possible to have brown dwarfs with masses above the canonical hydrogen burning limit which would reside in the upper-left part of Figure \ref{fig:schematic}. In what follows we shall explore the properties of these ``overmassive brown dwarfs,'' evolutionary pathways by which they may arise, and potential observational signatures. 

\subsection{Properties}

\begin{figure*}
\centering
\includegraphics[width=5.5in]{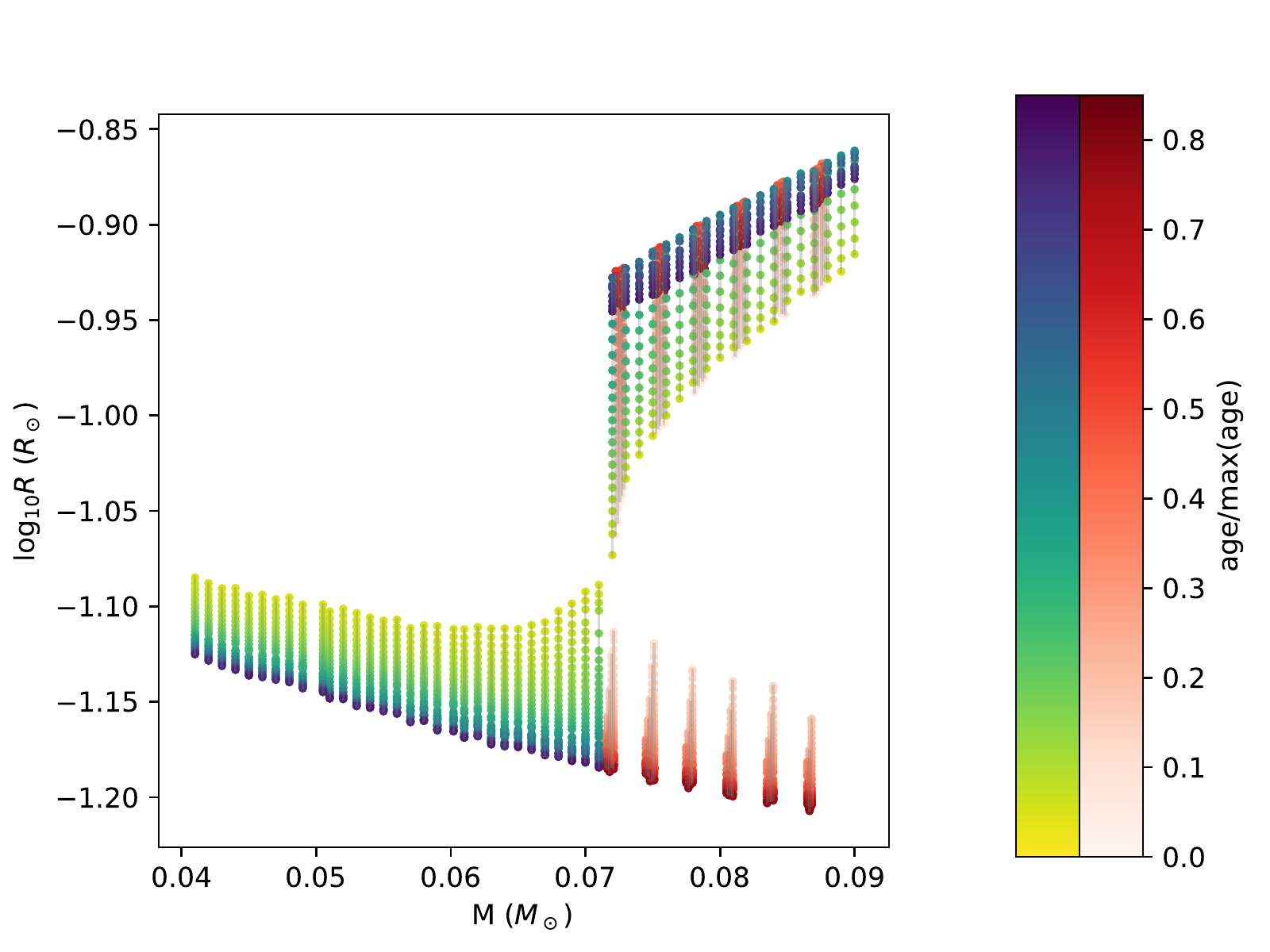}
\caption{The mass-radius relation. Ordinary MESA models are shown in yellow-purple, and modified MESA models where mass has been artificially added to a high-mass brown dwarf are shown in red. The evolution of each model is shown, where the color represents the fractional age of the model before it is eventually terminated by becoming too dense and cold to be adequately described by MESA's equation of state. As expected, at the hydrogen burning limit, $\approx 0.07 M_\odot$, the radius of the ordinary MESA models shows a sudden increase as the objects transition from brown dwarfs to main sequence stars. Some of the modified models (shown in red) become ordinary stars, with paths similar to the yellow-purple tracks of stars with masses above $\approx 0.07 M_\odot$, while those with sufficiently degenerate initial conditions behave like a continuation of the sequence of ordinary brown dwarfs. These are overmassive brown dwarfs, hydrogen-rich degenerate objects that do not burn hydrogen despite exceeding the hydrogen burning limit.}
\label{fig:radius}
\end{figure*}

\begin{figure*}
\centering
\includegraphics[width=5.5in]{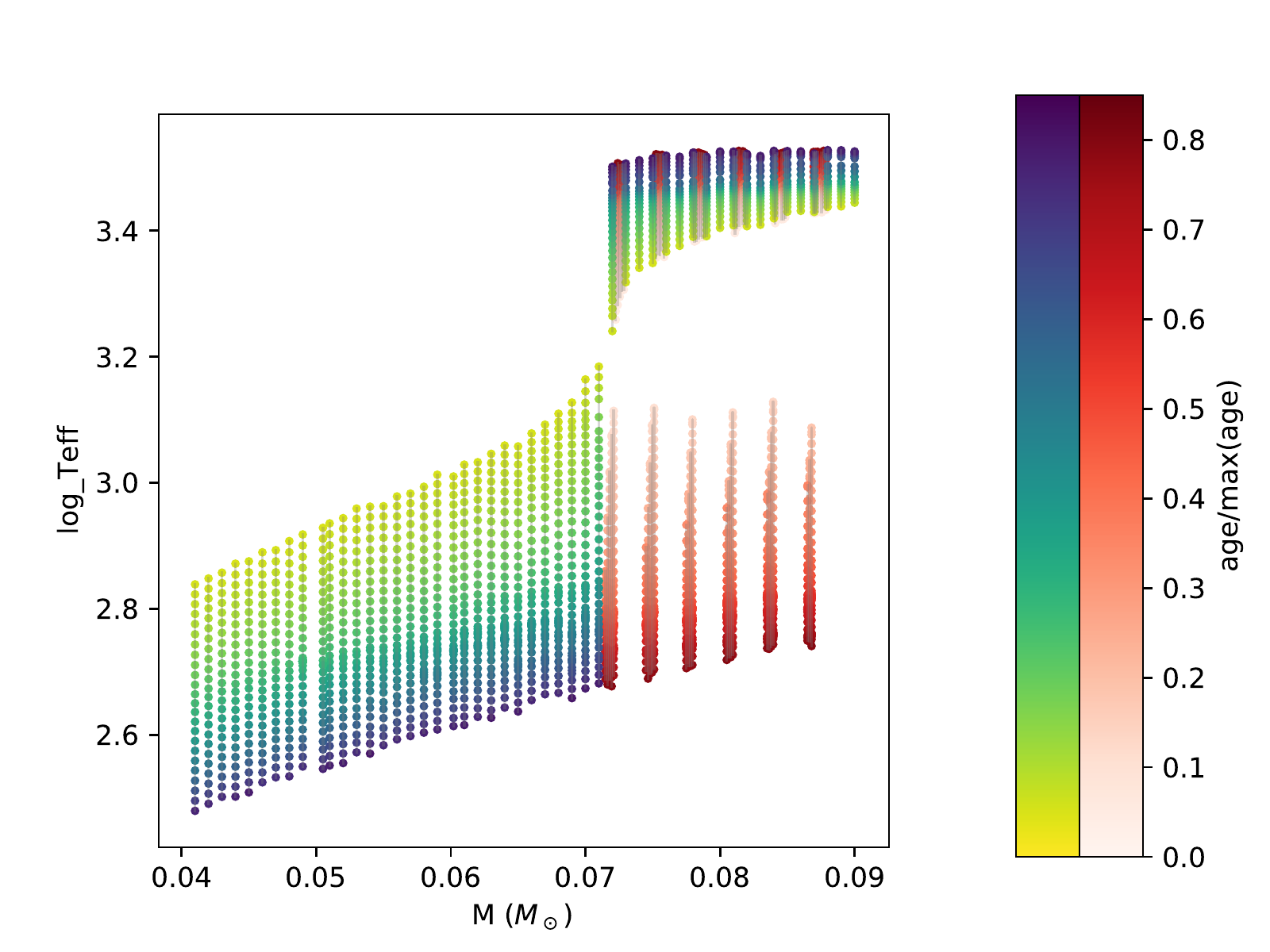}
\caption{The mass-effective temperature relation. This diagram is constructed in the same way as Figure \ref{fig:radius}, and in fact shows essentially the same behavior. Mass added to a sufficiently degenerate brown dwarf below the hydrogen burning limit results in stars that behave just like brown dwarfs, except they are more massive than the conventional hydrogen burning limit.}
\label{fig:Teff}
\end{figure*}

In order to understand the properties of the overmassive brown dwarfs predicted based on Figure \ref{fig:schematic}, we set out first to verify that such objects could be produced in MESA. Our general strategy is to start with a brown dwarf with a mass just below the hydrogen burning limit, and allow it to evolve and cool to a lower value of $\psi$. At different points along this trajectory in $\psi$, we will instantaneously artificially add mass to the model. Once this mass is added, the model quickly adjusts and begins its usual long-term evolution. If the value of $\psi$ at which we add mass to the model is sufficiently low, i.e. if it passes the abscissa of the minimum in the $L_N=L_S$ curve, the model continues to cool even after a substantial amount of mass is added, producing an overmassive brown dwarf. If instead the model ends up in the $L_N>L_S$ region of $\psi-M$ space, it quickly becomes a main sequence star. In this section we will explain this procedure in more detail, and show the trajectories of these modified MESA models in Figures \ref{fig:radius} and \ref{fig:Teff}.

Based on the isochrones produced by running MESA for a few dozen closely-spaced initial masses (see Figure \ref{fig:isochrone}), we found that the hydrogen burning limit is about $M_\mathrm{HBL} = 0.071 M_\odot$, as far as MESA is concerned in this configuration for this set of parameters. In order to produce MESA models with properties resembling the overmassive brown dwarfs we expect to exist in the upper-left part of the $M-\psi$ diagram, we re-ran one of the highest-mass MESA model that produces an ordinary brown dwarf, namely the $M=0.07 M_\odot$ model, many times. Each time, we terminated the model prematurely by setting the MESA parameter {\verb eta_center_limit }. We used 10 different values starting at 1 and increasing in increments of 0.2 dex. This parameter will terminate the MESA run when $\eta = 1/\psi$ exceeds the given value, so this is equivalent to allowing the brown dwarf model to evolve to various points evenly-spaced on the logarithmic x-axis of Figure \ref{fig:isochrone}. At the point of termination, the MESA model is saved.

Starting from these saved models, we next added 6 different masses to each model by setting {\verb relax_initial_mass = .true.} and {\verb new_mass } to values exceeding $0.07 M_\odot$ in increments of $0.003 M_\odot$. In addition, the new values of the mass were incremented by $-0.00015 M_\odot$ for each different value of $\eta$ for visual clarity when plotting these models. The results are shown in Figures \ref{fig:radius} and \ref{fig:Teff}. The yellow-purple points are models that were not modified by this procedure, i.e. the same set of MESA runs used to produce Figure \ref{fig:isochrone}, while the red points show the models that have undergone this procedure of premature termination at a prescribed value of $\psi$, and the addition of some mass. Each model is shown as a series of points corresponding to different ages. For a given model, the points lie along a single vertical line corresponding either to the initial stellar mass in the case of unmodified models or the new stellar mass for the modified models. To plot brown dwarfs and stars with the same set of colors, the colors correspond to fractional rather than absolute ages. The last 15\% of each model's life is not shown, since the objects that reach the main sequence eventually undergo post-main-sequence evolution when they deplete their supply of hydrogen, and their trajectories in this diagram obscure the hydrogen-rich phase on which we are focusing.

The normal, unmodified models, the same ones used to produce Figure \ref{fig:isochrone}, show an extremely clear transition between brown dwarfs with $M < M_\mathrm{HBL}$ and stars with $M>M_\mathrm{HBL}$ in both radius and effective temperature. This agrees well with standard brown dwarf evolutionary models, and has been suggested as a means of observationally determining the value of $M_\mathrm{HBL}$ \citep{burrows_dependence_2011, dieterich_solar_2014}. The modified models shown in red all lie above $M_\mathrm{HBL}$, and show a bifurcation, with some moving to lower $T_\mathrm{eff}$ and $R$, and others to higher values. On one side of this bifurcation, the warmer, larger-radii models follow the evolution of the unmodified models in this mass-range, behaving exactly as if they had always been at this new mass, i.e. like main sequence stars. This is because the additional mass placed them either in the $L_N>L_S$ region of $\psi-M$ space, or even in the pre-main-sequence region of parameter space (to the right of the stable branch of the $L_N=L_S$ curve). Meanwhile on the other side of the bifurcation, the modified models that cool and contract appear for all intents and purposes to be brown dwarfs despite having masses exceeding $M_\mathrm{HBL}$. Their properties are a simple continuation of the sequence of brown dwarfs, despite their non-trivial evolutionary history.

With this simple experiment, we have demonstrated that the qualitative picture we inferred by examining the analytic model is borne out in the far more sophisticated MESA models: overmassive brown dwarfs can be produced in MESA. These stellar objects, which happen to find themselves cooler than, i.e. to the left of, the unstable branch of the $L_N=L_S$ curve, behave in every way like brown dwarfs despite exceeding the hydrogen burning limit. In the next section we discuss plausible ways that the universe may produce these objects.

\subsection{Evolution}
Thinking in terms of the $\psi-M$ diagram, we recall that newly-formed objects begin their life on the right hand side of this diagram. In order to make the jump across the region where $L_N>L_S$ (see Figure \ref{fig:schematic}), the object would need to suddenly cool and shrink. It seems unlikely that any realistic astrophysical process would have this effect. It follows that the most plausible route to forming overmassive brown dwarfs is for an ordinary brown dwarf, i.e. below the hydrogen burning limit, to gain mass. The key is that the brown dwarf must be sufficiently old that its value of $\psi$ falls below $\psi_\mathrm{min}$, the abscissa of the hydrogen burning limit. Based on Figure \ref{fig:isochrone}, $\psi_\mathrm{min} \approx 0.1$, both for the analytic models and the MESA models. The isochrones suggest that the brown dwarf must be older than $\sim 3 \times 10^9\ \mathrm{yr}$ for $\psi$ to be less than $\psi_\mathrm{min}$.

Not only must $\psi$ be less than $\psi_\mathrm{min}$ when new material is added to the brown dwarf, but $\psi$ must remain to the left of the low-$\psi$ branch of the $L_N=L_S$ curve as the mass increases. This may be challenging because the process of adding mass to the star is likely to heat it up and increase $\psi$. One could imagine this process either as steady accretion or collision(s) with other objects. In the case of the former, the accretion luminosity is
\begin{equation}
L_a =\frac{GM}{R} \frac{dM}{dt}
\end{equation}
If we assert that the accretion luminosity must be small compared to the luminosity along the low-$\psi$ branch of the $L_N=L_S$ curve, we can say that the accretion rate must not exceed
\begin{equation}
\label{eq:mdotlimit}
\frac{dM}{dt} \la 4.7 \times 10^{-3} \frac{M_\odot}{ \mathrm{Gyr}} \left(\frac{R}{0.1 R_\odot} \right) \left(\frac{M}{0.07 M_\odot} \right)^{-1} \left(\frac{L_S}{10^{-4} L_\odot} \right)^{-1} 
\end{equation}
The timescale for this accretion rate to increase the mass of the brown dwarf from $0.07 M_\odot$ to $0.08 M_\odot$ is 2 Gyr, implying that such objects could in principle exist in today's universe. 

We caution that this condition, namely $L_a \la L_S$ is at best a rough estimate. The energy associated with accretion may be able to efficiently radiate away without substantially heating the brown dwarf, depending on the geometry and other details of the process \citep[e.g.][]{prialnik_outcome_1985, hartmann_disk_1997}. Indeed substantially higher accretion rates have been posited that allow brown dwarfs to grow above $\sim 0.1 M_\odot$ \citep{lenzuni_formation_1992, hansen_origin_1999}, though these calculations assumed zero metallicity. While this may modestly relax the constraints on the plausible physical mechanisms for producing overmassive brown dwarfs, it is still the case that the brown dwarf needs to cool for a few Gyr before new mass can be added. This condition can also be relaxed if, rather than forming rapidly, the original brown dwarf grows slowly over the entire course of its formation as suggested by \citet{gold_early_1984}, though this also seems implausible given the short timescales associated with gas dynamics.

For now we will proceed under the assumption that Equation \eqref{eq:mdotlimit} must be satisfied to keep the brown dwarf degenerate. This rate of accretion is, however, non-trivial to supply to a brown dwarf. The Bondi accretion rate \citep{bondi_mechanism_1944} for a background number density $n$ and relative velocity $v$ is 
\begin{equation}
\frac{dM}{dt} = 7.2 \times 10^{-9} \frac{M_\odot}{\mathrm{Gyr}} \left( \frac{M}{0.07 M_\odot} \right)^2 \left( \frac{n}{1\  \mathrm{cm}^{-3}} \right) \left( \frac{v}{10\ \mathrm{km}\ \mathrm{s}^{-1}} \right)^{-3},
\end{equation}
so the timescale for a brown dwarf to grow at a rate approaching the limit of Equation \eqref{eq:mdotlimit} from anything resembling the Milky Way's interstellar medium is orders of magnitude longer than the age of the universe. Interestingly the conditions in an AGN accretion disk, with $n\sim10^9\ \mathrm{cm}^{-3}$ and $v\sim c_s \sim 100\ \mathrm{km}\ \mathrm{s}^{-1}$ yield about the right order of magnitude for feeding the brown dwarf near the limit imposed by Equation \eqref{eq:mdotlimit}. Suitable conditions may also arise in dense structures of molecular clouds, but these conditions will generally persist for far shorter than the Gyr timescales necessary to substantially grow the brown dwarf's mass.

The most promising mechanism to substantially grow the brown dwarf's mass on a slow enough timescale to avoid violating Condition \eqref{eq:mdotlimit}, is via Roche lobe overflow in binary mass transfer. This is a well-studied theoretical problem with applications to stellar population synthesis, variable stars, and accretion onto compact objects. First, we define the mass ratio $q=M_\mathrm{donor}/M_\mathrm{accretor}$, where $M_\mathrm{accretor}$ is the mass of the (denser) star that is accreting mass from its companion, and $M_\mathrm{donor}$ is the mass of the star that is filling its Roche lobe and losing mass as a result. Because the size of the Roche lobe itself depends on $q$, transferring mass and decreasing $q$ tends to reduce the size of the Roche lobe. If this decrease happens too quickly, the mass loss rate can run away and proceed on a dynamical timescale. Roughly speaking, stable mass transfer requires that $q \la 5/6$, with some modest dependence on how the donor star reacts to the mass loss; if the donor star can contract faster than its Roche lobe size decreases, it could transfer mass stably at somewhat larger $q$ values.

Assuming the mass loss proceeds stably the question becomes how it is driven. A common scenario is that the donor star has begun to fill its Roche lobe as the result of post main sequence stellar evolution, and in this case the mass transfer proceeds on a nuclear timescale of the donor star. However, if $q\la5/6$, and we are attempting to add mass to a brown dwarf with initial mass $M_\mathrm{accretor} \approx M_\mathrm{HBL} \approx 0.07 M_\odot$, there is no nuclear timescale, and in fact the donor object is cooling and contracting. Something else must therefore be invoked to get the donor star to fill its Roche lobe.

An obvious candidate is for the binary to lose angular momentum due to the radiation of gravitational waves. This possibility was first pointed out by \citet{kraft_binary_1962}, and employed in simple models by \citet{paczynski_gravitational_1967} and \citet{faulkner_ultrashort-period_1971}. In these models, the time derivative of the angular momentum $J$ is set equal to the rate at which angular momentum is lost by gravitational waves $\dot{J}_\mathrm{GW}$, where 
\begin{equation}
J = G M_\mathrm{donor} M_\mathrm{accretor} \sqrt{G a/M_\mathrm{tot}},
\end{equation}
$M_\mathrm{tot}=M_\mathrm{donor}+M_\mathrm{accretor}$, $a$ is the orbital separation between the two objects, and, from \citet{peters_gravitational_1964} or \citet{landau_classical_1975},
\begin{equation}
\dot{J}_\mathrm{GW} = -\frac{32}{5 c^5}  M_\mathrm{donor}^2 M_\mathrm{accretor}^2 \sqrt{\frac{G^7 M_\mathrm{tot}}{a^7}}.
\end{equation}
As presently-formulated, this problem appears to be a single evolution equation, $dJ/dt = \dot{J}_\mathrm{GW},$ with three independent unknown variables: $M_\mathrm{donor}$, $M_\mathrm{accretor}$, and $a$. We can of course employ the reasonable assumption that mass is conserved in the system to re-express all the masses as appropriate combinations of $q$ and the constant $M_\mathrm{tot}$, i.e. $M_\mathrm{donor} = M_\mathrm{tot} q (1+q)^{-1}$ and $M_\mathrm{accretor} = M_\mathrm{tot} (1+q)^{-1}$.

\begin{figure}
\centering
\includegraphics[width=3.7in]{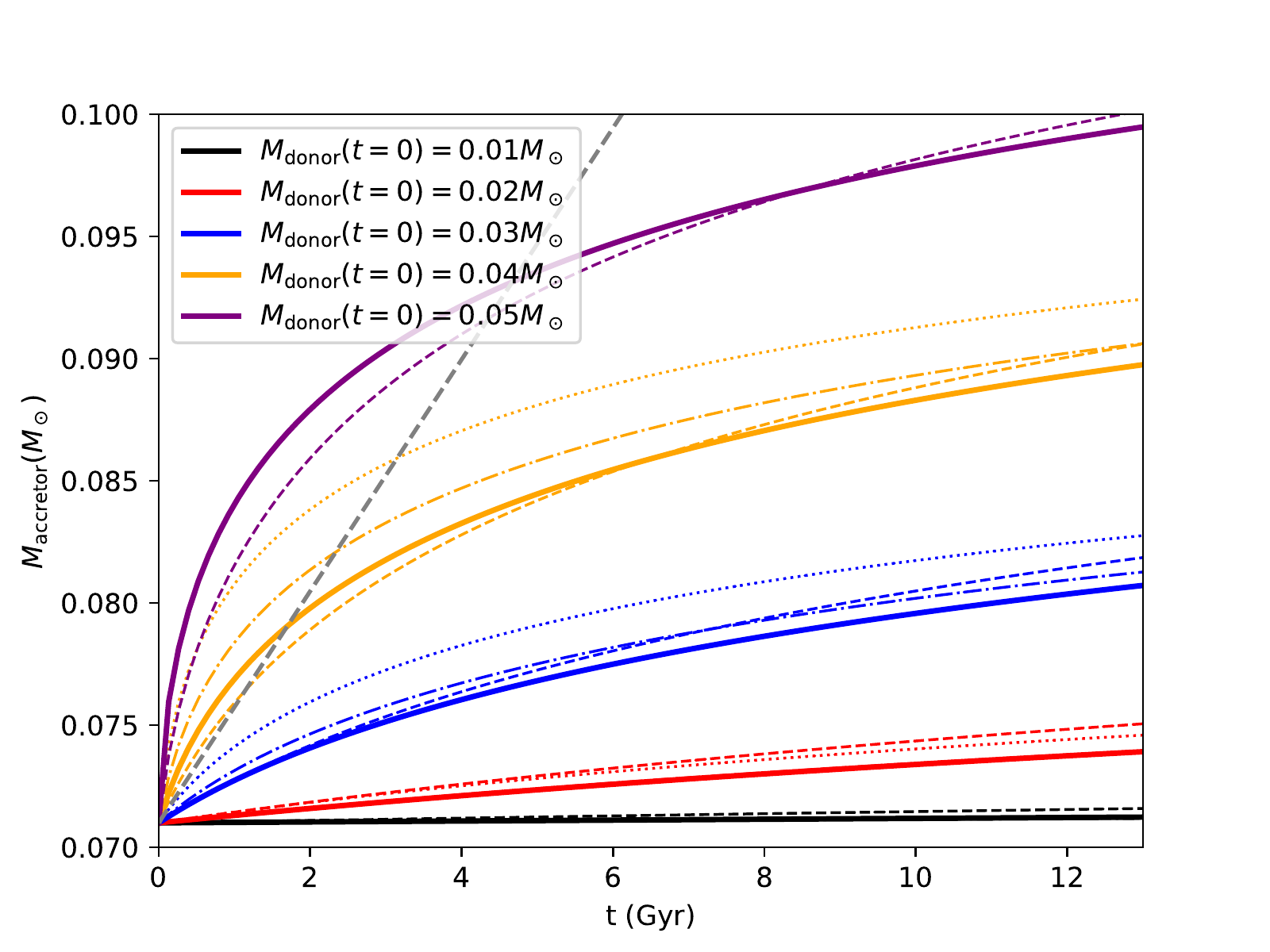}
\caption{Growth of a brown dwarf via gravitational wave regulated Roche lobe overflow. The mass of the accreting brown dwarf is shown as a function of time for different initial companion masses (different colors), and for slightly different assumptions about the mass-radius relation (different linestyles). These are $R_0=10^{-1.08}R_\odot, n=-0.1 $ (solid), $R_0=10^{-1.03}R_\odot, n=0.1 $ (dashed), $R_0=10^{-1.13}R_\odot, n=-0.2 $ (dash-dotted), and $R_0=10^{-1.2}R_\odot, n=-0.28 $ (dotted), covering the range of values for these parameters as the brown dwarfs modeled with MESA age. In addition, a straight grey dashed line shows roughly the maximum growth rate of the accreting brown dwarf such that its interior remains sufficiently degenerate to keep the object an overmassive brown dwarf rather than a new star.}
\label{fig:Maccretor}
\end{figure}

Finally, to eliminate $a$ from the evolution equation, we follow \citet{faulkner_ultrashort-period_1971} and make the assumption that there is a powerlaw mass-radius relation that applies for the donor star, namely $R_\mathrm{donor} = \mathcal{R} M_\mathrm{donor}^n$, and that throughout the evolution $R_\mathrm{donor} = R_{L,\mathrm{donor}}$, the radius of a sphere whose volume is the same as the Roche volume. \citet{eggleton_approximations_1983} derived a convenient numerical expression for this quantity as a function of $a$ and $q$, namely
\begin{equation}
\label{eq:eggleton}
\frac{R_{L,\mathrm{donor}}}{a} = \frac{0.49 q^{2/3} }{0.6 q^{2/3} + \log(1+q^{1/3})}.
\end{equation}
Following this chain of relations, all values of $a$ may be replaced with functions of $M_\mathrm{donor}$, which may in turn be replaced with functions of $q$, leaving an evolution equation for $q$ alone, depending on the total mass of the system, and the normalization and slope of the mass-radius relationship.
\begin{equation}
\label{eq:dqdT}
\frac{dq}{dT} = \frac{-1}{F'(q)} (1+q)^{7n/2-4} q^{2-7n/2} \mathcal{G}(q)^{7/2},
\end{equation}
where $\mathcal{G}(q)$ is simply the right-hand side of Equation \eqref{eq:eggleton}, and $F'(q) = dF/dq$ is the derivative of 
\begin{equation}
F = (1+q)^{-2-n/2} q^{1+n/2} \mathcal{G}(q)^{-1/2}.
\end{equation}
Equation \eqref{eq:dqdT} has also been non-dimensionalized by re-scaling the time variable as
\begin{equation}
t = T \mathcal{R}^4 \frac{5 c^5}{32} G^{-3} M_\mathrm{tot}^{4n-3}.
\end{equation}
Equation \eqref{eq:dqdT} can be evolved numerically for any initial $q$ and any value of $n$, then redimensionalized to obtain, e.g. $M_\mathrm{accretor}(t)$. We do this for a variety of donor mass values, and reasonable values for $n$ and $\mathcal{R}$. For convenience we vary $R_0$ instead of $\mathcal{R}$, where $R_0 = \mathcal{R}/(0.071 M_\odot)^n$, the value of the donor star's radius if the mass-radius relation were extrapolated to a mass of $0.071 M_\odot$. The result is shown in Figure \ref{fig:Maccretor}.

Over the range of values we have chosen for the mass-radius relationship, the growth of mass for the accreting brown dwarf is relatively insensitive to these parameters. The primary driver controlling how quickly mass is transferred is the total mass of the system. For an initial brown dwarf mass of $0.07 M_\odot$, the maximum mass the companion brown dwarf may have to maintain stable mass transfer is not much larger than $0.05 M_\odot$. In this case, the mass of the accreting brown dwarf could conceivably grow by about 50\%, up to about $0.1 M_\odot$. However, in addition to various scenarios for the growth of the accreting brown dwarf, Figure \ref{fig:Maccretor} shows a straight diagonal line representing the mass growth of a brown dwarf marginally obeying condition \eqref{eq:mdotlimit}, namely that the accretion rate should be low enough that the accretion luminosity does not exceed the intrinsic luminosity of the object. For the largest values of $M_\mathrm{donor}$, this line is crossed, indicating that the accretion rates early in the evolution of these systems are high enough to potentially heat up the accreting brown dwarf and push it into the regime of ordinary stellar evolution for an object of that mass. However, once the initial donor mass falls below about $0.04 M_\odot$, the mass transfer proceeds sufficiently slowly at all times that the brown dwarf may grow in mass by $\sim 10\%$, exceeding $M_\mathrm{HBL}$ while remaining a brown dwarf in every other respect.

The problem of binary mass transfer is certainly more complicated than this simple model captures. There may be additional physics involved, including tides, spin, and magnetic fields, in addition to differences in the physics of brown dwarfs imposed by having the donor star fill its Roche lobe and having both stars rotating. \citep[e.g.][]{king_evolution_1988, hurley_evolution_2002, knigge_evolution_2011, paxton_modules_2015} We defer a fuller treatment of this problem to later work; for now it suffices to show that there is a plausible means by which overmassive brown dwarfs may exist in the present-day universe. 

Other possibilities for adding mass to brown dwarfs include capture of stellar wind material from a massive companion, the complete or partial tidal disruption of neighboring objects, and collisions. Although each of these scenarios entails substantial uncertainty, we estimate that each one is fairly unlikely to produce the bulk of the putative overmassive brown dwarf population. Capture of stellar wind material occurs when the brown dwarf is in a binary system with a more massive star that is losing mass in a stellar wind. If mass is accreted following the prescription of \citet{bondi_mechanism_1944}, given a mass loss rate from the donor star from \citet{nieuwenhuijzen_parametrization_1990}, and assuming that the wind velocity is of order the escape velocity from the surface of the donor star, we find that the maximum mass accreted by the brown dwarf over the course of the life of the donor star never exceeds a few times $10^{-5} M_\odot$. This maximum occurs when the separation is as close as possible without running into the Roche limit, and the mass of the donor is large enough for the mass transfer rate to approach the limit of Condition \eqref{eq:mdotlimit}. At higher masses, the accretion rate is too great to keep the brown dwarf cool, and at closer separations the donor star loses mass directly via tidal perturbation. This intersection occurs at about $M_\mathrm{donor} \approx 13 M_\odot$ and $a \approx 10^{11}\ \mathrm{cm}$. Fundamentally this mechanism will have a difficult time transferring enough mass to the brown dwarf to make a difference because in order for the mass transfer rate to be large enough to be relevant, the donor star must be so massive that its lifetime is far shorter than the 2 Gyr necessary to transfer $\sim 10^{-2} M_\odot$. 

It is also worth considering tidal disruption events, wherein some object passes sufficiently close to the brown dwarf that tidal forces rip the donor object apart. Tidal disruptions are well-studied in the case of stars interacting with supermassive black holes at the centers of galaxies \citep[e.g.][]{rees_tidal_1988, guillochon_hydrodynamical_2013}. Since brown dwarfs just below the hydrogen burning limit lie at a minimum in the mass-radius relation (see e.g. Figure \ref{fig:radius}) they may in principle tidally disrupt other objects. In the MESA models just below the hydrogen burning limit, the mean density of the brown dwarf reaches about 200 g cm$^{-3}$. This implies a tidal radius of about 
\begin{equation}
r_t \approx 3 \times 10^{10}\ \mathrm{cm} \left(\frac{\rho_\mathrm{sat}}{1\ \mathrm{g}\ \mathrm{cm}^{-3}},\right)^{-1/3}
\end{equation}
for satellites of mean density $\rho_\mathrm{sat}$. 

Tidal disruptions occur on a dynamical timescale, with the accretion potentially spread out by the viscous and collisional dynamics of the bound tidal stream. Nonetheless, the timescale for this process is quite short. In order to feed the brown dwarf sufficiently slowly, one could instead imagine a process analogous to that presented in \citet{macleod_spoon-feeding_2013}. In that scenario, an evolved star on an eccentric orbit is partially tidally disrupted during each pericenter passage, and as a result its radius expands preparing it for another partial tidal disruption on its next orbit. The evolutionary timescale of an evolved star is too short for our purposes, but we could imagine instead that the companion star is on the main sequence with a mass of order $1 M_\odot$. Such a star has too large a radius to fit within $r_t$, but the tidal forces may be sufficient to transfer part of the outer layers of the star without such a close pericenter passage \citep{macleod_spoon-feeding_2013}. Repeated partial tidal disruptions over Gyr would likely also circularize the orbit, which would likely lead to a more destructive encounter, so for the moment we consider the binary brown dwarf scenario to be more plausible.

\section{Discussion}
\label{sec:discussion}

We have established that brown dwarfs with masses exceeding the hydrogen burning limit may in principle exist in the universe today, and the most promising means to create them is Roche lobe overflow in compact binary brown dwarf systems. We now turn to the question of the implications, both in interpreting observational data, and empirically constraining the frequency of overmassive brown dwarfs.

Binary brown dwarf systems are highly prized for their power in calibrating the evolutionary models mentioned in the introduction. Since it is possible to measure the mass of one or both components of a binary dynamically, relationships between stellar mass and spectral type can be determined directly, and degeneracies in model predictions of stellar properties may be broken \citep[see][]{dupuy_individual_2017}. Binary brown dwarfs are doubly important in the context of overmassive brown dwarfs, since binaries are likely necessary for their formation, and definitively confirming their existence independent of spectral models will likely require a dynamical mass measurement.

The fraction of brown dwarfs in visually-resolved binaries with separations exceeding about 3 AU is approximately 10\% \citep{burgasser_binaries_2007}. The true binary fraction may be somewhat higher, but based on a small sample of brown dwarfs monitored with radial velocity, \citet{joergens_binary_2008} estimate that the true binary fraction is unlikely to exceed 20\%, consistent with the trend that lower-mass stars have lower binary fractions \citep[e.g.][]{duquennoy_multiplicity_1991,yuan_stellar_2015}. There is also some evidence \citep{allers_brown_2012} that brown dwarf binaries preferentially have mass ratios close to unity, which poses an additional barrier to our stable binary mass transfer scenario. This suggests that the ultracompact binaries necessary to produce overmassive brown dwarfs are rare, with the bulk of brown dwarf binaries being too wide to transfer mass in the course of their lifetimes, too close to equal mass ratios, and brown dwarf binaries as a whole being reasonably uncommon.

We can make a very rough estimate of the frequency of Roche lobe overflow events among the brown dwarf population as follows. Differentiating the definition of the total angular momentum of the binary system (though this is likely just a lower limit on the angular momentum loss, e.g. \citet{knigge_evolution_2011}), we see that in the absence of mass transfer $\dot{a}/a = 2\dot{J}/J$. If we assume gravitational waves are the dominant sink of orbital angular momentum and plug in the definitions of $J$ and $\dot{J}_\mathrm{GW}$, we can find how far the binary shrinks over some time $t$ analytically. In particular, if the binary ends up at a final separation $a_f$, it must have started at a separation of 
\begin{equation}
a_0 = \left(a_f^4 + \frac{256 G^3}{5 c^5} M_1 M_2 (M_1+M_2) t \right)^{1/4}
\end{equation}
Setting the final separation to be equal to the radius at which the donor star begins to overflow its Roche lobe, namely $a_f = R_\mathrm{donor}/\mathcal{G}(q)$ in the parlance of the previous section, and fixing $t$ to be the typical age of stars in the solar neighborhood, i.e. about 5 Gyr, we can determine the separation $a_0$ out to which some initial population of binaries would be interacting via Roche lobe overflow today. Next, we can assume some simple distribution for the initial separation of the binaries. This will depend on their formation mechanism and is poorly constrained by existing data \citep[see][]{fontanive_constraining_2018}, so for simplicity we adopt a log-uniform distribution, not dissimilar to the separation distribution of M dwarfs \citep{fischer_multiplicity_1992} extending from $a_f$ all the way through the empirical maximum separation of $a_\mathrm{max} \sim 30\ \mathrm{AU}$. The fraction of brown dwarfs interacting via Roche lobe overflow $f_\mathrm{RLOF}$ would then be 
\begin{equation}
f_\mathrm{RLOF} = f_\mathrm{bin}\frac{\ln(a_0/a_f)}{\ln(a_\mathrm{max}/a_f)}.
\end{equation}
For binaries where the accretor is 0.06-0.07 $M_\odot$ and the donor is about $0.05 M_\odot$, we find $f_\mathrm{RLOF} \sim 10^{-2}$ assuming $f_\mathrm{bin}=20\%$.

There are only about $10^4$ brown dwarfs known in all \citep{burgasser_brown_2015}, and these are largely discovered via wide-field surveys, e.g. WISE \citep{wright_wide-field_2010}. It is possible that $\sim 10^2$ overmassive brown dwarfs are hidden among this sample if $f_\mathrm{RLOF} \sim 10^{-2}$, depending on the intrinsic distribution of separations. Regardless of $f_\mathrm{RLOF}$, it is currently quite difficult to determine whether a particular brown dwarf is in this phase. Binaries which are too compact to separate by their visual separation may still be identified in a few ways: the Doppler effect, eclipses, photometric variability, and through large discrepancies between expectations for single stars and observational data. 

The Doppler effect requires expensive spectroscopic monitoring of these faint stars, and so has been restricted to reasonably small samples. However, binaries interacting via Roche lobe overflow have sufficiently short periods and large velocities that these systems should be relatively easy to find, with no need for long-term monitoring or extremely high spectroscopic precision. LSST\footnote{https://www.lsst.org/} is expected to discover $\sim 7$ eclipsing binary brown dwarfs \citep[see chapter 6.9 of][]{lsst_science_collaboration_lsst_2009}. These systems will have immense constraining power on evolutionary models, and with luck $\sim 1$ of these systems may be a RLOF system. Sufficiently precise photometry may also reveal tight binaries \citep{faigler_photometric_2011, millholland_supervised_2017}, though it is unclear what the yield from LSST will be.

Another possibility for uncovering binaries is identifying outliers in observational relations appropriate for single brown dwarfs. A number of examples are discussed in some detail in \citet{dupuy_individual_2017}, including 2MASS
J0920+3517B \citep{kirkpatrick_67_2000, burgasser_unified_2006} an object which routinely appears as an outlier in the scaling relations presented in that work, and interpreted as an unresolved binary system. We caution though that overmassive brown dwarfs may also appear in observational relations outside the region expected for conventional brown dwarfs. 

Another interesting consequence of our current observational understanding of brown dwarf binaries is that binaries that undergo unstable mass transfer are likely to be more common than those undergoing the stable mass transfer with $q \la 5/6$ necessary to produce overmassive brown dwarfs. Binaries that undergo unstable mass transfer should very quickly, on a dynamical timescale, genuinely produce a new star provided the combined mass of the system exceeds the hydrogen burning limit. This sort of process has been considered as the dominant source of star formation in the distant future \citep{adams_dying_1997} long after even the lowest-mass stars have died \citep{laughlin_end_1997}. However, nothing prevents events like this from occurring in the present-day universe as well. While the event itself may be short, the newly-extant star where only a faint brown dwarf had existed just prior may be detectable in forthcoming transient surveys such as LSST. The frequency of such events may provide an efficient complementary alternative to spectroscopic monitoring as a means to constrain the population of brown dwarf binaries. Dynamical mass transfer may also leave remnants of the donor object in analogy to ``black widow'' pulsars \citep{bailes_transformation_2011} which may be detected in searches for planets around M dwarfs such as MEarth \citep{nutzman_design_2008, berta_transit_2012}.

\section{Summary}
\label{sec:summary}

According to conventional wisdom the dividing line between stars and brown dwarfs is very clear: there is a mass, the hydrogen burning limit, below which hydrostatic objects cannot support long-term hydrogen burning in their cores. Instead they are supported by degeneracy pressure as they cool and contract. These are brown dwarfs. Above this critical mass, objects may reach a steady state where they are supported against collapse, at least in part, by thermal pressure powered by steady hydrogen burning in their core. In this work we have shown that there are in principle objects in the universe that behave for all intents and purposes like brown dwarfs, but have masses exceeding the hydrogen burning limit.

Brown dwarfs in this category, which we call overmassive brown dwarfs, may exist so long as their central degeneracy is large. When this is the case, the object's surface luminosity exceeds the nuclear luminosity it can generate, and the object steadily cools becoming more degenerate in the process. If the object's degeneracy is insufficiently large, its nuclear luminosity will exceed its surface luminosity, forcing the object to quickly increase its surface luminosity, and bringing it back to the main sequence of hydrogen burning and a track of ordinary stellar evolution.

Practically speaking, in order to produce a overmassive brown dwarf, an old ($\ga 3$ Gyr) ordinary brown dwarf that has cooled far enough for its degeneracy to exceed some critical value must gain mass. The rate of mass gain must be large enough to substantially alter the brown dwarf's mass, but not so large that the liberation of gravitational energy associated with the accretion heats up the star's interior and decreases its central degeneracy below the critical value. We estimate that this requires an accretion rate below about $4.7\times10^{-3} M_\odot\ \mathrm{Gyr}^{-1}$, which means it takes at least 2 Gyr to increase the object's mass  by $0.01 M_\odot$, a substantial amount for a $\approx 0.07 M_\odot$ object. In the lifetime of the universe, an overmassive brown dwarf could in principle reach about $0.12 M_\odot$ if it can grow continuously at the maximum rate for 10 Gyr following the $\sim 3$ Gyr of cooling necessary to become sufficiently degenerate.

A number of physical processes may allow mass transfer on about the right timescale. We consider the most promising to be Roche lobe overflow in a compact brown dwarf-brown dwarf binary driven by the radiation of gravitational waves. This scenario, and the other plausible alternatives that we have considered, is likely to be rare, but exactly how rare depends on poorly-constrained properties of brown dwarf binaries. LSST will likely yield a handful of eclipsing binary brown dwarfs which could include an example of a pair interacting via Roche lobe overflow. Photometric variability from non-eclipsing binaries, and short-term low-precision radial velocity surveys are also potential routes to discover systems undergoing Roche lobe overflow and overmassive brown dwarfs. 

\section*{Acknowledgements}

This work was supported in part by an ITC Fellowship (JCF) and grants from the Breakthrough Prize Foundation and the John Templeton Foundation (AL). This work has benefited from helpful conversations with Morgan MacLeod, Jieun Choi, Sayantan Auddy, Brad Hansen, and Adam Burrows. We would like to thank the anonymous referee for comments that have led to a clearer and improved manuscript. In addition to MESA, we have used a variety of free resources including matplotlib \citep{hunter_matplotlib_2007}, numpy and scipy \citep{oliphant_python_2007}, NASA ADS and the arXiv.

\bibliography{/Users/jforbes/updatingzotlib}

\begin{thebibliography}{83}%
\makeatletter
\providecommand \@ifxundefined [1]{%
 \@ifx{#1\undefined}
}%
\providecommand \@ifnum [1]{%
 \ifnum #1\expandafter \@firstoftwo
 \else \expandafter \@secondoftwo
 \fi
}%
\providecommand \@ifx [1]{%
 \ifx #1\expandafter \@firstoftwo
 \else \expandafter \@secondoftwo
 \fi
}%
\providecommand \natexlab [1]{#1}%
\providecommand \enquote  [1]{``#1''}%
\providecommand \bibnamefont  [1]{#1}%
\providecommand \bibfnamefont [1]{#1}%
\providecommand \citenamefont [1]{#1}%
\providecommand \href@noop [0]{\@secondoftwo}%
\providecommand \href [0]{\begingroup \@sanitize@url \@href}%
\providecommand \@href[1]{\@@startlink{#1}\@@href}%
\providecommand \@@href[1]{\endgroup#1\@@endlink}%
\providecommand \@sanitize@url [0]{\catcode `\\12\catcode `\$12\catcode
  `\&12\catcode `\#12\catcode `\^12\catcode `\_12\catcode `\%12\relax}%
\providecommand \@@startlink[1]{}%
\providecommand \@@endlink[0]{}%
\providecommand \url  [0]{\begingroup\@sanitize@url \@url }%
\providecommand \@url [1]{\endgroup\@href {#1}{\urlprefix }}%
\providecommand \urlprefix  [0]{URL }%
\providecommand \Eprint [0]{\href }%
\providecommand \doibase [0]{http://dx.doi.org/}%
\providecommand \selectlanguage [0]{\@gobble}%
\providecommand \bibinfo  [0]{\@secondoftwo}%
\providecommand \bibfield  [0]{\@secondoftwo}%
\providecommand \translation [1]{[#1]}%
\providecommand \BibitemOpen [0]{}%
\providecommand \bibitemStop [0]{}%
\providecommand \bibitemNoStop [0]{.\EOS\space}%
\providecommand \EOS [0]{\spacefactor3000\relax}%
\providecommand \BibitemShut  [1]{\csname bibitem#1\endcsname}%
\let\auto@bib@innerbib\@empty
\bibitem [{\citenamefont {Chabrier}(2003)}]{chabrier_galactic_2003}%
  \BibitemOpen
  \bibfield  {author} {\bibinfo {author} {\bibfnamefont {G.}~\bibnamefont
  {Chabrier}},\ }\href {\doibase 10.1086/376392} {\bibfield  {journal}
  {\bibinfo  {journal} {Publications of the Astronomical Society of the
  Pacific}\ }\textbf {\bibinfo {volume} {115}},\ \bibinfo {pages} {763}
  (\bibinfo {year} {2003})}\BibitemShut {NoStop}%
\bibitem [{\citenamefont {Kumar}(1963)}]{kumar_structure_1963}%
  \BibitemOpen
  \bibfield  {author} {\bibinfo {author} {\bibfnamefont {S.~S.}\ \bibnamefont
  {Kumar}},\ }\href {\doibase 10.1086/147589} {\bibfield  {journal} {\bibinfo
  {journal} {The Astrophysical Journal}\ }\textbf {\bibinfo {volume} {137}},\
  \bibinfo {pages} {1121} (\bibinfo {year} {1963})}\BibitemShut {NoStop}%
\bibitem [{\citenamefont {Hayashi}\ and\ \citenamefont
  {Nakano}(1963)}]{hayashi_evolution_1963}%
  \BibitemOpen
  \bibfield  {author} {\bibinfo {author} {\bibfnamefont {C.}~\bibnamefont
  {Hayashi}}\ and\ \bibinfo {author} {\bibfnamefont {T.}~\bibnamefont
  {Nakano}},\ }\href {\doibase 10.1143/PTP.30.460} {\bibfield  {journal}
  {\bibinfo  {journal} {Progress of Theoretical Physics}\ }\textbf {\bibinfo
  {volume} {30}},\ \bibinfo {pages} {460} (\bibinfo {year} {1963})}\BibitemShut
  {NoStop}%
\bibitem [{\citenamefont {Rebolo}\ \emph {et~al.}(1995)\citenamefont {Rebolo},
  \citenamefont {Zapatero~Osorio},\ and\ \citenamefont
  {Mart\'in}}]{rebolo_discovery_1995}%
  \BibitemOpen
  \bibfield  {author} {\bibinfo {author} {\bibfnamefont {R.}~\bibnamefont
  {Rebolo}}, \bibinfo {author} {\bibfnamefont {M.~R.}\ \bibnamefont
  {Zapatero~Osorio}}, \ and\ \bibinfo {author} {\bibfnamefont {E.~L.}\
  \bibnamefont {Mart\'in}},\ }\href {\doibase 10.1038/377129a0} {\bibfield
  {journal} {\bibinfo  {journal} {Nature}\ }\textbf {\bibinfo {volume} {377}},\
  \bibinfo {pages} {129} (\bibinfo {year} {1995})}\BibitemShut {NoStop}%
\bibitem [{\citenamefont {Laughlin}\ and\ \citenamefont
  {Bodenheimer}(1993)}]{laughlin_luminosity_1993}%
  \BibitemOpen
  \bibfield  {author} {\bibinfo {author} {\bibfnamefont {G.}~\bibnamefont
  {Laughlin}}\ and\ \bibinfo {author} {\bibfnamefont {P.}~\bibnamefont
  {Bodenheimer}},\ }\href {\doibase 10.1086/172203} {\bibfield  {journal}
  {\bibinfo  {journal} {The Astrophysical Journal}\ }\textbf {\bibinfo {volume}
  {403}},\ \bibinfo {pages} {303} (\bibinfo {year} {1993})}\BibitemShut
  {NoStop}%
\bibitem [{\citenamefont {Burrows}\ \emph {et~al.}(1993)\citenamefont
  {Burrows}, \citenamefont {Hubbard}, \citenamefont {Saumon},\ and\
  \citenamefont {Lunine}}]{burrows_expanded_1993}%
  \BibitemOpen
  \bibfield  {author} {\bibinfo {author} {\bibfnamefont {A.}~\bibnamefont
  {Burrows}}, \bibinfo {author} {\bibfnamefont {W.~B.}\ \bibnamefont
  {Hubbard}}, \bibinfo {author} {\bibfnamefont {D.}~\bibnamefont {Saumon}}, \
  and\ \bibinfo {author} {\bibfnamefont {J.~I.}\ \bibnamefont {Lunine}},\
  }\href {\doibase 10.1086/172427} {\bibfield  {journal} {\bibinfo  {journal}
  {The Astrophysical Journal}\ }\textbf {\bibinfo {volume} {406}},\ \bibinfo
  {pages} {158} (\bibinfo {year} {1993})}\BibitemShut {NoStop}%
\bibitem [{\citenamefont {Saumon}\ \emph {et~al.}(1994)\citenamefont {Saumon},
  \citenamefont {Bergeron}, \citenamefont {Lunine}, \citenamefont {Hubbard},\
  and\ \citenamefont {Burrows}}]{saumon_cool_1994}%
  \BibitemOpen
  \bibfield  {author} {\bibinfo {author} {\bibfnamefont {D.}~\bibnamefont
  {Saumon}}, \bibinfo {author} {\bibfnamefont {P.}~\bibnamefont {Bergeron}},
  \bibinfo {author} {\bibfnamefont {J.~I.}\ \bibnamefont {Lunine}}, \bibinfo
  {author} {\bibfnamefont {W.~B.}\ \bibnamefont {Hubbard}}, \ and\ \bibinfo
  {author} {\bibfnamefont {A.}~\bibnamefont {Burrows}},\ }\href {\doibase
  10.1086/173892} {\bibfield  {journal} {\bibinfo  {journal} {The Astrophysical
  Journal}\ }\textbf {\bibinfo {volume} {424}},\ \bibinfo {pages} {333}
  (\bibinfo {year} {1994})}\BibitemShut {NoStop}%
\bibitem [{\citenamefont {Baraffe}\ \emph {et~al.}(1995)\citenamefont
  {Baraffe}, \citenamefont {Chabrier}, \citenamefont {Allard},\ and\
  \citenamefont {Hauschildt}}]{baraffe_new_1995}%
  \BibitemOpen
  \bibfield  {author} {\bibinfo {author} {\bibfnamefont {I.}~\bibnamefont
  {Baraffe}}, \bibinfo {author} {\bibfnamefont {G.}~\bibnamefont {Chabrier}},
  \bibinfo {author} {\bibfnamefont {F.}~\bibnamefont {Allard}}, \ and\ \bibinfo
  {author} {\bibfnamefont {P.~H.}\ \bibnamefont {Hauschildt}},\ }\href
  {\doibase 10.1086/187924} {\bibfield  {journal} {\bibinfo  {journal} {The
  Astrophysical Journal Letters}\ }\textbf {\bibinfo {volume} {446}},\ \bibinfo
  {pages} {L35} (\bibinfo {year} {1995})}\BibitemShut {NoStop}%
\bibitem [{\citenamefont {Tsuji}\ \emph {et~al.}(1996)\citenamefont {Tsuji},
  \citenamefont {Ohnaka}, \citenamefont {Aoki},\ and\ \citenamefont
  {Nakajima}}]{tsuji_evolution_1996}%
  \BibitemOpen
  \bibfield  {author} {\bibinfo {author} {\bibfnamefont {T.}~\bibnamefont
  {Tsuji}}, \bibinfo {author} {\bibfnamefont {K.}~\bibnamefont {Ohnaka}},
  \bibinfo {author} {\bibfnamefont {W.}~\bibnamefont {Aoki}}, \ and\ \bibinfo
  {author} {\bibfnamefont {T.}~\bibnamefont {Nakajima}},\ }\href@noop {}
  {\bibfield  {journal} {\bibinfo  {journal} {Astronomy and Astrophysics}\
  }\textbf {\bibinfo {volume} {308}},\ \bibinfo {pages} {L29} (\bibinfo {year}
  {1996})}\BibitemShut {NoStop}%
\bibitem [{\citenamefont {Burrows}\ \emph {et~al.}(1997)\citenamefont
  {Burrows}, \citenamefont {Marley}, \citenamefont {Hubbard}, \citenamefont
  {Lunine}, \citenamefont {Guillot}, \citenamefont {Saumon}, \citenamefont
  {Freedman}, \citenamefont {Sudarsky},\ and\ \citenamefont
  {Sharp}}]{burrows_nongray_1997}%
  \BibitemOpen
  \bibfield  {author} {\bibinfo {author} {\bibfnamefont {A.}~\bibnamefont
  {Burrows}}, \bibinfo {author} {\bibfnamefont {M.}~\bibnamefont {Marley}},
  \bibinfo {author} {\bibfnamefont {W.~B.}\ \bibnamefont {Hubbard}}, \bibinfo
  {author} {\bibfnamefont {J.~I.}\ \bibnamefont {Lunine}}, \bibinfo {author}
  {\bibfnamefont {T.}~\bibnamefont {Guillot}}, \bibinfo {author} {\bibfnamefont
  {D.}~\bibnamefont {Saumon}}, \bibinfo {author} {\bibfnamefont
  {R.}~\bibnamefont {Freedman}}, \bibinfo {author} {\bibfnamefont
  {D.}~\bibnamefont {Sudarsky}}, \ and\ \bibinfo {author} {\bibfnamefont
  {C.}~\bibnamefont {Sharp}},\ }\href {\doibase 10.1086/305002} {\bibfield
  {journal} {\bibinfo  {journal} {The Astrophysical Journal}\ }\textbf
  {\bibinfo {volume} {491}},\ \bibinfo {pages} {856} (\bibinfo {year}
  {1997})}\BibitemShut {NoStop}%
\bibitem [{\citenamefont {Allard}\ \emph {et~al.}(1997)\citenamefont {Allard},
  \citenamefont {Hauschildt}, \citenamefont {Alexander},\ and\ \citenamefont
  {Starrfield}}]{allard_model_1997}%
  \BibitemOpen
  \bibfield  {author} {\bibinfo {author} {\bibfnamefont {F.}~\bibnamefont
  {Allard}}, \bibinfo {author} {\bibfnamefont {P.~H.}\ \bibnamefont
  {Hauschildt}}, \bibinfo {author} {\bibfnamefont {D.~R.}\ \bibnamefont
  {Alexander}}, \ and\ \bibinfo {author} {\bibfnamefont {S.}~\bibnamefont
  {Starrfield}},\ }\href {\doibase 10.1146/annurev.astro.35.1.137} {\bibfield
  {journal} {\bibinfo  {journal} {Annual Review of Astronomy and Astrophysics}\
  }\textbf {\bibinfo {volume} {35}},\ \bibinfo {pages} {137} (\bibinfo {year}
  {1997})}\BibitemShut {NoStop}%
\bibitem [{\citenamefont {Chabrier}\ and\ \citenamefont
  {Baraffe}(1997)}]{chabrier_structure_1997}%
  \BibitemOpen
  \bibfield  {author} {\bibinfo {author} {\bibfnamefont {G.}~\bibnamefont
  {Chabrier}}\ and\ \bibinfo {author} {\bibfnamefont {I.}~\bibnamefont
  {Baraffe}},\ }\href@noop {} {\bibfield  {journal} {\bibinfo  {journal}
  {Astronomy and Astrophysics}\ }\textbf {\bibinfo {volume} {327}},\ \bibinfo
  {pages} {1039} (\bibinfo {year} {1997})}\BibitemShut {NoStop}%
\bibitem [{\citenamefont {Chabrier}\ \emph {et~al.}(2000)\citenamefont
  {Chabrier}, \citenamefont {Baraffe}, \citenamefont {Allard},\ and\
  \citenamefont {Hauschildt}}]{chabrier_evolutionary_2000}%
  \BibitemOpen
  \bibfield  {author} {\bibinfo {author} {\bibfnamefont {G.}~\bibnamefont
  {Chabrier}}, \bibinfo {author} {\bibfnamefont {I.}~\bibnamefont {Baraffe}},
  \bibinfo {author} {\bibfnamefont {F.}~\bibnamefont {Allard}}, \ and\ \bibinfo
  {author} {\bibfnamefont {P.}~\bibnamefont {Hauschildt}},\ }\href {\doibase
  10.1086/309513} {\bibfield  {journal} {\bibinfo  {journal} {The Astrophysical
  Journal}\ }\textbf {\bibinfo {volume} {542}},\ \bibinfo {pages} {464}
  (\bibinfo {year} {2000})}\BibitemShut {NoStop}%
\bibitem [{\citenamefont {Burrows}\ \emph {et~al.}(2001)\citenamefont
  {Burrows}, \citenamefont {Hubbard}, \citenamefont {Lunine},\ and\
  \citenamefont {Liebert}}]{burrows_theory_2001}%
  \BibitemOpen
  \bibfield  {author} {\bibinfo {author} {\bibfnamefont {A.}~\bibnamefont
  {Burrows}}, \bibinfo {author} {\bibfnamefont {W.~B.}\ \bibnamefont
  {Hubbard}}, \bibinfo {author} {\bibfnamefont {J.~I.}\ \bibnamefont {Lunine}},
  \ and\ \bibinfo {author} {\bibfnamefont {J.}~\bibnamefont {Liebert}},\ }\href
  {\doibase 10.1103/RevModPhys.73.719} {\bibfield  {journal} {\bibinfo
  {journal} {Reviews of Modern Physics}\ }\textbf {\bibinfo {volume} {73}},\
  \bibinfo {pages} {719} (\bibinfo {year} {2001})}\BibitemShut {NoStop}%
\bibitem [{\citenamefont {Baraffe}\ \emph {et~al.}(2003)\citenamefont
  {Baraffe}, \citenamefont {Chabrier}, \citenamefont {Barman}, \citenamefont
  {Allard},\ and\ \citenamefont {Hauschildt}}]{baraffe_evolutionary_2003}%
  \BibitemOpen
  \bibfield  {author} {\bibinfo {author} {\bibfnamefont {I.}~\bibnamefont
  {Baraffe}}, \bibinfo {author} {\bibfnamefont {G.}~\bibnamefont {Chabrier}},
  \bibinfo {author} {\bibfnamefont {T.~S.}\ \bibnamefont {Barman}}, \bibinfo
  {author} {\bibfnamefont {F.}~\bibnamefont {Allard}}, \ and\ \bibinfo {author}
  {\bibfnamefont {P.~H.}\ \bibnamefont {Hauschildt}},\ }\href {\doibase
  10.1051/0004-6361:20030252} {\bibfield  {journal} {\bibinfo  {journal}
  {Astronomy and Astrophysics}\ }\textbf {\bibinfo {volume} {402}},\ \bibinfo
  {pages} {701} (\bibinfo {year} {2003})}\BibitemShut {NoStop}%
\bibitem [{\citenamefont {Saumon}\ and\ \citenamefont
  {Marley}(2008)}]{saumon_evolution_2008}%
  \BibitemOpen
  \bibfield  {author} {\bibinfo {author} {\bibfnamefont {D.}~\bibnamefont
  {Saumon}}\ and\ \bibinfo {author} {\bibfnamefont {M.~S.}\ \bibnamefont
  {Marley}},\ }\href {\doibase 10.1086/592734} {\bibfield  {journal} {\bibinfo
  {journal} {The Astrophysical Journal}\ }\textbf {\bibinfo {volume} {689}},\
  \bibinfo {pages} {1327} (\bibinfo {year} {2008})}\BibitemShut {NoStop}%
\bibitem [{\citenamefont {MacDonald}\ and\ \citenamefont
  {Mullan}(2009)}]{macdonald_structural_2009}%
  \BibitemOpen
  \bibfield  {author} {\bibinfo {author} {\bibfnamefont {J.}~\bibnamefont
  {MacDonald}}\ and\ \bibinfo {author} {\bibfnamefont {D.~J.}\ \bibnamefont
  {Mullan}},\ }\href {\doibase 10.1088/0004-637X/700/1/387} {\bibfield
  {journal} {\bibinfo  {journal} {The Astrophysical Journal}\ }\textbf
  {\bibinfo {volume} {700}},\ \bibinfo {pages} {387} (\bibinfo {year}
  {2009})}\BibitemShut {NoStop}%
\bibitem [{\citenamefont {Baraffe}\ \emph {et~al.}(2015)\citenamefont
  {Baraffe}, \citenamefont {Homeier}, \citenamefont {Allard},\ and\
  \citenamefont {Chabrier}}]{baraffe_new_2015}%
  \BibitemOpen
  \bibfield  {author} {\bibinfo {author} {\bibfnamefont {I.}~\bibnamefont
  {Baraffe}}, \bibinfo {author} {\bibfnamefont {D.}~\bibnamefont {Homeier}},
  \bibinfo {author} {\bibfnamefont {F.}~\bibnamefont {Allard}}, \ and\ \bibinfo
  {author} {\bibfnamefont {G.}~\bibnamefont {Chabrier}},\ }\href {\doibase
  10.1051/0004-6361/201425481} {\bibfield  {journal} {\bibinfo  {journal}
  {Astronomy and Astrophysics}\ }\textbf {\bibinfo {volume} {577}},\ \bibinfo
  {pages} {A42} (\bibinfo {year} {2015})}\BibitemShut {NoStop}%
\bibitem [{\citenamefont {Ackerman}\ and\ \citenamefont
  {Marley}(2001)}]{ackerman_precipitating_2001}%
  \BibitemOpen
  \bibfield  {author} {\bibinfo {author} {\bibfnamefont {A.~S.}\ \bibnamefont
  {Ackerman}}\ and\ \bibinfo {author} {\bibfnamefont {M.~S.}\ \bibnamefont
  {Marley}},\ }\href {\doibase 10.1086/321540} {\bibfield  {journal} {\bibinfo
  {journal} {The Astrophysical Journal}\ }\textbf {\bibinfo {volume} {556}},\
  \bibinfo {pages} {872} (\bibinfo {year} {2001})}\BibitemShut {NoStop}%
\bibitem [{\citenamefont {Cushing}\ \emph {et~al.}(2008)\citenamefont
  {Cushing}, \citenamefont {Marley}, \citenamefont {Saumon}, \citenamefont
  {Kelly}, \citenamefont {Vacca}, \citenamefont {Rayner}, \citenamefont
  {Freedman}, \citenamefont {Lodders},\ and\ \citenamefont
  {Roellig}}]{cushing_atmospheric_2008}%
  \BibitemOpen
  \bibfield  {author} {\bibinfo {author} {\bibfnamefont {M.~C.}\ \bibnamefont
  {Cushing}}, \bibinfo {author} {\bibfnamefont {M.~S.}\ \bibnamefont {Marley}},
  \bibinfo {author} {\bibfnamefont {D.}~\bibnamefont {Saumon}}, \bibinfo
  {author} {\bibfnamefont {B.~C.}\ \bibnamefont {Kelly}}, \bibinfo {author}
  {\bibfnamefont {W.~D.}\ \bibnamefont {Vacca}}, \bibinfo {author}
  {\bibfnamefont {J.~T.}\ \bibnamefont {Rayner}}, \bibinfo {author}
  {\bibfnamefont {R.~S.}\ \bibnamefont {Freedman}}, \bibinfo {author}
  {\bibfnamefont {K.}~\bibnamefont {Lodders}}, \ and\ \bibinfo {author}
  {\bibfnamefont {T.~L.}\ \bibnamefont {Roellig}},\ }\href {\doibase
  10.1086/526489} {\bibfield  {journal} {\bibinfo  {journal} {The Astrophysical
  Journal}\ }\textbf {\bibinfo {volume} {678}},\ \bibinfo {pages} {1372}
  (\bibinfo {year} {2008})}\BibitemShut {NoStop}%
\bibitem [{\citenamefont {Morley}\ \emph {et~al.}(2012)\citenamefont {Morley},
  \citenamefont {Fortney}, \citenamefont {Marley}, \citenamefont {Visscher},
  \citenamefont {Saumon},\ and\ \citenamefont
  {Leggett}}]{morley_neglected_2012}%
  \BibitemOpen
  \bibfield  {author} {\bibinfo {author} {\bibfnamefont {C.~V.}\ \bibnamefont
  {Morley}}, \bibinfo {author} {\bibfnamefont {J.~J.}\ \bibnamefont {Fortney}},
  \bibinfo {author} {\bibfnamefont {M.~S.}\ \bibnamefont {Marley}}, \bibinfo
  {author} {\bibfnamefont {C.}~\bibnamefont {Visscher}}, \bibinfo {author}
  {\bibfnamefont {D.}~\bibnamefont {Saumon}}, \ and\ \bibinfo {author}
  {\bibfnamefont {S.~K.}\ \bibnamefont {Leggett}},\ }\href {\doibase
  10.1088/0004-637X/756/2/172} {\bibfield  {journal} {\bibinfo  {journal} {The
  Astrophysical Journal}\ }\textbf {\bibinfo {volume} {756}},\ \bibinfo {pages}
  {172} (\bibinfo {year} {2012})}\BibitemShut {NoStop}%
\bibitem [{\citenamefont {Marley}\ and\ \citenamefont
  {Robinson}(2015)}]{marley_cool_2015}%
  \BibitemOpen
  \bibfield  {author} {\bibinfo {author} {\bibfnamefont {M.~S.}\ \bibnamefont
  {Marley}}\ and\ \bibinfo {author} {\bibfnamefont {T.~D.}\ \bibnamefont
  {Robinson}},\ }\href {\doibase 10.1146/annurev-astro-082214-122522}
  {\bibfield  {journal} {\bibinfo  {journal} {Annual Review of Astronomy and
  Astrophysics}\ }\textbf {\bibinfo {volume} {53}},\ \bibinfo {pages} {279}
  (\bibinfo {year} {2015})}\BibitemShut {NoStop}%
\bibitem [{\citenamefont {Salpeter}(1992)}]{salpeter_minimum_1992}%
  \BibitemOpen
  \bibfield  {author} {\bibinfo {author} {\bibfnamefont {E.~E.}\ \bibnamefont
  {Salpeter}},\ }\href {\doibase 10.1086/171502} {\bibfield  {journal}
  {\bibinfo  {journal} {The Astrophysical Journal}\ }\textbf {\bibinfo {volume}
  {393}},\ \bibinfo {pages} {258} (\bibinfo {year} {1992})}\BibitemShut
  {NoStop}%
\bibitem [{\citenamefont {Hansen}(1999)}]{hansen_origin_1999}%
  \BibitemOpen
  \bibfield  {author} {\bibinfo {author} {\bibfnamefont {B.~M.~S.}\
  \bibnamefont {Hansen}},\ }\href {\doibase 10.1086/312023} {\bibfield
  {journal} {\bibinfo  {journal} {The Astrophysical Journal Letters}\ }\textbf
  {\bibinfo {volume} {517}},\ \bibinfo {pages} {L39} (\bibinfo {year}
  {1999})}\BibitemShut {NoStop}%
\bibitem [{\citenamefont {{Lynden-Bell}}\ and\ \citenamefont
  {Tout}(2001)}]{lynden-bell_russell_2001}%
  \BibitemOpen
  \bibfield  {author} {\bibinfo {author} {\bibfnamefont {D.}~\bibnamefont
  {{Lynden-Bell}}}\ and\ \bibinfo {author} {\bibfnamefont {C.~A.}\ \bibnamefont
  {Tout}},\ }\href {\doibase 10.1086/322454} {\bibfield  {journal} {\bibinfo
  {journal} {The Astrophysical Journal}\ }\textbf {\bibinfo {volume} {558}},\
  \bibinfo {pages} {1} (\bibinfo {year} {2001})}\BibitemShut {NoStop}%
\bibitem [{\citenamefont {Alcock}\ \emph {et~al.}(2000)\citenamefont {Alcock},
  \citenamefont {Allsman}, \citenamefont {Alves}, \citenamefont {Axelrod},
  \citenamefont {Becker}, \citenamefont {Bennett}, \citenamefont {Cook},
  \citenamefont {Dalal}, \citenamefont {Drake}, \citenamefont {Freeman},
  \citenamefont {Geha}, \citenamefont {Griest}, \citenamefont {Lehner},
  \citenamefont {Marshall}, \citenamefont {Minniti}, \citenamefont {Nelson},
  \citenamefont {Peterson}, \citenamefont {Popowski}, \citenamefont {Pratt},
  \citenamefont {Quinn}, \citenamefont {Stubbs}, \citenamefont {Sutherland},
  \citenamefont {Tomaney}, \citenamefont {Vandehei},\ and\ \citenamefont
  {Welch}}]{alcock_macho_2000}%
  \BibitemOpen
  \bibfield  {author} {\bibinfo {author} {\bibfnamefont {C.}~\bibnamefont
  {Alcock}}, \bibinfo {author} {\bibfnamefont {R.~A.}\ \bibnamefont {Allsman}},
  \bibinfo {author} {\bibfnamefont {D.~R.}\ \bibnamefont {Alves}}, \bibinfo
  {author} {\bibfnamefont {T.~S.}\ \bibnamefont {Axelrod}}, \bibinfo {author}
  {\bibfnamefont {A.~C.}\ \bibnamefont {Becker}}, \bibinfo {author}
  {\bibfnamefont {D.~P.}\ \bibnamefont {Bennett}}, \bibinfo {author}
  {\bibfnamefont {K.~H.}\ \bibnamefont {Cook}}, \bibinfo {author}
  {\bibfnamefont {N.}~\bibnamefont {Dalal}}, \bibinfo {author} {\bibfnamefont
  {A.~J.}\ \bibnamefont {Drake}}, \bibinfo {author} {\bibfnamefont {K.~C.}\
  \bibnamefont {Freeman}}, \bibinfo {author} {\bibfnamefont {M.}~\bibnamefont
  {Geha}}, \bibinfo {author} {\bibfnamefont {K.}~\bibnamefont {Griest}},
  \bibinfo {author} {\bibfnamefont {M.~J.}\ \bibnamefont {Lehner}}, \bibinfo
  {author} {\bibfnamefont {S.~L.}\ \bibnamefont {Marshall}}, \bibinfo {author}
  {\bibfnamefont {D.}~\bibnamefont {Minniti}}, \bibinfo {author} {\bibfnamefont
  {C.~A.}\ \bibnamefont {Nelson}}, \bibinfo {author} {\bibfnamefont {B.~A.}\
  \bibnamefont {Peterson}}, \bibinfo {author} {\bibfnamefont {P.}~\bibnamefont
  {Popowski}}, \bibinfo {author} {\bibfnamefont {M.~R.}\ \bibnamefont {Pratt}},
  \bibinfo {author} {\bibfnamefont {P.~J.}\ \bibnamefont {Quinn}}, \bibinfo
  {author} {\bibfnamefont {C.~W.}\ \bibnamefont {Stubbs}}, \bibinfo {author}
  {\bibfnamefont {W.}~\bibnamefont {Sutherland}}, \bibinfo {author}
  {\bibfnamefont {A.~B.}\ \bibnamefont {Tomaney}}, \bibinfo {author}
  {\bibfnamefont {T.}~\bibnamefont {Vandehei}}, \ and\ \bibinfo {author}
  {\bibfnamefont {D.}~\bibnamefont {Welch}},\ }\href {\doibase 10.1086/309512}
  {\bibfield  {journal} {\bibinfo  {journal} {The Astrophysical Journal}\
  }\textbf {\bibinfo {volume} {542}},\ \bibinfo {pages} {281} (\bibinfo {year}
  {2000})}\BibitemShut {NoStop}%
\bibitem [{\citenamefont {Tisserand}\ \emph {et~al.}(2007)\citenamefont
  {Tisserand}, \citenamefont {Le~Guillou}, \citenamefont {Afonso},
  \citenamefont {Albert}, \citenamefont {Andersen}, \citenamefont {Ansari},
  \citenamefont {Aubourg}, \citenamefont {Bareyre}, \citenamefont {Beaulieu},
  \citenamefont {Charlot}, \citenamefont {Coutures}, \citenamefont {Ferlet},
  \citenamefont {Fouqu\'e}, \citenamefont {Glicenstein}, \citenamefont
  {Goldman}, \citenamefont {Gould}, \citenamefont {Graff}, \citenamefont
  {Gros}, \citenamefont {Haissinski}, \citenamefont {Hamadache}, \citenamefont
  {{de Kat}}, \citenamefont {Lasserre}, \citenamefont {Lesquoy}, \citenamefont
  {Loup}, \citenamefont {Magneville}, \citenamefont {Marquette}, \citenamefont
  {Maurice}, \citenamefont {Maury}, \citenamefont {Milsztajn}, \citenamefont
  {Moniez}, \citenamefont {{Palanque-Delabrouille}}, \citenamefont {Perdereau},
  \citenamefont {Rahal}, \citenamefont {Rich}, \citenamefont {Spiro},
  \citenamefont {{Vidal-Madjar}}, \citenamefont {Vigroux}, \citenamefont
  {Zylberajch},\ and\ \citenamefont {{EROS-2
  Collaboration}}}]{tisserand_limits_2007}%
  \BibitemOpen
  \bibfield  {author} {\bibinfo {author} {\bibfnamefont {P.}~\bibnamefont
  {Tisserand}}, \bibinfo {author} {\bibfnamefont {L.}~\bibnamefont
  {Le~Guillou}}, \bibinfo {author} {\bibfnamefont {C.}~\bibnamefont {Afonso}},
  \bibinfo {author} {\bibfnamefont {J.~N.}\ \bibnamefont {Albert}}, \bibinfo
  {author} {\bibfnamefont {J.}~\bibnamefont {Andersen}}, \bibinfo {author}
  {\bibfnamefont {R.}~\bibnamefont {Ansari}}, \bibinfo {author} {\bibfnamefont
  {E.}~\bibnamefont {Aubourg}}, \bibinfo {author} {\bibfnamefont
  {P.}~\bibnamefont {Bareyre}}, \bibinfo {author} {\bibfnamefont {J.~P.}\
  \bibnamefont {Beaulieu}}, \bibinfo {author} {\bibfnamefont {X.}~\bibnamefont
  {Charlot}}, \bibinfo {author} {\bibfnamefont {C.}~\bibnamefont {Coutures}},
  \bibinfo {author} {\bibfnamefont {R.}~\bibnamefont {Ferlet}}, \bibinfo
  {author} {\bibfnamefont {P.}~\bibnamefont {Fouqu\'e}}, \bibinfo {author}
  {\bibfnamefont {J.~F.}\ \bibnamefont {Glicenstein}}, \bibinfo {author}
  {\bibfnamefont {B.}~\bibnamefont {Goldman}}, \bibinfo {author} {\bibfnamefont
  {A.}~\bibnamefont {Gould}}, \bibinfo {author} {\bibfnamefont
  {D.}~\bibnamefont {Graff}}, \bibinfo {author} {\bibfnamefont
  {M.}~\bibnamefont {Gros}}, \bibinfo {author} {\bibfnamefont {J.}~\bibnamefont
  {Haissinski}}, \bibinfo {author} {\bibfnamefont {C.}~\bibnamefont
  {Hamadache}}, \bibinfo {author} {\bibfnamefont {J.}~\bibnamefont {{de Kat}}},
  \bibinfo {author} {\bibfnamefont {T.}~\bibnamefont {Lasserre}}, \bibinfo
  {author} {\bibfnamefont {E.}~\bibnamefont {Lesquoy}}, \bibinfo {author}
  {\bibfnamefont {C.}~\bibnamefont {Loup}}, \bibinfo {author} {\bibfnamefont
  {C.}~\bibnamefont {Magneville}}, \bibinfo {author} {\bibfnamefont {J.~B.}\
  \bibnamefont {Marquette}}, \bibinfo {author} {\bibfnamefont {E.}~\bibnamefont
  {Maurice}}, \bibinfo {author} {\bibfnamefont {A.}~\bibnamefont {Maury}},
  \bibinfo {author} {\bibfnamefont {A.}~\bibnamefont {Milsztajn}}, \bibinfo
  {author} {\bibfnamefont {M.}~\bibnamefont {Moniez}}, \bibinfo {author}
  {\bibfnamefont {N.}~\bibnamefont {{Palanque-Delabrouille}}}, \bibinfo
  {author} {\bibfnamefont {O.}~\bibnamefont {Perdereau}}, \bibinfo {author}
  {\bibfnamefont {Y.~R.}\ \bibnamefont {Rahal}}, \bibinfo {author}
  {\bibfnamefont {J.}~\bibnamefont {Rich}}, \bibinfo {author} {\bibfnamefont
  {M.}~\bibnamefont {Spiro}}, \bibinfo {author} {\bibfnamefont
  {A.}~\bibnamefont {{Vidal-Madjar}}}, \bibinfo {author} {\bibfnamefont
  {L.}~\bibnamefont {Vigroux}}, \bibinfo {author} {\bibfnamefont
  {S.}~\bibnamefont {Zylberajch}}, \ and\ \bibinfo {author} {\bibnamefont
  {{EROS-2 Collaboration}}},\ }\href {\doibase 10.1051/0004-6361:20066017}
  {\bibfield  {journal} {\bibinfo  {journal} {Astronomy and Astrophysics}\
  }\textbf {\bibinfo {volume} {469}},\ \bibinfo {pages} {387} (\bibinfo {year}
  {2007})}\BibitemShut {NoStop}%
\bibitem [{\citenamefont {Auddy}\ \emph {et~al.}(2016)\citenamefont {Auddy},
  \citenamefont {Basu},\ and\ \citenamefont {Valluri}}]{auddy_analytic_2016}%
  \BibitemOpen
  \bibfield  {author} {\bibinfo {author} {\bibfnamefont {S.}~\bibnamefont
  {Auddy}}, \bibinfo {author} {\bibfnamefont {S.}~\bibnamefont {Basu}}, \ and\
  \bibinfo {author} {\bibfnamefont {S.~R.}\ \bibnamefont {Valluri}},\ }\href
  {\doibase 10.1155/2016/5743272} {\bibfield  {journal} {\bibinfo  {journal}
  {Advances in Astronomy}\ }\textbf {\bibinfo {volume} {2016}},\ \bibinfo
  {pages} {574327} (\bibinfo {year} {2016})}\BibitemShut {NoStop}%
\bibitem [{\citenamefont {Burrows}\ and\ \citenamefont
  {Liebert}(1993)}]{burrows_science_1993}%
  \BibitemOpen
  \bibfield  {author} {\bibinfo {author} {\bibfnamefont {A.}~\bibnamefont
  {Burrows}}\ and\ \bibinfo {author} {\bibfnamefont {J.}~\bibnamefont
  {Liebert}},\ }\href {\doibase 10.1103/RevModPhys.65.301} {\bibfield
  {journal} {\bibinfo  {journal} {Reviews of Modern Physics}\ }\textbf
  {\bibinfo {volume} {65}},\ \bibinfo {pages} {301} (\bibinfo {year}
  {1993})}\BibitemShut {NoStop}%
\bibitem [{\citenamefont {Paxton}\ \emph {et~al.}(2011)\citenamefont {Paxton},
  \citenamefont {Bildsten}, \citenamefont {Dotter}, \citenamefont {Herwig},
  \citenamefont {Lesaffre},\ and\ \citenamefont
  {Timmes}}]{paxton_modules_2011}%
  \BibitemOpen
  \bibfield  {author} {\bibinfo {author} {\bibfnamefont {B.}~\bibnamefont
  {Paxton}}, \bibinfo {author} {\bibfnamefont {L.}~\bibnamefont {Bildsten}},
  \bibinfo {author} {\bibfnamefont {A.}~\bibnamefont {Dotter}}, \bibinfo
  {author} {\bibfnamefont {F.}~\bibnamefont {Herwig}}, \bibinfo {author}
  {\bibfnamefont {P.}~\bibnamefont {Lesaffre}}, \ and\ \bibinfo {author}
  {\bibfnamefont {F.}~\bibnamefont {Timmes}},\ }\href {\doibase
  10.1088/0067-0049/192/1/3} {\bibfield  {journal} {\bibinfo  {journal} {The
  Astrophysical Journal Supplement Series}\ }\textbf {\bibinfo {volume}
  {192}},\ \bibinfo {pages} {3} (\bibinfo {year} {2011})}\BibitemShut {NoStop}%
\bibitem [{\citenamefont {Paxton}\ \emph {et~al.}(2013)\citenamefont {Paxton},
  \citenamefont {Cantiello}, \citenamefont {Arras}, \citenamefont {Bildsten},
  \citenamefont {Brown}, \citenamefont {Dotter}, \citenamefont {Mankovich},
  \citenamefont {Montgomery}, \citenamefont {Stello}, \citenamefont {Timmes},\
  and\ \citenamefont {Townsend}}]{paxton_modules_2013}%
  \BibitemOpen
  \bibfield  {author} {\bibinfo {author} {\bibfnamefont {B.}~\bibnamefont
  {Paxton}}, \bibinfo {author} {\bibfnamefont {M.}~\bibnamefont {Cantiello}},
  \bibinfo {author} {\bibfnamefont {P.}~\bibnamefont {Arras}}, \bibinfo
  {author} {\bibfnamefont {L.}~\bibnamefont {Bildsten}}, \bibinfo {author}
  {\bibfnamefont {E.~F.}\ \bibnamefont {Brown}}, \bibinfo {author}
  {\bibfnamefont {A.}~\bibnamefont {Dotter}}, \bibinfo {author} {\bibfnamefont
  {C.}~\bibnamefont {Mankovich}}, \bibinfo {author} {\bibfnamefont {M.~H.}\
  \bibnamefont {Montgomery}}, \bibinfo {author} {\bibfnamefont
  {D.}~\bibnamefont {Stello}}, \bibinfo {author} {\bibfnamefont {F.~X.}\
  \bibnamefont {Timmes}}, \ and\ \bibinfo {author} {\bibfnamefont
  {R.}~\bibnamefont {Townsend}},\ }\href {\doibase 10.1088/0067-0049/208/1/4}
  {\bibfield  {journal} {\bibinfo  {journal} {The Astrophysical Journal
  Supplement Series}\ }\textbf {\bibinfo {volume} {208}},\ \bibinfo {pages} {4}
  (\bibinfo {year} {2013})}\BibitemShut {NoStop}%
\bibitem [{\citenamefont {Paxton}\ \emph {et~al.}(2015)\citenamefont {Paxton},
  \citenamefont {Marchant}, \citenamefont {Schwab}, \citenamefont {Bauer},
  \citenamefont {Bildsten}, \citenamefont {Cantiello}, \citenamefont {Dessart},
  \citenamefont {Farmer}, \citenamefont {Hu}, \citenamefont {Langer},
  \citenamefont {Townsend}, \citenamefont {Townsley},\ and\ \citenamefont
  {Timmes}}]{paxton_modules_2015}%
  \BibitemOpen
  \bibfield  {author} {\bibinfo {author} {\bibfnamefont {B.}~\bibnamefont
  {Paxton}}, \bibinfo {author} {\bibfnamefont {P.}~\bibnamefont {Marchant}},
  \bibinfo {author} {\bibfnamefont {J.}~\bibnamefont {Schwab}}, \bibinfo
  {author} {\bibfnamefont {E.~B.}\ \bibnamefont {Bauer}}, \bibinfo {author}
  {\bibfnamefont {L.}~\bibnamefont {Bildsten}}, \bibinfo {author}
  {\bibfnamefont {M.}~\bibnamefont {Cantiello}}, \bibinfo {author}
  {\bibfnamefont {L.}~\bibnamefont {Dessart}}, \bibinfo {author} {\bibfnamefont
  {R.}~\bibnamefont {Farmer}}, \bibinfo {author} {\bibfnamefont
  {H.}~\bibnamefont {Hu}}, \bibinfo {author} {\bibfnamefont {N.}~\bibnamefont
  {Langer}}, \bibinfo {author} {\bibfnamefont {R.~H.~D.}\ \bibnamefont
  {Townsend}}, \bibinfo {author} {\bibfnamefont {D.~M.}\ \bibnamefont
  {Townsley}}, \ and\ \bibinfo {author} {\bibfnamefont {F.~X.}\ \bibnamefont
  {Timmes}},\ }\href {\doibase 10.1088/0067-0049/220/1/15} {\bibfield
  {journal} {\bibinfo  {journal} {The Astrophysical Journal Supplement Series}\
  }\textbf {\bibinfo {volume} {220}},\ \bibinfo {pages} {15} (\bibinfo {year}
  {2015})}\BibitemShut {NoStop}%
\bibitem [{\citenamefont {Cooke}\ \emph {et~al.}(2016)\citenamefont {Cooke},
  \citenamefont {Pettini}, \citenamefont {Nollett},\ and\ \citenamefont
  {Jorgenson}}]{cooke_primordial_2016}%
  \BibitemOpen
  \bibfield  {author} {\bibinfo {author} {\bibfnamefont {R.~J.}\ \bibnamefont
  {Cooke}}, \bibinfo {author} {\bibfnamefont {M.}~\bibnamefont {Pettini}},
  \bibinfo {author} {\bibfnamefont {K.~M.}\ \bibnamefont {Nollett}}, \ and\
  \bibinfo {author} {\bibfnamefont {R.}~\bibnamefont {Jorgenson}},\ }\href
  {\doibase 10.3847/0004-637X/830/2/148} {\bibfield  {journal} {\bibinfo
  {journal} {The Astrophysical Journal}\ }\textbf {\bibinfo {volume} {830}},\
  \bibinfo {pages} {148} (\bibinfo {year} {2016})}\BibitemShut {NoStop}%
\bibitem [{\citenamefont {Choi}\ \emph {et~al.}(2016)\citenamefont {Choi},
  \citenamefont {Dotter}, \citenamefont {Conroy}, \citenamefont {Cantiello},
  \citenamefont {Paxton},\ and\ \citenamefont {Johnson}}]{choi_mesa_2016}%
  \BibitemOpen
  \bibfield  {author} {\bibinfo {author} {\bibfnamefont {J.}~\bibnamefont
  {Choi}}, \bibinfo {author} {\bibfnamefont {A.}~\bibnamefont {Dotter}},
  \bibinfo {author} {\bibfnamefont {C.}~\bibnamefont {Conroy}}, \bibinfo
  {author} {\bibfnamefont {M.}~\bibnamefont {Cantiello}}, \bibinfo {author}
  {\bibfnamefont {B.}~\bibnamefont {Paxton}}, \ and\ \bibinfo {author}
  {\bibfnamefont {B.~D.}\ \bibnamefont {Johnson}},\ }\href {\doibase
  10.3847/0004-637X/823/2/102} {\bibfield  {journal} {\bibinfo  {journal} {The
  Astrophysical Journal}\ }\textbf {\bibinfo {volume} {823}},\ \bibinfo {pages}
  {102} (\bibinfo {year} {2016})}\BibitemShut {NoStop}%
\bibitem [{\citenamefont {Laughlin}\ \emph {et~al.}(1997)\citenamefont
  {Laughlin}, \citenamefont {Bodenheimer},\ and\ \citenamefont
  {Adams}}]{laughlin_end_1997}%
  \BibitemOpen
  \bibfield  {author} {\bibinfo {author} {\bibfnamefont {G.}~\bibnamefont
  {Laughlin}}, \bibinfo {author} {\bibfnamefont {P.}~\bibnamefont
  {Bodenheimer}}, \ and\ \bibinfo {author} {\bibfnamefont {F.~C.}\ \bibnamefont
  {Adams}},\ }\href@noop {} {\bibfield  {journal} {\bibinfo  {journal} {The
  Astrophysical Journal}\ }\textbf {\bibinfo {volume} {482}},\ \bibinfo {pages}
  {420} (\bibinfo {year} {1997})}\BibitemShut {NoStop}%
\bibitem [{\citenamefont {Ferguson}\ \emph {et~al.}(2005)\citenamefont
  {Ferguson}, \citenamefont {Alexander}, \citenamefont {Allard}, \citenamefont
  {Barman}, \citenamefont {Bodnarik}, \citenamefont {Hauschildt}, \citenamefont
  {{Heffner-Wong}},\ and\ \citenamefont
  {Tamanai}}]{ferguson_low-temperature_2005}%
  \BibitemOpen
  \bibfield  {author} {\bibinfo {author} {\bibfnamefont {J.~W.}\ \bibnamefont
  {Ferguson}}, \bibinfo {author} {\bibfnamefont {D.~R.}\ \bibnamefont
  {Alexander}}, \bibinfo {author} {\bibfnamefont {F.}~\bibnamefont {Allard}},
  \bibinfo {author} {\bibfnamefont {T.}~\bibnamefont {Barman}}, \bibinfo
  {author} {\bibfnamefont {J.~G.}\ \bibnamefont {Bodnarik}}, \bibinfo {author}
  {\bibfnamefont {P.~H.}\ \bibnamefont {Hauschildt}}, \bibinfo {author}
  {\bibfnamefont {A.}~\bibnamefont {{Heffner-Wong}}}, \ and\ \bibinfo {author}
  {\bibfnamefont {A.}~\bibnamefont {Tamanai}},\ }\href {\doibase
  10.1086/428642} {\bibfield  {journal} {\bibinfo  {journal} {The Astrophysical
  Journal}\ }\textbf {\bibinfo {volume} {623}},\ \bibinfo {pages} {585}
  (\bibinfo {year} {2005})}\BibitemShut {NoStop}%
\bibitem [{\citenamefont {Burrows}\ \emph {et~al.}(2011)\citenamefont
  {Burrows}, \citenamefont {Heng},\ and\ \citenamefont
  {Nampaisarn}}]{burrows_dependence_2011}%
  \BibitemOpen
  \bibfield  {author} {\bibinfo {author} {\bibfnamefont {A.}~\bibnamefont
  {Burrows}}, \bibinfo {author} {\bibfnamefont {K.}~\bibnamefont {Heng}}, \
  and\ \bibinfo {author} {\bibfnamefont {T.}~\bibnamefont {Nampaisarn}},\
  }\href {\doibase 10.1088/0004-637X/736/1/47} {\bibfield  {journal} {\bibinfo
  {journal} {The Astrophysical Journal}\ }\textbf {\bibinfo {volume} {736}},\
  \bibinfo {pages} {47} (\bibinfo {year} {2011})}\BibitemShut {NoStop}%
\bibitem [{\citenamefont {Dieterich}\ \emph {et~al.}(2014)\citenamefont
  {Dieterich}, \citenamefont {Henry}, \citenamefont {Jao}, \citenamefont
  {Winters}, \citenamefont {Hosey}, \citenamefont {Riedel},\ and\ \citenamefont
  {Subasavage}}]{dieterich_solar_2014}%
  \BibitemOpen
  \bibfield  {author} {\bibinfo {author} {\bibfnamefont {S.~B.}\ \bibnamefont
  {Dieterich}}, \bibinfo {author} {\bibfnamefont {T.~J.}\ \bibnamefont
  {Henry}}, \bibinfo {author} {\bibfnamefont {W.-C.}\ \bibnamefont {Jao}},
  \bibinfo {author} {\bibfnamefont {J.~G.}\ \bibnamefont {Winters}}, \bibinfo
  {author} {\bibfnamefont {A.~D.}\ \bibnamefont {Hosey}}, \bibinfo {author}
  {\bibfnamefont {A.~R.}\ \bibnamefont {Riedel}}, \ and\ \bibinfo {author}
  {\bibfnamefont {J.~P.}\ \bibnamefont {Subasavage}},\ }\href {\doibase
  10.1088/0004-6256/147/5/94} {\bibfield  {journal} {\bibinfo  {journal} {The
  Astronomical Journal}\ }\textbf {\bibinfo {volume} {147}},\ \bibinfo {pages}
  {94} (\bibinfo {year} {2014})}\BibitemShut {NoStop}%
\bibitem [{\citenamefont {Prialnik}\ and\ \citenamefont
  {Livio}(1985)}]{prialnik_outcome_1985}%
  \BibitemOpen
  \bibfield  {author} {\bibinfo {author} {\bibfnamefont {D.}~\bibnamefont
  {Prialnik}}\ and\ \bibinfo {author} {\bibfnamefont {M.}~\bibnamefont
  {Livio}},\ }\href {\doibase 10.1093/mnras/216.1.37} {\bibfield  {journal}
  {\bibinfo  {journal} {Monthly Notices of the Royal Astronomical Society}\
  }\textbf {\bibinfo {volume} {216}},\ \bibinfo {pages} {37} (\bibinfo {year}
  {1985})}\BibitemShut {NoStop}%
\bibitem [{\citenamefont {Hartmann}\ \emph {et~al.}(1997)\citenamefont
  {Hartmann}, \citenamefont {Cassen},\ and\ \citenamefont
  {Kenyon}}]{hartmann_disk_1997}%
  \BibitemOpen
  \bibfield  {author} {\bibinfo {author} {\bibfnamefont {L.}~\bibnamefont
  {Hartmann}}, \bibinfo {author} {\bibfnamefont {P.}~\bibnamefont {Cassen}}, \
  and\ \bibinfo {author} {\bibfnamefont {S.~J.}\ \bibnamefont {Kenyon}},\
  }\href {\doibase 10.1086/303547} {\bibfield  {journal} {\bibinfo  {journal}
  {The Astrophysical Journal}\ }\textbf {\bibinfo {volume} {475}},\ \bibinfo
  {pages} {770} (\bibinfo {year} {1997})}\BibitemShut {NoStop}%
\bibitem [{\citenamefont {Lenzuni}\ \emph {et~al.}(1992)\citenamefont
  {Lenzuni}, \citenamefont {Chernoff},\ and\ \citenamefont
  {Salpeter}}]{lenzuni_formation_1992}%
  \BibitemOpen
  \bibfield  {author} {\bibinfo {author} {\bibfnamefont {P.}~\bibnamefont
  {Lenzuni}}, \bibinfo {author} {\bibfnamefont {D.~F.}\ \bibnamefont
  {Chernoff}}, \ and\ \bibinfo {author} {\bibfnamefont {E.~E.}\ \bibnamefont
  {Salpeter}},\ }\href {\doibase 10.1086/171501} {\bibfield  {journal}
  {\bibinfo  {journal} {The Astrophysical Journal}\ }\textbf {\bibinfo {volume}
  {393}},\ \bibinfo {pages} {232} (\bibinfo {year} {1992})}\BibitemShut
  {NoStop}%
\bibitem [{\citenamefont {Gold}\ \emph {et~al.}(1984)\citenamefont {Gold},
  \citenamefont {Zinnecker}, \citenamefont {McCrea}, \citenamefont {Tayler},
  \citenamefont {Woolfson},\ and\ \citenamefont {Gough}}]{gold_early_1984}%
  \BibitemOpen
  \bibfield  {author} {\bibinfo {author} {\bibfnamefont {T.}~\bibnamefont
  {Gold}}, \bibinfo {author} {\bibfnamefont {H.}~\bibnamefont {Zinnecker}},
  \bibinfo {author} {\bibfnamefont {W.~H.}\ \bibnamefont {McCrea}}, \bibinfo
  {author} {\bibfnamefont {R.~J.}\ \bibnamefont {Tayler}}, \bibinfo {author}
  {\bibfnamefont {M.~M.}\ \bibnamefont {Woolfson}}, \ and\ \bibinfo {author}
  {\bibfnamefont {D.~O.}\ \bibnamefont {Gough}},\ }\href@noop {} {\bibfield
  {journal} {\bibinfo  {journal} {Philosophical Transactions of the Royal
  Society of London. Series A, Mathematical and Physical Sciences}\ }\textbf
  {\bibinfo {volume} {313}},\ \bibinfo {pages} {39} (\bibinfo {year}
  {1984})}\BibitemShut {NoStop}%
\bibitem [{\citenamefont {Bondi}\ and\ \citenamefont
  {Hoyle}(1944)}]{bondi_mechanism_1944}%
  \BibitemOpen
  \bibfield  {author} {\bibinfo {author} {\bibfnamefont {H.}~\bibnamefont
  {Bondi}}\ and\ \bibinfo {author} {\bibfnamefont {F.}~\bibnamefont {Hoyle}},\
  }\href {\doibase 10.1093/mnras/104.5.273} {\bibfield  {journal} {\bibinfo
  {journal} {Monthly Notices of the Royal Astronomical Society}\ }\textbf
  {\bibinfo {volume} {104}},\ \bibinfo {pages} {273} (\bibinfo {year}
  {1944})}\BibitemShut {NoStop}%
\bibitem [{\citenamefont {Kraft}\ \emph {et~al.}(1962)\citenamefont {Kraft},
  \citenamefont {Mathews},\ and\ \citenamefont
  {Greenstein}}]{kraft_binary_1962}%
  \BibitemOpen
  \bibfield  {author} {\bibinfo {author} {\bibfnamefont {R.~P.}\ \bibnamefont
  {Kraft}}, \bibinfo {author} {\bibfnamefont {J.}~\bibnamefont {Mathews}}, \
  and\ \bibinfo {author} {\bibfnamefont {J.~L.}\ \bibnamefont {Greenstein}},\
  }\href {\doibase 10.1086/147381} {\bibfield  {journal} {\bibinfo  {journal}
  {The Astrophysical Journal}\ }\textbf {\bibinfo {volume} {136}},\ \bibinfo
  {pages} {312} (\bibinfo {year} {1962})}\BibitemShut {NoStop}%
\bibitem [{\citenamefont {Paczy\'nski}(1967)}]{paczynski_gravitational_1967}%
  \BibitemOpen
  \bibfield  {author} {\bibinfo {author} {\bibfnamefont {B.}~\bibnamefont
  {Paczy\'nski}},\ }\href@noop {} {\bibfield  {journal} {\bibinfo  {journal}
  {Acta Astronomica}\ }\textbf {\bibinfo {volume} {17}},\ \bibinfo {pages}
  {287} (\bibinfo {year} {1967})}\BibitemShut {NoStop}%
\bibitem [{\citenamefont {Faulkner}(1971)}]{faulkner_ultrashort-period_1971}%
  \BibitemOpen
  \bibfield  {author} {\bibinfo {author} {\bibfnamefont {J.}~\bibnamefont
  {Faulkner}},\ }\href {\doibase 10.1086/180848} {\bibfield  {journal}
  {\bibinfo  {journal} {The Astrophysical Journal Letters}\ }\textbf {\bibinfo
  {volume} {170}},\ \bibinfo {pages} {L99} (\bibinfo {year}
  {1971})}\BibitemShut {NoStop}%
\bibitem [{\citenamefont {Peters}(1964)}]{peters_gravitational_1964}%
  \BibitemOpen
  \bibfield  {author} {\bibinfo {author} {\bibfnamefont {P.~C.}\ \bibnamefont
  {Peters}},\ }\href {\doibase 10.1103/PhysRev.136.B1224} {\bibfield  {journal}
  {\bibinfo  {journal} {Physical Review}\ }\textbf {\bibinfo {volume} {136}},\
  \bibinfo {pages} {B1224} (\bibinfo {year} {1964})}\BibitemShut {NoStop}%
\bibitem [{\citenamefont {Landau}\ and\ \citenamefont
  {Lifshitz}(1975)}]{landau_classical_1975}%
  \BibitemOpen
  \bibfield  {author} {\bibinfo {author} {\bibfnamefont {L.~D.}\ \bibnamefont
  {Landau}}\ and\ \bibinfo {author} {\bibfnamefont {E.~M.}\ \bibnamefont
  {Lifshitz}},\ }\href@noop {} {\emph {\bibinfo {title} {The Classical Theory
  of Fields}}}\ (\bibinfo {year} {1975})\BibitemShut {NoStop}%
\bibitem [{\citenamefont {Eggleton}(1983)}]{eggleton_approximations_1983}%
  \BibitemOpen
  \bibfield  {author} {\bibinfo {author} {\bibfnamefont {P.~P.}\ \bibnamefont
  {Eggleton}},\ }\href {\doibase 10.1086/160960} {\bibfield  {journal}
  {\bibinfo  {journal} {The Astrophysical Journal}\ }\textbf {\bibinfo {volume}
  {268}},\ \bibinfo {pages} {368} (\bibinfo {year} {1983})}\BibitemShut
  {NoStop}%
\bibitem [{\citenamefont {King}(1988)}]{king_evolution_1988}%
  \BibitemOpen
  \bibfield  {author} {\bibinfo {author} {\bibfnamefont {A.~R.}\ \bibnamefont
  {King}},\ }\href@noop {} {\bibfield  {journal} {\bibinfo  {journal}
  {Quarterly Journal of the Royal Astronomical Society}\ }\textbf {\bibinfo
  {volume} {29}},\ \bibinfo {pages} {1} (\bibinfo {year} {1988})}\BibitemShut
  {NoStop}%
\bibitem [{\citenamefont {Hurley}\ \emph {et~al.}(2002)\citenamefont {Hurley},
  \citenamefont {Tout},\ and\ \citenamefont {Pols}}]{hurley_evolution_2002}%
  \BibitemOpen
  \bibfield  {author} {\bibinfo {author} {\bibfnamefont {J.~R.}\ \bibnamefont
  {Hurley}}, \bibinfo {author} {\bibfnamefont {C.~A.}\ \bibnamefont {Tout}}, \
  and\ \bibinfo {author} {\bibfnamefont {O.~R.}\ \bibnamefont {Pols}},\ }\href
  {\doibase 10.1046/j.1365-8711.2002.05038.x} {\bibfield  {journal} {\bibinfo
  {journal} {Monthly Notices of the Royal Astronomical Society}\ }\textbf
  {\bibinfo {volume} {329}},\ \bibinfo {pages} {897} (\bibinfo {year}
  {2002})}\BibitemShut {NoStop}%
\bibitem [{\citenamefont {Knigge}\ \emph {et~al.}(2011)\citenamefont {Knigge},
  \citenamefont {Baraffe},\ and\ \citenamefont
  {Patterson}}]{knigge_evolution_2011}%
  \BibitemOpen
  \bibfield  {author} {\bibinfo {author} {\bibfnamefont {C.}~\bibnamefont
  {Knigge}}, \bibinfo {author} {\bibfnamefont {I.}~\bibnamefont {Baraffe}}, \
  and\ \bibinfo {author} {\bibfnamefont {J.}~\bibnamefont {Patterson}},\ }\href
  {\doibase 10.1088/0067-0049/194/2/28} {\bibfield  {journal} {\bibinfo
  {journal} {The Astrophysical Journal Supplement Series}\ }\textbf {\bibinfo
  {volume} {194}},\ \bibinfo {pages} {28} (\bibinfo {year} {2011})}\BibitemShut
  {NoStop}%
\bibitem [{\citenamefont {Nieuwenhuijzen}\ and\ \citenamefont {{de
  Jager}}(1990)}]{nieuwenhuijzen_parametrization_1990}%
  \BibitemOpen
  \bibfield  {author} {\bibinfo {author} {\bibfnamefont {H.}~\bibnamefont
  {Nieuwenhuijzen}}\ and\ \bibinfo {author} {\bibfnamefont {C.}~\bibnamefont
  {{de Jager}}},\ }\href@noop {} {\bibfield  {journal} {\bibinfo  {journal}
  {Astronomy and Astrophysics}\ }\textbf {\bibinfo {volume} {231}},\ \bibinfo
  {pages} {134} (\bibinfo {year} {1990})}\BibitemShut {NoStop}%
\bibitem [{\citenamefont {Rees}(1988)}]{rees_tidal_1988}%
  \BibitemOpen
  \bibfield  {author} {\bibinfo {author} {\bibfnamefont {M.~J.}\ \bibnamefont
  {Rees}},\ }\href {\doibase 10.1038/333523a0} {\bibfield  {journal} {\bibinfo
  {journal} {Nature}\ }\textbf {\bibinfo {volume} {333}},\ \bibinfo {pages}
  {523} (\bibinfo {year} {1988})}\BibitemShut {NoStop}%
\bibitem [{\citenamefont {Guillochon}\ and\ \citenamefont
  {{Ramirez-Ruiz}}(2013)}]{guillochon_hydrodynamical_2013}%
  \BibitemOpen
  \bibfield  {author} {\bibinfo {author} {\bibfnamefont {J.}~\bibnamefont
  {Guillochon}}\ and\ \bibinfo {author} {\bibfnamefont {E.}~\bibnamefont
  {{Ramirez-Ruiz}}},\ }\href {\doibase 10.1088/0004-637X/767/1/25} {\bibfield
  {journal} {\bibinfo  {journal} {The Astrophysical Journal}\ }\textbf
  {\bibinfo {volume} {767}},\ \bibinfo {pages} {25} (\bibinfo {year}
  {2013})}\BibitemShut {NoStop}%
\bibitem [{\citenamefont {MacLeod}\ \emph {et~al.}(2013)\citenamefont
  {MacLeod}, \citenamefont {{Ramirez-Ruiz}}, \citenamefont {Grady},\ and\
  \citenamefont {Guillochon}}]{macleod_spoon-feeding_2013}%
  \BibitemOpen
  \bibfield  {author} {\bibinfo {author} {\bibfnamefont {M.}~\bibnamefont
  {MacLeod}}, \bibinfo {author} {\bibfnamefont {E.}~\bibnamefont
  {{Ramirez-Ruiz}}}, \bibinfo {author} {\bibfnamefont {S.}~\bibnamefont
  {Grady}}, \ and\ \bibinfo {author} {\bibfnamefont {J.}~\bibnamefont
  {Guillochon}},\ }\href {\doibase 10.1088/0004-637X/777/2/133} {\bibfield
  {journal} {\bibinfo  {journal} {The Astrophysical Journal}\ }\textbf
  {\bibinfo {volume} {777}},\ \bibinfo {pages} {133} (\bibinfo {year}
  {2013})}\BibitemShut {NoStop}%
\bibitem [{\citenamefont {Dupuy}\ and\ \citenamefont
  {Liu}(2017)}]{dupuy_individual_2017}%
  \BibitemOpen
  \bibfield  {author} {\bibinfo {author} {\bibfnamefont {T.~J.}\ \bibnamefont
  {Dupuy}}\ and\ \bibinfo {author} {\bibfnamefont {M.~C.}\ \bibnamefont
  {Liu}},\ }\href {\doibase 10.3847/1538-4365/aa5e4c} {\bibfield  {journal}
  {\bibinfo  {journal} {The Astrophysical Journal Supplement Series}\ }\textbf
  {\bibinfo {volume} {231}},\ \bibinfo {pages} {15} (\bibinfo {year}
  {2017})}\BibitemShut {NoStop}%
\bibitem [{\citenamefont {Burgasser}(2007)}]{burgasser_binaries_2007}%
  \BibitemOpen
  \bibfield  {author} {\bibinfo {author} {\bibfnamefont {A.~J.}\ \bibnamefont
  {Burgasser}},\ }\href {\doibase 10.1086/511027} {\bibfield  {journal}
  {\bibinfo  {journal} {The Astrophysical Journal}\ }\textbf {\bibinfo {volume}
  {659}},\ \bibinfo {pages} {655} (\bibinfo {year} {2007})}\BibitemShut
  {NoStop}%
\bibitem [{\citenamefont {Joergens}(2008)}]{joergens_binary_2008}%
  \BibitemOpen
  \bibfield  {author} {\bibinfo {author} {\bibfnamefont {V.}~\bibnamefont
  {Joergens}},\ }\href {\doibase 10.1051/0004-6361:200810413} {\bibfield
  {journal} {\bibinfo  {journal} {Astronomy and Astrophysics}\ }\textbf
  {\bibinfo {volume} {492}},\ \bibinfo {pages} {545} (\bibinfo {year}
  {2008})}\BibitemShut {NoStop}%
\bibitem [{\citenamefont {Duquennoy}\ and\ \citenamefont
  {Mayor}(1991)}]{duquennoy_multiplicity_1991}%
  \BibitemOpen
  \bibfield  {author} {\bibinfo {author} {\bibfnamefont {A.}~\bibnamefont
  {Duquennoy}}\ and\ \bibinfo {author} {\bibfnamefont {M.}~\bibnamefont
  {Mayor}},\ }\href@noop {} {\bibfield  {journal} {\bibinfo  {journal}
  {Astronomy and Astrophysics}\ }\textbf {\bibinfo {volume} {248}},\ \bibinfo
  {pages} {485} (\bibinfo {year} {1991})}\BibitemShut {NoStop}%
\bibitem [{\citenamefont {Yuan}\ \emph {et~al.}(2015)\citenamefont {Yuan},
  \citenamefont {Liu}, \citenamefont {Xiang}, \citenamefont {Huang},
  \citenamefont {Chen}, \citenamefont {Wu}, \citenamefont {Hou},\ and\
  \citenamefont {Zhang}}]{yuan_stellar_2015}%
  \BibitemOpen
  \bibfield  {author} {\bibinfo {author} {\bibfnamefont {H.}~\bibnamefont
  {Yuan}}, \bibinfo {author} {\bibfnamefont {X.}~\bibnamefont {Liu}}, \bibinfo
  {author} {\bibfnamefont {M.}~\bibnamefont {Xiang}}, \bibinfo {author}
  {\bibfnamefont {Y.}~\bibnamefont {Huang}}, \bibinfo {author} {\bibfnamefont
  {B.}~\bibnamefont {Chen}}, \bibinfo {author} {\bibfnamefont {Y.}~\bibnamefont
  {Wu}}, \bibinfo {author} {\bibfnamefont {Y.}~\bibnamefont {Hou}}, \ and\
  \bibinfo {author} {\bibfnamefont {Y.}~\bibnamefont {Zhang}},\ }\href
  {\doibase 10.1088/0004-637X/799/2/135} {\bibfield  {journal} {\bibinfo
  {journal} {The Astrophysical Journal}\ }\textbf {\bibinfo {volume} {799}},\
  \bibinfo {pages} {135} (\bibinfo {year} {2015})}\BibitemShut {NoStop}%
\bibitem [{\citenamefont {Allers}(2012)}]{allers_brown_2012}%
  \BibitemOpen
  \bibfield  {author} {\bibinfo {author} {\bibfnamefont {K.~N.}\ \bibnamefont
  {Allers}},\ }in\ \href {\doibase 10.1017/S1743921311027086} {\emph {\bibinfo
  {booktitle} {From {{Interacting Binaries}} to {{Exoplanets}}: {{Essential
  Modeling Tools}}}}},\ Vol.\ \bibinfo {volume} {282}\ (\bibinfo {year}
  {2012})\ pp.\ \bibinfo {pages} {105--110}\BibitemShut {NoStop}%
\bibitem [{\citenamefont {Fontanive}\ \emph {et~al.}(2018)\citenamefont
  {Fontanive}, \citenamefont {Biller}, \citenamefont {Bonavita},\ and\
  \citenamefont {Allers}}]{fontanive_constraining_2018}%
  \BibitemOpen
  \bibfield  {author} {\bibinfo {author} {\bibfnamefont {C.}~\bibnamefont
  {Fontanive}}, \bibinfo {author} {\bibfnamefont {B.}~\bibnamefont {Biller}},
  \bibinfo {author} {\bibfnamefont {M.}~\bibnamefont {Bonavita}}, \ and\
  \bibinfo {author} {\bibfnamefont {K.}~\bibnamefont {Allers}},\ }\href
  {\doibase 10.1093/mnras/sty1682} {\bibfield  {journal} {\bibinfo  {journal}
  {Monthly Notices of the Royal Astronomical Society}\ }\textbf {\bibinfo
  {volume} {479}},\ \bibinfo {pages} {2702} (\bibinfo {year}
  {2018})}\BibitemShut {NoStop}%
\bibitem [{\citenamefont {Fischer}\ and\ \citenamefont
  {Marcy}(1992)}]{fischer_multiplicity_1992}%
  \BibitemOpen
  \bibfield  {author} {\bibinfo {author} {\bibfnamefont {D.~A.}\ \bibnamefont
  {Fischer}}\ and\ \bibinfo {author} {\bibfnamefont {G.~W.}\ \bibnamefont
  {Marcy}},\ }\href {\doibase 10.1086/171708} {\bibfield  {journal} {\bibinfo
  {journal} {The Astrophysical Journal}\ }\textbf {\bibinfo {volume} {396}},\
  \bibinfo {pages} {178} (\bibinfo {year} {1992})}\BibitemShut {NoStop}%
\bibitem [{\citenamefont {Burgasser}\ \emph {et~al.}(2015)\citenamefont
  {Burgasser}, \citenamefont {Logsdon}, \citenamefont {Gagn\'e}, \citenamefont
  {Bochanski}, \citenamefont {Faherty}, \citenamefont {West}, \citenamefont
  {Mamajek}, \citenamefont {Schmidt},\ and\ \citenamefont
  {Cruz}}]{burgasser_brown_2015}%
  \BibitemOpen
  \bibfield  {author} {\bibinfo {author} {\bibfnamefont {A.~J.}\ \bibnamefont
  {Burgasser}}, \bibinfo {author} {\bibfnamefont {S.~E.}\ \bibnamefont
  {Logsdon}}, \bibinfo {author} {\bibfnamefont {J.}~\bibnamefont {Gagn\'e}},
  \bibinfo {author} {\bibfnamefont {J.~J.}\ \bibnamefont {Bochanski}}, \bibinfo
  {author} {\bibfnamefont {J.~K.}\ \bibnamefont {Faherty}}, \bibinfo {author}
  {\bibfnamefont {A.~A.}\ \bibnamefont {West}}, \bibinfo {author}
  {\bibfnamefont {E.~E.}\ \bibnamefont {Mamajek}}, \bibinfo {author}
  {\bibfnamefont {S.~J.}\ \bibnamefont {Schmidt}}, \ and\ \bibinfo {author}
  {\bibfnamefont {K.~L.}\ \bibnamefont {Cruz}},\ }\href {\doibase
  10.1088/0067-0049/220/1/18} {\bibfield  {journal} {\bibinfo  {journal} {The
  Astrophysical Journal Supplement Series}\ }\textbf {\bibinfo {volume}
  {220}},\ \bibinfo {pages} {18} (\bibinfo {year} {2015})}\BibitemShut
  {NoStop}%
\bibitem [{\citenamefont {Wright}\ \emph {et~al.}(2010)\citenamefont {Wright},
  \citenamefont {Eisenhardt}, \citenamefont {Mainzer}, \citenamefont {Ressler},
  \citenamefont {Cutri}, \citenamefont {Jarrett}, \citenamefont {Kirkpatrick},
  \citenamefont {Padgett}, \citenamefont {McMillan}, \citenamefont {Skrutskie},
  \citenamefont {Stanford}, \citenamefont {Cohen}, \citenamefont {Walker},
  \citenamefont {Mather}, \citenamefont {Leisawitz}, \citenamefont {Gautier},
  \citenamefont {McLean}, \citenamefont {Benford}, \citenamefont {Lonsdale},
  \citenamefont {Blain}, \citenamefont {Mendez}, \citenamefont {Irace},
  \citenamefont {Duval}, \citenamefont {Liu}, \citenamefont {Royer},
  \citenamefont {Heinrichsen}, \citenamefont {Howard}, \citenamefont {Shannon},
  \citenamefont {Kendall}, \citenamefont {Walsh}, \citenamefont {Larsen},
  \citenamefont {Cardon}, \citenamefont {Schick}, \citenamefont {Schwalm},
  \citenamefont {Abid}, \citenamefont {Fabinsky}, \citenamefont {Naes},\ and\
  \citenamefont {Tsai}}]{wright_wide-field_2010}%
  \BibitemOpen
  \bibfield  {author} {\bibinfo {author} {\bibfnamefont {E.~L.}\ \bibnamefont
  {Wright}}, \bibinfo {author} {\bibfnamefont {P.~R.~M.}\ \bibnamefont
  {Eisenhardt}}, \bibinfo {author} {\bibfnamefont {A.~K.}\ \bibnamefont
  {Mainzer}}, \bibinfo {author} {\bibfnamefont {M.~E.}\ \bibnamefont
  {Ressler}}, \bibinfo {author} {\bibfnamefont {R.~M.}\ \bibnamefont {Cutri}},
  \bibinfo {author} {\bibfnamefont {T.}~\bibnamefont {Jarrett}}, \bibinfo
  {author} {\bibfnamefont {J.~D.}\ \bibnamefont {Kirkpatrick}}, \bibinfo
  {author} {\bibfnamefont {D.}~\bibnamefont {Padgett}}, \bibinfo {author}
  {\bibfnamefont {R.~S.}\ \bibnamefont {McMillan}}, \bibinfo {author}
  {\bibfnamefont {M.}~\bibnamefont {Skrutskie}}, \bibinfo {author}
  {\bibfnamefont {S.~A.}\ \bibnamefont {Stanford}}, \bibinfo {author}
  {\bibfnamefont {M.}~\bibnamefont {Cohen}}, \bibinfo {author} {\bibfnamefont
  {R.~G.}\ \bibnamefont {Walker}}, \bibinfo {author} {\bibfnamefont {J.~C.}\
  \bibnamefont {Mather}}, \bibinfo {author} {\bibfnamefont {D.}~\bibnamefont
  {Leisawitz}}, \bibinfo {author} {\bibfnamefont {T.~N.}\ \bibnamefont
  {Gautier}, \bibfnamefont {III}}, \bibinfo {author} {\bibfnamefont
  {I.}~\bibnamefont {McLean}}, \bibinfo {author} {\bibfnamefont
  {D.}~\bibnamefont {Benford}}, \bibinfo {author} {\bibfnamefont {C.~J.}\
  \bibnamefont {Lonsdale}}, \bibinfo {author} {\bibfnamefont {A.}~\bibnamefont
  {Blain}}, \bibinfo {author} {\bibfnamefont {B.}~\bibnamefont {Mendez}},
  \bibinfo {author} {\bibfnamefont {W.~R.}\ \bibnamefont {Irace}}, \bibinfo
  {author} {\bibfnamefont {V.}~\bibnamefont {Duval}}, \bibinfo {author}
  {\bibfnamefont {F.}~\bibnamefont {Liu}}, \bibinfo {author} {\bibfnamefont
  {D.}~\bibnamefont {Royer}}, \bibinfo {author} {\bibfnamefont
  {I.}~\bibnamefont {Heinrichsen}}, \bibinfo {author} {\bibfnamefont
  {J.}~\bibnamefont {Howard}}, \bibinfo {author} {\bibfnamefont
  {M.}~\bibnamefont {Shannon}}, \bibinfo {author} {\bibfnamefont
  {M.}~\bibnamefont {Kendall}}, \bibinfo {author} {\bibfnamefont {A.~L.}\
  \bibnamefont {Walsh}}, \bibinfo {author} {\bibfnamefont {M.}~\bibnamefont
  {Larsen}}, \bibinfo {author} {\bibfnamefont {J.~G.}\ \bibnamefont {Cardon}},
  \bibinfo {author} {\bibfnamefont {S.}~\bibnamefont {Schick}}, \bibinfo
  {author} {\bibfnamefont {M.}~\bibnamefont {Schwalm}}, \bibinfo {author}
  {\bibfnamefont {M.}~\bibnamefont {Abid}}, \bibinfo {author} {\bibfnamefont
  {B.}~\bibnamefont {Fabinsky}}, \bibinfo {author} {\bibfnamefont
  {L.}~\bibnamefont {Naes}}, \ and\ \bibinfo {author} {\bibfnamefont {C.-W.}\
  \bibnamefont {Tsai}},\ }\href {\doibase 10.1088/0004-6256/140/6/1868}
  {\bibfield  {journal} {\bibinfo  {journal} {The Astronomical Journal}\
  }\textbf {\bibinfo {volume} {140}},\ \bibinfo {pages} {1868} (\bibinfo {year}
  {2010})}\BibitemShut {NoStop}%
\bibitem [{\citenamefont {{LSST Science Collaboration}}\ \emph
  {et~al.}(2009)\citenamefont {{LSST Science Collaboration}}, \citenamefont
  {Abell}, \citenamefont {Allison}, \citenamefont {Anderson}, \citenamefont
  {Andrew}, \citenamefont {Angel}, \citenamefont {Armus}, \citenamefont
  {Arnett}, \citenamefont {Asztalos}, \citenamefont {Axelrod}, \citenamefont
  {Bailey}, \citenamefont {Ballantyne}, \citenamefont {Bankert}, \citenamefont
  {Barkhouse}, \citenamefont {Barr}, \citenamefont {Barrientos}, \citenamefont
  {Barth}, \citenamefont {Bartlett}, \citenamefont {Becker}, \citenamefont
  {Becla}, \citenamefont {Beers}, \citenamefont {Bernstein}, \citenamefont
  {Biswas}, \citenamefont {Blanton}, \citenamefont {Bloom}, \citenamefont
  {Bochanski}, \citenamefont {Boeshaar}, \citenamefont {Borne}, \citenamefont
  {Bradac}, \citenamefont {Brandt}, \citenamefont {Bridge}, \citenamefont
  {Brown}, \citenamefont {Brunner}, \citenamefont {Bullock}, \citenamefont
  {Burgasser}, \citenamefont {Burge}, \citenamefont {Burke}, \citenamefont
  {Cargile}, \citenamefont {Chandrasekharan}, \citenamefont {Chartas},
  \citenamefont {Chesley}, \citenamefont {Chu}, \citenamefont {Cinabro},
  \citenamefont {Claire}, \citenamefont {Claver}, \citenamefont {Clowe},
  \citenamefont {Connolly}, \citenamefont {Cook}, \citenamefont {Cooke},
  \citenamefont {Cooray}, \citenamefont {Covey}, \citenamefont {Culliton},
  \citenamefont {{de Jong}}, \citenamefont {{de Vries}}, \citenamefont
  {Debattista}, \citenamefont {Delgado}, \citenamefont {Dell'Antonio},
  \citenamefont {Dhital}, \citenamefont {Di~Stefano}, \citenamefont
  {Dickinson}, \citenamefont {Dilday}, \citenamefont {Djorgovski},
  \citenamefont {Dobler}, \citenamefont {Donalek}, \citenamefont
  {{Dubois-Felsmann}}, \citenamefont {Durech}, \citenamefont {Eliasdottir},
  \citenamefont {Eracleous}, \citenamefont {Eyer}, \citenamefont {Falco},
  \citenamefont {Fan}, \citenamefont {Fassnacht}, \citenamefont {Ferguson},
  \citenamefont {Fernandez}, \citenamefont {Fields}, \citenamefont
  {Finkbeiner}, \citenamefont {Figueroa}, \citenamefont {Fox}, \citenamefont
  {Francke}, \citenamefont {Frank}, \citenamefont {Frieman}, \citenamefont
  {Fromenteau}, \citenamefont {Furqan}, \citenamefont {Galaz}, \citenamefont
  {{Gal-Yam}}, \citenamefont {Garnavich}, \citenamefont {Gawiser},
  \citenamefont {Geary}, \citenamefont {Gee}, \citenamefont {Gibson},
  \citenamefont {Gilmore}, \citenamefont {Grace}, \citenamefont {Green},
  \citenamefont {Gressler}, \citenamefont {Grillmair}, \citenamefont {Habib},
  \citenamefont {Haggerty}, \citenamefont {Hamuy}, \citenamefont {Harris},
  \citenamefont {Hawley}, \citenamefont {Heavens}, \citenamefont {Hebb},
  \citenamefont {Henry}, \citenamefont {Hileman}, \citenamefont {Hilton},
  \citenamefont {Hoadley}, \citenamefont {Holberg}, \citenamefont {Holman},
  \citenamefont {Howell}, \citenamefont {Infante}, \citenamefont {Ivezic},
  \citenamefont {Jacoby}, \citenamefont {Jain}, \citenamefont {{R}},
  \citenamefont {{Jedicke}}, \citenamefont {Jee}, \citenamefont
  {Garrett~Jernigan}, \citenamefont {Jha}, \citenamefont {Johnston},
  \citenamefont {Jones}, \citenamefont {Juric}, \citenamefont {Kaasalainen},
  \citenamefont {{Styliani}}, \citenamefont {{Kafka}}, \citenamefont {Kahn},
  \citenamefont {Kaib}, \citenamefont {Kalirai}, \citenamefont {Kantor},
  \citenamefont {Kasliwal}, \citenamefont {Keeton}, \citenamefont {Kessler},
  \citenamefont {Knezevic}, \citenamefont {Kowalski}, \citenamefont
  {Krabbendam}, \citenamefont {Krughoff}, \citenamefont {Kulkarni},
  \citenamefont {Kuhlman}, \citenamefont {Lacy}, \citenamefont {Lepine},
  \citenamefont {Liang}, \citenamefont {Lien}, \citenamefont {Lira},
  \citenamefont {Long}, \citenamefont {Lorenz}, \citenamefont {Lotz},
  \citenamefont {Lupton}, \citenamefont {Lutz}, \citenamefont {Macri},
  \citenamefont {Mahabal}, \citenamefont {Mandelbaum}, \citenamefont
  {Marshall}, \citenamefont {May}, \citenamefont {McGehee}, \citenamefont
  {Meadows}, \citenamefont {Meert}, \citenamefont {Milani}, \citenamefont
  {Miller}, \citenamefont {Miller}, \citenamefont {Mills}, \citenamefont
  {Minniti}, \citenamefont {Monet}, \citenamefont {Mukadam}, \citenamefont
  {Nakar}, \citenamefont {Neill}, \citenamefont {Newman}, \citenamefont
  {Nikolaev}, \citenamefont {Nordby}, \citenamefont {O'Connor}, \citenamefont
  {Oguri}, \citenamefont {Oliver}, \citenamefont {Olivier}, \citenamefont
  {Olsen}, \citenamefont {Olsen}, \citenamefont {Olszewski}, \citenamefont
  {Oluseyi}, \citenamefont {Padilla}, \citenamefont {Parker}, \citenamefont
  {Pepper}, \citenamefont {Peterson}, \citenamefont {Petry}, \citenamefont
  {Pinto}, \citenamefont {Pizagno}, \citenamefont {Popescu}, \citenamefont
  {Prsa}, \citenamefont {Radcka}, \citenamefont {Raddick}, \citenamefont
  {Rasmussen}, \citenamefont {Rau}, \citenamefont {Rho}, \citenamefont
  {Rhoads}, \citenamefont {Richards}, \citenamefont {Ridgway}, \citenamefont
  {Robertson}, \citenamefont {Roskar}, \citenamefont {Saha}, \citenamefont
  {Sarajedini}, \citenamefont {Scannapieco}, \citenamefont {Schalk},
  \citenamefont {Schindler}, \citenamefont {Schmidt}, \citenamefont {Schmidt},
  \citenamefont {Schneider}, \citenamefont {Schumacher}, \citenamefont
  {Scranton}, \citenamefont {Sebag}, \citenamefont {Seppala}, \citenamefont
  {Shemmer}, \citenamefont {Simon}, \citenamefont {Sivertz}, \citenamefont
  {Smith}, \citenamefont {Allyn~Smith}, \citenamefont {Smith}, \citenamefont
  {Spitz}, \citenamefont {Stanford}, \citenamefont {Stassun}, \citenamefont
  {Strader}, \citenamefont {Strauss}, \citenamefont {Stubbs}, \citenamefont
  {Sweeney}, \citenamefont {Szalay}, \citenamefont {Szkody}, \citenamefont
  {Takada}, \citenamefont {Thorman}, \citenamefont {Trilling}, \citenamefont
  {Trimble}, \citenamefont {Tyson}, \citenamefont {Van~Berg}, \citenamefont
  {Vanden~Berk}, \citenamefont {VanderPlas}, \citenamefont {Verde},
  \citenamefont {Vrsnak}, \citenamefont {Walkowicz}, \citenamefont {Wandelt},
  \citenamefont {Wang}, \citenamefont {Wang}, \citenamefont {Warner},
  \citenamefont {Wechsler}, \citenamefont {West}, \citenamefont {Wiecha},
  \citenamefont {Williams}, \citenamefont {Willman}, \citenamefont {Wittman},
  \citenamefont {Wolff}, \citenamefont {{Wood-Vasey}}, \citenamefont {Wozniak},
  \citenamefont {Young}, \citenamefont {Zentner},\ and\ \citenamefont
  {Zhan}}]{lsst_science_collaboration_lsst_2009}%
  \BibitemOpen
  \bibfield  {author} {\bibinfo {author} {\bibnamefont {{LSST Science
  Collaboration}}}, \bibinfo {author} {\bibfnamefont {P.~A.}\ \bibnamefont
  {Abell}}, \bibinfo {author} {\bibfnamefont {J.}~\bibnamefont {Allison}},
  \bibinfo {author} {\bibfnamefont {S.~F.}\ \bibnamefont {Anderson}}, \bibinfo
  {author} {\bibfnamefont {J.~R.}\ \bibnamefont {Andrew}}, \bibinfo {author}
  {\bibfnamefont {J.~R.~P.}\ \bibnamefont {Angel}}, \bibinfo {author}
  {\bibfnamefont {L.}~\bibnamefont {Armus}}, \bibinfo {author} {\bibfnamefont
  {D.}~\bibnamefont {Arnett}}, \bibinfo {author} {\bibfnamefont {S.~J.}\
  \bibnamefont {Asztalos}}, \bibinfo {author} {\bibfnamefont {T.~S.}\
  \bibnamefont {Axelrod}}, \bibinfo {author} {\bibfnamefont {S.}~\bibnamefont
  {Bailey}}, \bibinfo {author} {\bibfnamefont {D.~R.}\ \bibnamefont
  {Ballantyne}}, \bibinfo {author} {\bibfnamefont {J.~R.}\ \bibnamefont
  {Bankert}}, \bibinfo {author} {\bibfnamefont {W.~A.}\ \bibnamefont
  {Barkhouse}}, \bibinfo {author} {\bibfnamefont {J.~D.}\ \bibnamefont {Barr}},
  \bibinfo {author} {\bibfnamefont {L.~F.}\ \bibnamefont {Barrientos}},
  \bibinfo {author} {\bibfnamefont {A.~J.}\ \bibnamefont {Barth}}, \bibinfo
  {author} {\bibfnamefont {J.~G.}\ \bibnamefont {Bartlett}}, \bibinfo {author}
  {\bibfnamefont {A.~C.}\ \bibnamefont {Becker}}, \bibinfo {author}
  {\bibfnamefont {J.}~\bibnamefont {Becla}}, \bibinfo {author} {\bibfnamefont
  {T.~C.}\ \bibnamefont {Beers}}, \bibinfo {author} {\bibfnamefont {J.~P.}\
  \bibnamefont {Bernstein}}, \bibinfo {author} {\bibfnamefont {R.}~\bibnamefont
  {Biswas}}, \bibinfo {author} {\bibfnamefont {M.~R.}\ \bibnamefont {Blanton}},
  \bibinfo {author} {\bibfnamefont {J.~S.}\ \bibnamefont {Bloom}}, \bibinfo
  {author} {\bibfnamefont {J.~J.}\ \bibnamefont {Bochanski}}, \bibinfo {author}
  {\bibfnamefont {P.}~\bibnamefont {Boeshaar}}, \bibinfo {author}
  {\bibfnamefont {K.~D.}\ \bibnamefont {Borne}}, \bibinfo {author}
  {\bibfnamefont {M.}~\bibnamefont {Bradac}}, \bibinfo {author} {\bibfnamefont
  {W.~N.}\ \bibnamefont {Brandt}}, \bibinfo {author} {\bibfnamefont {C.~R.}\
  \bibnamefont {Bridge}}, \bibinfo {author} {\bibfnamefont {M.~E.}\
  \bibnamefont {Brown}}, \bibinfo {author} {\bibfnamefont {R.~J.}\ \bibnamefont
  {Brunner}}, \bibinfo {author} {\bibfnamefont {J.~S.}\ \bibnamefont
  {Bullock}}, \bibinfo {author} {\bibfnamefont {A.~J.}\ \bibnamefont
  {Burgasser}}, \bibinfo {author} {\bibfnamefont {J.~H.}\ \bibnamefont
  {Burge}}, \bibinfo {author} {\bibfnamefont {D.~L.}\ \bibnamefont {Burke}},
  \bibinfo {author} {\bibfnamefont {P.~A.}\ \bibnamefont {Cargile}}, \bibinfo
  {author} {\bibfnamefont {S.}~\bibnamefont {Chandrasekharan}}, \bibinfo
  {author} {\bibfnamefont {G.}~\bibnamefont {Chartas}}, \bibinfo {author}
  {\bibfnamefont {S.~R.}\ \bibnamefont {Chesley}}, \bibinfo {author}
  {\bibfnamefont {Y.-H.}\ \bibnamefont {Chu}}, \bibinfo {author} {\bibfnamefont
  {D.}~\bibnamefont {Cinabro}}, \bibinfo {author} {\bibfnamefont {M.~W.}\
  \bibnamefont {Claire}}, \bibinfo {author} {\bibfnamefont {C.~F.}\
  \bibnamefont {Claver}}, \bibinfo {author} {\bibfnamefont {D.}~\bibnamefont
  {Clowe}}, \bibinfo {author} {\bibfnamefont {A.~J.}\ \bibnamefont {Connolly}},
  \bibinfo {author} {\bibfnamefont {K.~H.}\ \bibnamefont {Cook}}, \bibinfo
  {author} {\bibfnamefont {J.}~\bibnamefont {Cooke}}, \bibinfo {author}
  {\bibfnamefont {A.}~\bibnamefont {Cooray}}, \bibinfo {author} {\bibfnamefont
  {K.~R.}\ \bibnamefont {Covey}}, \bibinfo {author} {\bibfnamefont {C.~S.}\
  \bibnamefont {Culliton}}, \bibinfo {author} {\bibfnamefont {R.}~\bibnamefont
  {{de Jong}}}, \bibinfo {author} {\bibfnamefont {W.~H.}\ \bibnamefont {{de
  Vries}}}, \bibinfo {author} {\bibfnamefont {V.~P.}\ \bibnamefont
  {Debattista}}, \bibinfo {author} {\bibfnamefont {F.}~\bibnamefont {Delgado}},
  \bibinfo {author} {\bibfnamefont {I.~P.}\ \bibnamefont {Dell'Antonio}},
  \bibinfo {author} {\bibfnamefont {S.}~\bibnamefont {Dhital}}, \bibinfo
  {author} {\bibfnamefont {R.}~\bibnamefont {Di~Stefano}}, \bibinfo {author}
  {\bibfnamefont {M.}~\bibnamefont {Dickinson}}, \bibinfo {author}
  {\bibfnamefont {B.}~\bibnamefont {Dilday}}, \bibinfo {author} {\bibfnamefont
  {S.~G.}\ \bibnamefont {Djorgovski}}, \bibinfo {author} {\bibfnamefont
  {G.}~\bibnamefont {Dobler}}, \bibinfo {author} {\bibfnamefont
  {C.}~\bibnamefont {Donalek}}, \bibinfo {author} {\bibfnamefont
  {G.}~\bibnamefont {{Dubois-Felsmann}}}, \bibinfo {author} {\bibfnamefont
  {J.}~\bibnamefont {Durech}}, \bibinfo {author} {\bibfnamefont
  {A.}~\bibnamefont {Eliasdottir}}, \bibinfo {author} {\bibfnamefont
  {M.}~\bibnamefont {Eracleous}}, \bibinfo {author} {\bibfnamefont
  {L.}~\bibnamefont {Eyer}}, \bibinfo {author} {\bibfnamefont {E.~E.}\
  \bibnamefont {Falco}}, \bibinfo {author} {\bibfnamefont {X.}~\bibnamefont
  {Fan}}, \bibinfo {author} {\bibfnamefont {C.~D.}\ \bibnamefont {Fassnacht}},
  \bibinfo {author} {\bibfnamefont {H.~C.}\ \bibnamefont {Ferguson}}, \bibinfo
  {author} {\bibfnamefont {Y.~R.}\ \bibnamefont {Fernandez}}, \bibinfo {author}
  {\bibfnamefont {B.~D.}\ \bibnamefont {Fields}}, \bibinfo {author}
  {\bibfnamefont {D.}~\bibnamefont {Finkbeiner}}, \bibinfo {author}
  {\bibfnamefont {E.~E.}\ \bibnamefont {Figueroa}}, \bibinfo {author}
  {\bibfnamefont {D.~B.}\ \bibnamefont {Fox}}, \bibinfo {author} {\bibfnamefont
  {H.}~\bibnamefont {Francke}}, \bibinfo {author} {\bibfnamefont {J.~S.}\
  \bibnamefont {Frank}}, \bibinfo {author} {\bibfnamefont {J.}~\bibnamefont
  {Frieman}}, \bibinfo {author} {\bibfnamefont {S.}~\bibnamefont {Fromenteau}},
  \bibinfo {author} {\bibfnamefont {M.}~\bibnamefont {Furqan}}, \bibinfo
  {author} {\bibfnamefont {G.}~\bibnamefont {Galaz}}, \bibinfo {author}
  {\bibfnamefont {A.}~\bibnamefont {{Gal-Yam}}}, \bibinfo {author}
  {\bibfnamefont {P.}~\bibnamefont {Garnavich}}, \bibinfo {author}
  {\bibfnamefont {E.}~\bibnamefont {Gawiser}}, \bibinfo {author} {\bibfnamefont
  {J.}~\bibnamefont {Geary}}, \bibinfo {author} {\bibfnamefont
  {P.}~\bibnamefont {Gee}}, \bibinfo {author} {\bibfnamefont {R.~R.}\
  \bibnamefont {Gibson}}, \bibinfo {author} {\bibfnamefont {K.}~\bibnamefont
  {Gilmore}}, \bibinfo {author} {\bibfnamefont {E.~A.}\ \bibnamefont {Grace}},
  \bibinfo {author} {\bibfnamefont {R.~F.}\ \bibnamefont {Green}}, \bibinfo
  {author} {\bibfnamefont {W.~J.}\ \bibnamefont {Gressler}}, \bibinfo {author}
  {\bibfnamefont {C.~J.}\ \bibnamefont {Grillmair}}, \bibinfo {author}
  {\bibfnamefont {S.}~\bibnamefont {Habib}}, \bibinfo {author} {\bibfnamefont
  {J.~S.}\ \bibnamefont {Haggerty}}, \bibinfo {author} {\bibfnamefont
  {M.}~\bibnamefont {Hamuy}}, \bibinfo {author} {\bibfnamefont {A.~W.}\
  \bibnamefont {Harris}}, \bibinfo {author} {\bibfnamefont {S.~L.}\
  \bibnamefont {Hawley}}, \bibinfo {author} {\bibfnamefont {A.~F.}\
  \bibnamefont {Heavens}}, \bibinfo {author} {\bibfnamefont {L.}~\bibnamefont
  {Hebb}}, \bibinfo {author} {\bibfnamefont {T.~J.}\ \bibnamefont {Henry}},
  \bibinfo {author} {\bibfnamefont {E.}~\bibnamefont {Hileman}}, \bibinfo
  {author} {\bibfnamefont {E.~J.}\ \bibnamefont {Hilton}}, \bibinfo {author}
  {\bibfnamefont {K.}~\bibnamefont {Hoadley}}, \bibinfo {author} {\bibfnamefont
  {J.~B.}\ \bibnamefont {Holberg}}, \bibinfo {author} {\bibfnamefont {M.~J.}\
  \bibnamefont {Holman}}, \bibinfo {author} {\bibfnamefont {S.~B.}\
  \bibnamefont {Howell}}, \bibinfo {author} {\bibfnamefont {L.}~\bibnamefont
  {Infante}}, \bibinfo {author} {\bibfnamefont {Z.}~\bibnamefont {Ivezic}},
  \bibinfo {author} {\bibfnamefont {S.~H.}\ \bibnamefont {Jacoby}}, \bibinfo
  {author} {\bibfnamefont {B.}~\bibnamefont {Jain}}, \bibinfo {author}
  {\bibnamefont {{R}}}, \bibinfo {author} {\bibnamefont {{Jedicke}}}, \bibinfo
  {author} {\bibfnamefont {M.~J.}\ \bibnamefont {Jee}}, \bibinfo {author}
  {\bibfnamefont {J.}~\bibnamefont {Garrett~Jernigan}}, \bibinfo {author}
  {\bibfnamefont {S.~W.}\ \bibnamefont {Jha}}, \bibinfo {author} {\bibfnamefont
  {K.~V.}\ \bibnamefont {Johnston}}, \bibinfo {author} {\bibfnamefont {R.~L.}\
  \bibnamefont {Jones}}, \bibinfo {author} {\bibfnamefont {M.}~\bibnamefont
  {Juric}}, \bibinfo {author} {\bibfnamefont {M.}~\bibnamefont {Kaasalainen}},
  \bibinfo {author} {\bibnamefont {{Styliani}}}, \bibinfo {author}
  {\bibnamefont {{Kafka}}}, \bibinfo {author} {\bibfnamefont {S.~M.}\
  \bibnamefont {Kahn}}, \bibinfo {author} {\bibfnamefont {N.~A.}\ \bibnamefont
  {Kaib}}, \bibinfo {author} {\bibfnamefont {J.}~\bibnamefont {Kalirai}},
  \bibinfo {author} {\bibfnamefont {J.}~\bibnamefont {Kantor}}, \bibinfo
  {author} {\bibfnamefont {M.~M.}\ \bibnamefont {Kasliwal}}, \bibinfo {author}
  {\bibfnamefont {C.~R.}\ \bibnamefont {Keeton}}, \bibinfo {author}
  {\bibfnamefont {R.}~\bibnamefont {Kessler}}, \bibinfo {author} {\bibfnamefont
  {Z.}~\bibnamefont {Knezevic}}, \bibinfo {author} {\bibfnamefont
  {A.}~\bibnamefont {Kowalski}}, \bibinfo {author} {\bibfnamefont {V.~L.}\
  \bibnamefont {Krabbendam}}, \bibinfo {author} {\bibfnamefont {K.~S.}\
  \bibnamefont {Krughoff}}, \bibinfo {author} {\bibfnamefont {S.}~\bibnamefont
  {Kulkarni}}, \bibinfo {author} {\bibfnamefont {S.}~\bibnamefont {Kuhlman}},
  \bibinfo {author} {\bibfnamefont {M.}~\bibnamefont {Lacy}}, \bibinfo {author}
  {\bibfnamefont {S.}~\bibnamefont {Lepine}}, \bibinfo {author} {\bibfnamefont
  {M.}~\bibnamefont {Liang}}, \bibinfo {author} {\bibfnamefont
  {A.}~\bibnamefont {Lien}}, \bibinfo {author} {\bibfnamefont {P.}~\bibnamefont
  {Lira}}, \bibinfo {author} {\bibfnamefont {K.~S.}\ \bibnamefont {Long}},
  \bibinfo {author} {\bibfnamefont {S.}~\bibnamefont {Lorenz}}, \bibinfo
  {author} {\bibfnamefont {J.~M.}\ \bibnamefont {Lotz}}, \bibinfo {author}
  {\bibfnamefont {R.~H.}\ \bibnamefont {Lupton}}, \bibinfo {author}
  {\bibfnamefont {J.}~\bibnamefont {Lutz}}, \bibinfo {author} {\bibfnamefont
  {L.~M.}\ \bibnamefont {Macri}}, \bibinfo {author} {\bibfnamefont {A.~A.}\
  \bibnamefont {Mahabal}}, \bibinfo {author} {\bibfnamefont {R.}~\bibnamefont
  {Mandelbaum}}, \bibinfo {author} {\bibfnamefont {P.}~\bibnamefont
  {Marshall}}, \bibinfo {author} {\bibfnamefont {M.}~\bibnamefont {May}},
  \bibinfo {author} {\bibfnamefont {P.~M.}\ \bibnamefont {McGehee}}, \bibinfo
  {author} {\bibfnamefont {B.~T.}\ \bibnamefont {Meadows}}, \bibinfo {author}
  {\bibfnamefont {A.}~\bibnamefont {Meert}}, \bibinfo {author} {\bibfnamefont
  {A.}~\bibnamefont {Milani}}, \bibinfo {author} {\bibfnamefont {C.~J.}\
  \bibnamefont {Miller}}, \bibinfo {author} {\bibfnamefont {M.}~\bibnamefont
  {Miller}}, \bibinfo {author} {\bibfnamefont {D.}~\bibnamefont {Mills}},
  \bibinfo {author} {\bibfnamefont {D.}~\bibnamefont {Minniti}}, \bibinfo
  {author} {\bibfnamefont {D.}~\bibnamefont {Monet}}, \bibinfo {author}
  {\bibfnamefont {A.~S.}\ \bibnamefont {Mukadam}}, \bibinfo {author}
  {\bibfnamefont {E.}~\bibnamefont {Nakar}}, \bibinfo {author} {\bibfnamefont
  {D.~R.}\ \bibnamefont {Neill}}, \bibinfo {author} {\bibfnamefont {J.~A.}\
  \bibnamefont {Newman}}, \bibinfo {author} {\bibfnamefont {S.}~\bibnamefont
  {Nikolaev}}, \bibinfo {author} {\bibfnamefont {M.}~\bibnamefont {Nordby}},
  \bibinfo {author} {\bibfnamefont {P.}~\bibnamefont {O'Connor}}, \bibinfo
  {author} {\bibfnamefont {M.}~\bibnamefont {Oguri}}, \bibinfo {author}
  {\bibfnamefont {J.}~\bibnamefont {Oliver}}, \bibinfo {author} {\bibfnamefont
  {S.~S.}\ \bibnamefont {Olivier}}, \bibinfo {author} {\bibfnamefont {J.~K.}\
  \bibnamefont {Olsen}}, \bibinfo {author} {\bibfnamefont {K.}~\bibnamefont
  {Olsen}}, \bibinfo {author} {\bibfnamefont {E.~W.}\ \bibnamefont
  {Olszewski}}, \bibinfo {author} {\bibfnamefont {H.}~\bibnamefont {Oluseyi}},
  \bibinfo {author} {\bibfnamefont {N.~D.}\ \bibnamefont {Padilla}}, \bibinfo
  {author} {\bibfnamefont {A.}~\bibnamefont {Parker}}, \bibinfo {author}
  {\bibfnamefont {J.}~\bibnamefont {Pepper}}, \bibinfo {author} {\bibfnamefont
  {J.~R.}\ \bibnamefont {Peterson}}, \bibinfo {author} {\bibfnamefont
  {C.}~\bibnamefont {Petry}}, \bibinfo {author} {\bibfnamefont {P.~A.}\
  \bibnamefont {Pinto}}, \bibinfo {author} {\bibfnamefont {J.~L.}\ \bibnamefont
  {Pizagno}}, \bibinfo {author} {\bibfnamefont {B.}~\bibnamefont {Popescu}},
  \bibinfo {author} {\bibfnamefont {A.}~\bibnamefont {Prsa}}, \bibinfo {author}
  {\bibfnamefont {V.}~\bibnamefont {Radcka}}, \bibinfo {author} {\bibfnamefont
  {M.~J.}\ \bibnamefont {Raddick}}, \bibinfo {author} {\bibfnamefont
  {A.}~\bibnamefont {Rasmussen}}, \bibinfo {author} {\bibfnamefont
  {A.}~\bibnamefont {Rau}}, \bibinfo {author} {\bibfnamefont {J.}~\bibnamefont
  {Rho}}, \bibinfo {author} {\bibfnamefont {J.~E.}\ \bibnamefont {Rhoads}},
  \bibinfo {author} {\bibfnamefont {G.~T.}\ \bibnamefont {Richards}}, \bibinfo
  {author} {\bibfnamefont {S.~T.}\ \bibnamefont {Ridgway}}, \bibinfo {author}
  {\bibfnamefont {B.~E.}\ \bibnamefont {Robertson}}, \bibinfo {author}
  {\bibfnamefont {R.}~\bibnamefont {Roskar}}, \bibinfo {author} {\bibfnamefont
  {A.}~\bibnamefont {Saha}}, \bibinfo {author} {\bibfnamefont {A.}~\bibnamefont
  {Sarajedini}}, \bibinfo {author} {\bibfnamefont {E.}~\bibnamefont
  {Scannapieco}}, \bibinfo {author} {\bibfnamefont {T.}~\bibnamefont {Schalk}},
  \bibinfo {author} {\bibfnamefont {R.}~\bibnamefont {Schindler}}, \bibinfo
  {author} {\bibfnamefont {S.}~\bibnamefont {Schmidt}}, \bibinfo {author}
  {\bibfnamefont {S.}~\bibnamefont {Schmidt}}, \bibinfo {author} {\bibfnamefont
  {D.~P.}\ \bibnamefont {Schneider}}, \bibinfo {author} {\bibfnamefont
  {G.}~\bibnamefont {Schumacher}}, \bibinfo {author} {\bibfnamefont
  {R.}~\bibnamefont {Scranton}}, \bibinfo {author} {\bibfnamefont
  {J.}~\bibnamefont {Sebag}}, \bibinfo {author} {\bibfnamefont {L.~G.}\
  \bibnamefont {Seppala}}, \bibinfo {author} {\bibfnamefont {O.}~\bibnamefont
  {Shemmer}}, \bibinfo {author} {\bibfnamefont {J.~D.}\ \bibnamefont {Simon}},
  \bibinfo {author} {\bibfnamefont {M.}~\bibnamefont {Sivertz}}, \bibinfo
  {author} {\bibfnamefont {H.~A.}\ \bibnamefont {Smith}}, \bibinfo {author}
  {\bibfnamefont {J.}~\bibnamefont {Allyn~Smith}}, \bibinfo {author}
  {\bibfnamefont {N.}~\bibnamefont {Smith}}, \bibinfo {author} {\bibfnamefont
  {A.~H.}\ \bibnamefont {Spitz}}, \bibinfo {author} {\bibfnamefont
  {A.}~\bibnamefont {Stanford}}, \bibinfo {author} {\bibfnamefont {K.~G.}\
  \bibnamefont {Stassun}}, \bibinfo {author} {\bibfnamefont {J.}~\bibnamefont
  {Strader}}, \bibinfo {author} {\bibfnamefont {M.~A.}\ \bibnamefont
  {Strauss}}, \bibinfo {author} {\bibfnamefont {C.~W.}\ \bibnamefont {Stubbs}},
  \bibinfo {author} {\bibfnamefont {D.~W.}\ \bibnamefont {Sweeney}}, \bibinfo
  {author} {\bibfnamefont {A.}~\bibnamefont {Szalay}}, \bibinfo {author}
  {\bibfnamefont {P.}~\bibnamefont {Szkody}}, \bibinfo {author} {\bibfnamefont
  {M.}~\bibnamefont {Takada}}, \bibinfo {author} {\bibfnamefont
  {P.}~\bibnamefont {Thorman}}, \bibinfo {author} {\bibfnamefont {D.~E.}\
  \bibnamefont {Trilling}}, \bibinfo {author} {\bibfnamefont {V.}~\bibnamefont
  {Trimble}}, \bibinfo {author} {\bibfnamefont {A.}~\bibnamefont {Tyson}},
  \bibinfo {author} {\bibfnamefont {R.}~\bibnamefont {Van~Berg}}, \bibinfo
  {author} {\bibfnamefont {D.}~\bibnamefont {Vanden~Berk}}, \bibinfo {author}
  {\bibfnamefont {J.}~\bibnamefont {VanderPlas}}, \bibinfo {author}
  {\bibfnamefont {L.}~\bibnamefont {Verde}}, \bibinfo {author} {\bibfnamefont
  {B.}~\bibnamefont {Vrsnak}}, \bibinfo {author} {\bibfnamefont {L.~M.}\
  \bibnamefont {Walkowicz}}, \bibinfo {author} {\bibfnamefont {B.~D.}\
  \bibnamefont {Wandelt}}, \bibinfo {author} {\bibfnamefont {S.}~\bibnamefont
  {Wang}}, \bibinfo {author} {\bibfnamefont {Y.}~\bibnamefont {Wang}}, \bibinfo
  {author} {\bibfnamefont {M.}~\bibnamefont {Warner}}, \bibinfo {author}
  {\bibfnamefont {R.~H.}\ \bibnamefont {Wechsler}}, \bibinfo {author}
  {\bibfnamefont {A.~A.}\ \bibnamefont {West}}, \bibinfo {author}
  {\bibfnamefont {O.}~\bibnamefont {Wiecha}}, \bibinfo {author} {\bibfnamefont
  {B.~F.}\ \bibnamefont {Williams}}, \bibinfo {author} {\bibfnamefont
  {B.}~\bibnamefont {Willman}}, \bibinfo {author} {\bibfnamefont
  {D.}~\bibnamefont {Wittman}}, \bibinfo {author} {\bibfnamefont {S.~C.}\
  \bibnamefont {Wolff}}, \bibinfo {author} {\bibfnamefont {W.~M.}\ \bibnamefont
  {{Wood-Vasey}}}, \bibinfo {author} {\bibfnamefont {P.}~\bibnamefont
  {Wozniak}}, \bibinfo {author} {\bibfnamefont {P.}~\bibnamefont {Young}},
  \bibinfo {author} {\bibfnamefont {A.}~\bibnamefont {Zentner}}, \ and\
  \bibinfo {author} {\bibfnamefont {H.}~\bibnamefont {Zhan}},\ }\href@noop {}
  {\bibfield  {journal} {\bibinfo  {journal} {ArXiv e-prints}\ }\textbf
  {\bibinfo {volume} {0912}},\ \bibinfo {pages} {arXiv:0912.0201} (\bibinfo
  {year} {2009})}\BibitemShut {NoStop}%
\bibitem [{\citenamefont {Faigler}\ and\ \citenamefont
  {Mazeh}(2011)}]{faigler_photometric_2011}%
  \BibitemOpen
  \bibfield  {author} {\bibinfo {author} {\bibfnamefont {S.}~\bibnamefont
  {Faigler}}\ and\ \bibinfo {author} {\bibfnamefont {T.}~\bibnamefont
  {Mazeh}},\ }\href {\doibase 10.1111/j.1365-2966.2011.19011.x} {\bibfield
  {journal} {\bibinfo  {journal} {Monthly Notices of the Royal Astronomical
  Society}\ }\textbf {\bibinfo {volume} {415}},\ \bibinfo {pages} {3921}
  (\bibinfo {year} {2011})}\BibitemShut {NoStop}%
\bibitem [{\citenamefont {Millholland}\ and\ \citenamefont
  {Laughlin}(2017)}]{millholland_supervised_2017}%
  \BibitemOpen
  \bibfield  {author} {\bibinfo {author} {\bibfnamefont {S.}~\bibnamefont
  {Millholland}}\ and\ \bibinfo {author} {\bibfnamefont {G.}~\bibnamefont
  {Laughlin}},\ }\href {\doibase 10.3847/1538-3881/aa7a0f} {\bibfield
  {journal} {\bibinfo  {journal} {The Astronomical Journal}\ }\textbf {\bibinfo
  {volume} {154}},\ \bibinfo {pages} {83} (\bibinfo {year} {2017})}\BibitemShut
  {NoStop}%
\bibitem [{\citenamefont {Kirkpatrick}\ \emph {et~al.}(2000)\citenamefont
  {Kirkpatrick}, \citenamefont {Reid}, \citenamefont {Liebert}, \citenamefont
  {Gizis}, \citenamefont {Burgasser}, \citenamefont {Monet}, \citenamefont
  {Dahn}, \citenamefont {Nelson},\ and\ \citenamefont
  {Williams}}]{kirkpatrick_67_2000}%
  \BibitemOpen
  \bibfield  {author} {\bibinfo {author} {\bibfnamefont {J.~D.}\ \bibnamefont
  {Kirkpatrick}}, \bibinfo {author} {\bibfnamefont {I.~N.}\ \bibnamefont
  {Reid}}, \bibinfo {author} {\bibfnamefont {J.}~\bibnamefont {Liebert}},
  \bibinfo {author} {\bibfnamefont {J.~E.}\ \bibnamefont {Gizis}}, \bibinfo
  {author} {\bibfnamefont {A.~J.}\ \bibnamefont {Burgasser}}, \bibinfo {author}
  {\bibfnamefont {D.~G.}\ \bibnamefont {Monet}}, \bibinfo {author}
  {\bibfnamefont {C.~C.}\ \bibnamefont {Dahn}}, \bibinfo {author}
  {\bibfnamefont {B.}~\bibnamefont {Nelson}}, \ and\ \bibinfo {author}
  {\bibfnamefont {R.~J.}\ \bibnamefont {Williams}},\ }\href {\doibase
  10.1086/301427} {\bibfield  {journal} {\bibinfo  {journal} {The Astronomical
  Journal}\ }\textbf {\bibinfo {volume} {120}},\ \bibinfo {pages} {447}
  (\bibinfo {year} {2000})}\BibitemShut {NoStop}%
\bibitem [{\citenamefont {Burgasser}\ \emph {et~al.}(2006)\citenamefont
  {Burgasser}, \citenamefont {Geballe}, \citenamefont {Leggett}, \citenamefont
  {Kirkpatrick},\ and\ \citenamefont {Golimowski}}]{burgasser_unified_2006}%
  \BibitemOpen
  \bibfield  {author} {\bibinfo {author} {\bibfnamefont {A.~J.}\ \bibnamefont
  {Burgasser}}, \bibinfo {author} {\bibfnamefont {T.~R.}\ \bibnamefont
  {Geballe}}, \bibinfo {author} {\bibfnamefont {S.~K.}\ \bibnamefont
  {Leggett}}, \bibinfo {author} {\bibfnamefont {J.~D.}\ \bibnamefont
  {Kirkpatrick}}, \ and\ \bibinfo {author} {\bibfnamefont {D.~A.}\ \bibnamefont
  {Golimowski}},\ }\href {\doibase 10.1086/498563} {\bibfield  {journal}
  {\bibinfo  {journal} {The Astrophysical Journal}\ }\textbf {\bibinfo {volume}
  {637}},\ \bibinfo {pages} {1067} (\bibinfo {year} {2006})}\BibitemShut
  {NoStop}%
\bibitem [{\citenamefont {Adams}\ and\ \citenamefont
  {Laughlin}(1997)}]{adams_dying_1997}%
  \BibitemOpen
  \bibfield  {author} {\bibinfo {author} {\bibfnamefont {F.~C.}\ \bibnamefont
  {Adams}}\ and\ \bibinfo {author} {\bibfnamefont {G.}~\bibnamefont
  {Laughlin}},\ }\href {\doibase 10.1103/RevModPhys.69.337} {\bibfield
  {journal} {\bibinfo  {journal} {Reviews of Modern Physics}\ }\textbf
  {\bibinfo {volume} {69}},\ \bibinfo {pages} {337} (\bibinfo {year}
  {1997})}\BibitemShut {NoStop}%
\bibitem [{\citenamefont {Bailes}\ \emph {et~al.}(2011)\citenamefont {Bailes},
  \citenamefont {Bates}, \citenamefont {Bhalerao}, \citenamefont {Bhat},
  \citenamefont {Burgay}, \citenamefont {{Burke-Spolaor}}, \citenamefont
  {D'Amico}, \citenamefont {Johnston}, \citenamefont {Keith}, \citenamefont
  {Kramer}, \citenamefont {Kulkarni}, \citenamefont {Levin}, \citenamefont
  {Lyne}, \citenamefont {Milia}, \citenamefont {Possenti}, \citenamefont
  {Spitler}, \citenamefont {Stappers},\ and\ \citenamefont {{van
  Straten}}}]{bailes_transformation_2011}%
  \BibitemOpen
  \bibfield  {author} {\bibinfo {author} {\bibfnamefont {M.}~\bibnamefont
  {Bailes}}, \bibinfo {author} {\bibfnamefont {S.~D.}\ \bibnamefont {Bates}},
  \bibinfo {author} {\bibfnamefont {V.}~\bibnamefont {Bhalerao}}, \bibinfo
  {author} {\bibfnamefont {N.~D.~R.}\ \bibnamefont {Bhat}}, \bibinfo {author}
  {\bibfnamefont {M.}~\bibnamefont {Burgay}}, \bibinfo {author} {\bibfnamefont
  {S.}~\bibnamefont {{Burke-Spolaor}}}, \bibinfo {author} {\bibfnamefont
  {N.}~\bibnamefont {D'Amico}}, \bibinfo {author} {\bibfnamefont
  {S.}~\bibnamefont {Johnston}}, \bibinfo {author} {\bibfnamefont {M.~J.}\
  \bibnamefont {Keith}}, \bibinfo {author} {\bibfnamefont {M.}~\bibnamefont
  {Kramer}}, \bibinfo {author} {\bibfnamefont {S.~R.}\ \bibnamefont
  {Kulkarni}}, \bibinfo {author} {\bibfnamefont {L.}~\bibnamefont {Levin}},
  \bibinfo {author} {\bibfnamefont {A.~G.}\ \bibnamefont {Lyne}}, \bibinfo
  {author} {\bibfnamefont {S.}~\bibnamefont {Milia}}, \bibinfo {author}
  {\bibfnamefont {A.}~\bibnamefont {Possenti}}, \bibinfo {author}
  {\bibfnamefont {L.}~\bibnamefont {Spitler}}, \bibinfo {author} {\bibfnamefont
  {B.}~\bibnamefont {Stappers}}, \ and\ \bibinfo {author} {\bibfnamefont
  {W.}~\bibnamefont {{van Straten}}},\ }\href {\doibase
  10.1126/science.1208890} {\bibfield  {journal} {\bibinfo  {journal}
  {Science}\ }\textbf {\bibinfo {volume} {333}},\ \bibinfo {pages} {1717}
  (\bibinfo {year} {2011})}\BibitemShut {NoStop}%
\bibitem [{\citenamefont {Nutzman}\ and\ \citenamefont
  {Charbonneau}(2008)}]{nutzman_design_2008}%
  \BibitemOpen
  \bibfield  {author} {\bibinfo {author} {\bibfnamefont {P.}~\bibnamefont
  {Nutzman}}\ and\ \bibinfo {author} {\bibfnamefont {D.}~\bibnamefont
  {Charbonneau}},\ }\href {\doibase 10.1086/533420} {\bibfield  {journal}
  {\bibinfo  {journal} {Publications of the Astronomical Society of the
  Pacific}\ }\textbf {\bibinfo {volume} {120}},\ \bibinfo {pages} {317}
  (\bibinfo {year} {2008})}\BibitemShut {NoStop}%
\bibitem [{\citenamefont {Berta}\ \emph {et~al.}(2012)\citenamefont {Berta},
  \citenamefont {Irwin}, \citenamefont {Charbonneau}, \citenamefont {Burke},\
  and\ \citenamefont {Falco}}]{berta_transit_2012}%
  \BibitemOpen
  \bibfield  {author} {\bibinfo {author} {\bibfnamefont {Z.~K.}\ \bibnamefont
  {Berta}}, \bibinfo {author} {\bibfnamefont {J.}~\bibnamefont {Irwin}},
  \bibinfo {author} {\bibfnamefont {D.}~\bibnamefont {Charbonneau}}, \bibinfo
  {author} {\bibfnamefont {C.~J.}\ \bibnamefont {Burke}}, \ and\ \bibinfo
  {author} {\bibfnamefont {E.~E.}\ \bibnamefont {Falco}},\ }\href {\doibase
  10.1088/0004-6256/144/5/145} {\bibfield  {journal} {\bibinfo  {journal} {The
  Astronomical Journal}\ }\textbf {\bibinfo {volume} {144}},\ \bibinfo {pages}
  {145} (\bibinfo {year} {2012})}\BibitemShut {NoStop}%
\bibitem [{\citenamefont {Hunter}(2007)}]{hunter_matplotlib_2007}%
  \BibitemOpen
  \bibfield  {author} {\bibinfo {author} {\bibfnamefont {J.~D.}\ \bibnamefont
  {Hunter}},\ }\href {\doibase 10.1109/MCSE.2007.55} {\bibfield  {journal}
  {\bibinfo  {journal} {Computing in Science Engineering}\ }\textbf {\bibinfo
  {volume} {9}},\ \bibinfo {pages} {90} (\bibinfo {year} {2007})}\BibitemShut
  {NoStop}%
\bibitem [{\citenamefont {Oliphant}(2007)}]{oliphant_python_2007}%
  \BibitemOpen
  \bibfield  {author} {\bibinfo {author} {\bibfnamefont {T.~E.}\ \bibnamefont
  {Oliphant}},\ }\href {\doibase 10.1109/MCSE.2007.58} {\bibfield  {journal}
  {\bibinfo  {journal} {Computing in Science and Engg.}\ }\textbf {\bibinfo
  {volume} {9}},\ \bibinfo {pages} {10} (\bibinfo {year} {2007})}\BibitemShut
  {NoStop}%
\bibitem [{\citenamefont {Chabrier}\ \emph {et~al.}(1992)\citenamefont
  {Chabrier}, \citenamefont {Saumon}, \citenamefont {Hubbard},\ and\
  \citenamefont {Lunine}}]{chabrier_molecular-metallic_1992}%
  \BibitemOpen
  \bibfield  {author} {\bibinfo {author} {\bibfnamefont {G.}~\bibnamefont
  {Chabrier}}, \bibinfo {author} {\bibfnamefont {D.}~\bibnamefont {Saumon}},
  \bibinfo {author} {\bibfnamefont {W.~B.}\ \bibnamefont {Hubbard}}, \ and\
  \bibinfo {author} {\bibfnamefont {J.~I.}\ \bibnamefont {Lunine}},\ }\href
  {\doibase 10.1086/171390} {\bibfield  {journal} {\bibinfo  {journal} {The
  Astrophysical Journal}\ }\textbf {\bibinfo {volume} {391}},\ \bibinfo {pages}
  {817} (\bibinfo {year} {1992})}\BibitemShut {NoStop}%
\bibitem [{\citenamefont {Knudson}\ \emph {et~al.}(2004)\citenamefont
  {Knudson}, \citenamefont {Hanson}, \citenamefont {Bailey}, \citenamefont
  {Hall}, \citenamefont {Asay},\ and\ \citenamefont
  {Deeney}}]{knudson_principal_2004}%
  \BibitemOpen
  \bibfield  {author} {\bibinfo {author} {\bibfnamefont {M.~D.}\ \bibnamefont
  {Knudson}}, \bibinfo {author} {\bibfnamefont {D.~L.}\ \bibnamefont {Hanson}},
  \bibinfo {author} {\bibfnamefont {J.~E.}\ \bibnamefont {Bailey}}, \bibinfo
  {author} {\bibfnamefont {C.~A.}\ \bibnamefont {Hall}}, \bibinfo {author}
  {\bibfnamefont {J.~R.}\ \bibnamefont {Asay}}, \ and\ \bibinfo {author}
  {\bibfnamefont {C.}~\bibnamefont {Deeney}},\ }\href {\doibase
  10.1103/PhysRevB.69.144209} {\bibfield  {journal} {\bibinfo  {journal}
  {Physical Review B}\ }\textbf {\bibinfo {volume} {69}},\ \bibinfo {pages}
  {144209} (\bibinfo {year} {2004})}\BibitemShut {NoStop}%
\bibitem [{\citenamefont {Morales}\ \emph {et~al.}(2010)\citenamefont
  {Morales}, \citenamefont {Pierleoni},\ and\ \citenamefont
  {Ceperley}}]{morales_equation_2010}%
  \BibitemOpen
  \bibfield  {author} {\bibinfo {author} {\bibfnamefont {M.~A.}\ \bibnamefont
  {Morales}}, \bibinfo {author} {\bibfnamefont {C.}~\bibnamefont {Pierleoni}},
  \ and\ \bibinfo {author} {\bibfnamefont {D.~M.}\ \bibnamefont {Ceperley}},\
  }\href {\doibase 10.1103/PhysRevE.81.021202} {\bibfield  {journal} {\bibinfo
  {journal} {Physical Review E}\ }\textbf {\bibinfo {volume} {81}},\ \bibinfo
  {pages} {021202} (\bibinfo {year} {2010})}\BibitemShut {NoStop}%
\bibitem [{\citenamefont {Knudson}\ and\ \citenamefont
  {Desjarlais}(2017)}]{knudson_high-precision_2017}%
  \BibitemOpen
  \bibfield  {author} {\bibinfo {author} {\bibfnamefont {M.~D.}\ \bibnamefont
  {Knudson}}\ and\ \bibinfo {author} {\bibfnamefont {M.~P.}\ \bibnamefont
  {Desjarlais}},\ }\href {\doibase 10.1103/PhysRevLett.118.035501} {\bibfield
  {journal} {\bibinfo  {journal} {Physical Review Letters}\ }\textbf {\bibinfo
  {volume} {118}},\ \bibinfo {pages} {035501} (\bibinfo {year}
  {2017})}\BibitemShut {NoStop}%
\bibitem [{\citenamefont {Fowler}\ \emph {et~al.}(1975)\citenamefont {Fowler},
  \citenamefont {Caughlan},\ and\ \citenamefont
  {Zimmerman}}]{fowler_thermonuclear_1975}%
  \BibitemOpen
  \bibfield  {author} {\bibinfo {author} {\bibfnamefont {W.~A.}\ \bibnamefont
  {Fowler}}, \bibinfo {author} {\bibfnamefont {G.~R.}\ \bibnamefont
  {Caughlan}}, \ and\ \bibinfo {author} {\bibfnamefont {B.~A.}\ \bibnamefont
  {Zimmerman}},\ }\href {\doibase 10.1146/annurev.aa.13.090175.000441}
  {\bibfield  {journal} {\bibinfo  {journal} {Annual Review of Astronomy and
  Astrophysics}\ }\textbf {\bibinfo {volume} {13}},\ \bibinfo {pages} {69}
  (\bibinfo {year} {1975})}\BibitemShut {NoStop}%
\bibitem [{\citenamefont {Graboske}\ \emph {et~al.}(1973)\citenamefont
  {Graboske}, \citenamefont {Dewitt}, \citenamefont {Grossman},\ and\
  \citenamefont {Cooper}}]{graboske_screening_1973}%
  \BibitemOpen
  \bibfield  {author} {\bibinfo {author} {\bibfnamefont {H.~C.}\ \bibnamefont
  {Graboske}}, \bibinfo {author} {\bibfnamefont {H.~E.}\ \bibnamefont
  {Dewitt}}, \bibinfo {author} {\bibfnamefont {A.~S.}\ \bibnamefont
  {Grossman}}, \ and\ \bibinfo {author} {\bibfnamefont {M.~S.}\ \bibnamefont
  {Cooper}},\ }\href {\doibase 10.1086/152062} {\bibfield  {journal} {\bibinfo
  {journal} {The Astrophysical Journal}\ }\textbf {\bibinfo {volume} {181}},\
  \bibinfo {pages} {457} (\bibinfo {year} {1973})}\BibitemShut {NoStop}%
\end{thebibliography}%

\appendix

\section{Analytic model of a low-mass star}

\subsection{Surface Luminosity}
The surface luminosity can be written $L_S= 4\pi R^2 \sigma T_\mathrm{eff}^4$, where $\sigma$ is the Stefan-Boltzmann constant, $R$ is the stellar radius, and $T_\mathrm{eff}$ is the effective temperature. Using a polytropic equation of state where the normalization is set accounting for the degeneracy pressure, \citet{auddy_analytic_2016} arrive at the relation
\begin{equation}
	R = 2.80858 \times 10^9\ \mathrm{cm} (M/M_\odot)^{-1/3} \mu_e^{-5/3} (1+\gamma+\alpha\psi),
\end{equation}
where $M$ is the total mass of the star, $1/\mu_e = X+Y/2$ is the number of nucleons per electron, with $X$ and $Y$ defined as the mass fractions of hydrogen and helium respectively, and $(1+\gamma+\alpha\psi)$ is a correction factor accounting for degeneracy. In particular $\psi$ is the ratio of the characteristic thermal energy per particle $k_B T$, the Boltzmann constant times the temperature, to the electron Fermi energy $\mu_F$. This is multiplied by the ratio $\alpha=5\mu_e/2\mu_1$, where $\mu_1 = ((1+x_{\mathrm{H}^+})X+Y/4)^{-1}$ is the mean molecular weight for a gas with an ionization fraction $x_{\mathrm{H}^+}$. Finally, $\gamma$ is a function of $\psi$
\begin{equation}
	\gamma = -\frac5{16}\psi \ln(1+e^{-1/\psi}) + \frac{15}8 \psi^2\left( \frac{\pi^2}{3} + \mathrm{Li}_2(-e^{-1/\psi}) \right),
\end{equation}
where $\mathrm{Li}_2$ is the polylogarithm function of order 2.

The effective temperature is derived by assuming that the photosphere lies at an optical depth of $\tau=2/3$ with a constant Rosseland mean opacity, $\kappa_R$. Combining this approximation with a precscribed entropy jump between the core and the surface from \citet{chabrier_molecular-metallic_1992} yields an effective pressure from which one can back out an effective density and temperature. Note that \citet{auddy_analytic_2016} use several different parameters to describe the jump in entropy, including a case with no change. This case is probably closest to what we observe in the MESA simulations, and is consistent with the modern result that the phase transition as postulated by \citet{chabrier_molecular-metallic_1992} does not occur \citep{knudson_principal_2004, morales_equation_2010, knudson_high-precision_2017}, but for simplicity and ease of comparison, we follow the default choice made by \citet{auddy_analytic_2016} in this discussion. For their fiducial case, we get
\begin{equation}
\label{eq:rho}
\left(\frac{\rho}{1\ \mathrm{g}\ \mathrm{cm}^{-3}}\right)^{1.4} = \frac{6.89811}{\kappa_R N_A k_B} \left( \frac{M}{M_\odot} \right)^{5/3} \frac{\mu_e^{10/3} \mu_2}{(1+\gamma+\alpha\psi)^2 \psi^{1.58}},
\end{equation}
where dimensional quantities on the right hand side are all evaluated in CGS units, $1/\mu_2 = X/2 + Y/4$, and
\begin{equation}
	T_\mathrm{eff} =2.57881 \times 10^4\ \mathrm{K}\ \kappa_R^{-2/7} \left(\frac{M}{M_\odot}\right)^{10/21} \frac{\mu_e^{20/21}\mu_2^{2/7}\psi^{8/7}}{(1+\gamma+\alpha\psi)^{4/7}},
	\label{eq:Teff}
\end{equation}
again with $\kappa_R$ in CGS units. These equations yield an expression\footnote{\citet{auddy_analytic_2016} evaluate $\mu_e$ at a particular value and absorb the result into the leading constant, but we retain it since it changes along with metallicity.} for $L_S$ as a function of $\kappa_R$, $\psi$, $M$, $X$, and $Y$,
\begin{equation}
	L_S = 0.6118 L_\odot \kappa_R^{-8/7} (M/M_\odot)^{26/21} \frac{\psi^{32/7}\mu_e^{10/21}\mu_2^{8/7}}{(1+\gamma+\alpha\psi)^{2/7}} .
\end{equation}
Given a particular value of $Z$, we assume that $Y=Y_p+f Z$, where $Y_p=0.24$ and $f=2$, leaving $X=1-Y-Z$. For now we leave $M$ and $\psi$ as free parameters. The opacity, however, is itself a function of density, temperature, and composition, i.e.
\begin{equation}
\kappa_R = \kappa_R(\rho,T,X,Z).
\label{eq:kappa}
\end{equation}
In the analytic models, $\kappa_R$ is left fixed, but here we attempt to make the model more self-consistent by iteratively searching for a value of $\kappa_R$ such that equations \eqref{eq:rho}, \eqref{eq:Teff}, and \eqref{eq:kappa} are simultaneously satisfied at the surface of the star. If the functional form of the opacity is a power-law in $\rho$ and $T$, these equations may be solved analytically. We iterate numerically instead in order to accomodate more general opacity laws. For consistency with the MESA models we use the tables from \citet{ferguson_low-temperature_2005} as well as simpler powerlaw scalings.

\subsection{Nuclear Luminoisty}
The surface luminosity must be balanced by the nuclear luminosity $L_N$ in a steady state. According to \citet{burrows_science_1993} and \citet{auddy_analytic_2016}, low-mass stars do not reach sufficient temperatures to burn all the way through the PPI chain of nuclear reactions, in particular because the $^3$He nuclei face a much higher Coulomb barrier. The first two reactions in PPI are $p + p \rightarrow d + e^+ + \bar{\nu}_e$ and $p + d \rightarrow ^3\mathrm{He} + \gamma$. The first reaction limits the rate, while the second takes place quickly and produces roughly 5 times the energy per reaction. 

The energy generation rate of each reaction per unit mass is an exponential function of temperature, namely \citep{fowler_thermonuclear_1975}
\begin{equation}
\dot{\epsilon}_{pp} = 2.5 \times 10^6 (\rho X^2 / T_6^{2/3} ) e^{-33.8 / T_6^{1/3}}\ \mathrm{erg}\ \mathrm{g}^{-1}\ \mathrm{s}^{-1},
\end{equation}
\begin{equation}
\dot{\epsilon}_{pd} = 1.4 \times 10^{24} (\rho XD / T_6^{2/3} ) e^{-37.2 / T_6^{1/3}}\ \mathrm{erg}\ \mathrm{g}^{-1}\ \mathrm{s}^{-1},
\end{equation}
where the temperature normalization is $T_6=T/(10^6\ \mathrm{K})$, and the deuterium fraction by mass is $D$. The fully-convective piecewise-constant-entropy nature of low-mass stars implies that the total nuclear reaction luminosity is \citep{burrows_science_1993},
\begin{equation}
L_N = \frac{2.4 \dot{\epsilon}_c }{(3u/2 + s)^{3/2}}(M/M_\odot) L_\odot
\end{equation}
where $\dot{\epsilon}_c$ is the energy generation rate per unit mass at the center of the star in CGS units. In deriving this equation, the energy generation rate in the innermmost part of the star has been integrated under the assumption that it may be approximated as a power law, $\dot{\epsilon}\propto T^s \rho^{u-1}$. \citet{burrows_science_1993} estimate that in this regime $u = 2+1.29\times H(0)/3$ and $s=33.8/(3 T_6^{1/3}) - 2/3 - 1.29\times H(0)$, where $H(0)\approx \min(0.977\Gamma^{1.29}, 1.06 \Gamma)$ is the log of the screening factor from \citet{graboske_screening_1973}, and $\Gamma = 0.227 (\rho/\mu_e)^{1/3}/T_6$ (so long as we have specialized to protons and deuterons for which the charge is 1).

To estimate $L_N$ we now need to know the value of $\dot{\epsilon}_c$. For this we simply take $\dot{\epsilon}_{pp} + \dot{\epsilon}_{pd}$, but to evaluate these we need the central temperature and density, and the hydrogen and deuterium abundances. For the former, we use the relevant equations from \citet{auddy_analytic_2016}, 
\begin{equation}
T_c = 7.68097 \times 10^8 (M/M_\odot)^{4/3} \psi \mu_e^{8/3} (1+\gamma+\alpha\psi)^{-2} \mathrm{K} ,
\end{equation}
\begin{equation}
\rho_c = 1.28412\times 10^5 (M/M_\odot)^2 \mu_e^5 (1+\gamma +\alpha\psi)^3 \mathrm{g}\ \mathrm{cm}^{-3}.
\end{equation}
The deuterium fraction is less obvious. In the long-run steady state, the $pp$ and $pd$ reactions are responsible for the continuous creation and destruction of deuterium. The equilibrium value of deuterium abundance is taken to be
\begin{equation}
D_\mathrm{eq} = 1.79 \times 10^{-18} X e^{3.4/T_6^{1/3}}  \frac{Q_{pd}}{Q_pp}
\end{equation}
This equation sets the rate of deuterium formation equal to the rate of destruction, where the reaction rates are taken to be equal to the energy generation rates divided by the energy produced per reaction, with $Q_{pd} = 5.494\ \mathrm{MeV}$ and $Q_{pp}=1.18\ \mathrm{MeV}$. The initial deuterium fraction, related to the primordial abundance, or at least the somewhat lower ISM abundance, is much higher than $D_\mathrm{eq}$, but this initial supply of deuterium is burned early in the life of the star or brown dwarf, and is unlikely to affects the long-run evolution relevant in this work.

\clearpage

\end{document}